\documentclass[onecolumn]{elsart3p}

\usepackage{graphics}
\usepackage{epsfig}
\usepackage{graphicx}
\usepackage[intlimits,fleqn]{amsmath}
\usepackage{amsfonts}
\usepackage{amssymb}
\usepackage{amscd}
\usepackage{latexsym}
\usepackage{bbm}
\usepackage[sort&compress]{natbib}
\usepackage{ifthen}
\usepackage{dcolumn}      
\usepackage{bm}           
\usepackage{psfig,pstricks,pst-grad,fancybox,graphics}
\usepackage{enumerate}
\usepackage[latin1]{inputenc} 

\newcommand{\ba}{\begin{eqnarray}}
\newcommand{\ea}{\end{eqnarray}}
\newcommand{\ban}{\begin{eqnarray*}}
\newcommand{\ean}{\end{eqnarray*}}

\newcommand{\braket}[2]{\mbox{$ \langle #1 | #2 \rangle $}}

\newcommand{\ket}[1]{\mbox{$ | #1 \rangle $}}
\newcommand{\bra}[1]{\mbox{$ \langle #1 | $}}

\newcommand{\one}{\leavevmode\hbox{\small1\normalsize\kern-.33em1}}

\newcommand{\beq}{\begin{equation}}
\newcommand{\eeq}{\end{equation}}
\newcommand{\beqa}{\begin{eqnarray}}
\newcommand{\eeqa}{\end{eqnarray}}

\newcommand{\Ket}[1]{\ensuremath{|#1\rangle}}
\newcommand{\BraKet}[2]{\ensuremath{\langle #1|#2\rangle}}
\newcommand{\KetBra}[1]{\ensuremath{| #1 \rangle \langle #1 |}}
\newcommand{\ketbra}[1]{\ensuremath{| #1 \rangle \langle #1 |}}

\newcommand{\Eins}{\ensuremath{\mathbbm 1}}
\newcommand{\eins}{\ensuremath{\mathbbm 1}}
\newcommand{\HH}{\ensuremath{\mathcal{H}}}
\newcommand{\BB}{\ensuremath{\mathcal{B}}}
\newcommand{\WW}{\ensuremath{\mathcal{W}}}
\newcommand{\VV}{\ensuremath{\mathcal{V}}}
\newcommand{\PP}{\ensuremath{\mathcal{P}}}
\newcommand{\QQ}{\ensuremath{\mathcal{Q}}}

\newcommand{\BE}{\begin{equation}}
\newcommand{\EE}{\end{equation}}
\newcommand{\be}{\begin{equation}}
\newcommand{\ee}{\end{equation}}
\newcommand{\bea}{\begin{eqnarray}}
\newcommand{\eea}{\end{eqnarray}}
\newcommand{\kommentar}[1]{}

\newcommand{\mean}[1]{\ensuremath{\langle #1 \rangle}}
\newcommand{\vr}{\ensuremath{\varrho}}

\newcommand{\exs}[1]{\ensuremath{\langle{#1}\rangle}}

\newcommand{\va}[1]{\ensuremath{(\Delta#1)^2}}
\newcommand{\ex}[1]{\ensuremath{\left\langle{#1}\right\rangle}}
\newcommand{\EQ}[1]{Eq.~(\ref{#1})}

\newcommand{\mf}[1]{\mathfrak{#1}}
\newcommand{\qqed}{\ensuremath{\hfill \Box}}
\newcommand{\II}{\ensuremath{\mathbbm I}}

\newcommand{\MM}{\mathcal{M}}

\newcommand{\openone}{\ensuremath{\mathbbm 1}}
\newcommand{\WGHZN}{\ensuremath{\mathcal{W}^{(GHZ_N)}}}

\newcommand{\WGHZNPRIME}{\ensuremath{\widehat{\mathcal{W}}^{(GHZ_N)}}}

\newcommand{\WCN}{\ensuremath{\mathcal{W}^{(C_N)}}}
\newcommand{\WCNPRIME}{\ensuremath{\widehat{\mathcal{W}}^{(C_N)}}}

\newcommand{\NN}{\mathcal{N}}

\bibpunct{[}{]}{,}{n}{}{}

\begin{document}

\begin{frontmatter}

\title{Entanglement detection}
\vspace{1cm}
\author{Otfried G{\"u}hne}
\address{Institut f\"ur Quantenoptik und Quanteninformation,
\"Osterreichische Akademie der Wissenschaften, \\
Technikerstra{\ss}e 21A, A-6020 Innsbruck, Austria}
\address{Institut f\"ur theoretische Physik, Universit\"at Innsbruck,
Technikerstra{\ss}e 25,
A-6020 Innsbruck, Austria}
\author{G\'eza T\'oth}
\address{Department of Theoretical Physics, 
The University of the Basque Country, P.O. Box 644, E-48080 Bilbao,
Spain}
\address{Ikerbasque-Basque Foundation for Science, Alameda Urquijo 36, E-48011 Bilbao, Spain}
\address{ICFO-The Institute of Photonic Sciences, Mediterranean Technology Park, E-08860
Castelldefels (Barcelona), Spain}
\address{Research Institute for Solid State Physics and Optics,
Hungarian Academy of Sciences, \\P.O. Box 49, H-1525 Budapest,
Hungary}

\begin{abstract}
How can one prove that a given state is entangled? In this paper we
review different methods that have been proposed for entanglement
detection. We first explain the basic elements of entanglement
theory for two or more particles and then entanglement verification
procedures such as Bell inequalities, entanglement witnesses, the
determination of nonlinear  properties of a quantum state via
measurements on several copies, and spin squeezing inequalities. An
emphasis is given to the theory and application of entanglement
witnesses. We also discuss several experiments, where  some  of the
presented methods have been implemented.
\end{abstract}



\end{frontmatter}

\tableofcontents

\section{Introduction}

Entanglement was first described by Einstein, Podolsky, and Rosen
\cite{PhysRev.47.777} and Schr\"odinger \cite{schroedingerkatze}
as a strange phenomenon of quantum mechanics, questioning the
completeness of the theory. Later, Bell recognized that entanglement
leads to experimentally testable deviations of quantum mechanics
from classical physics \cite{Bell64}. Finally, with the advent of
quantum information theory, entanglement was recognized as a resource,
enabling tasks like quantum cryptography \cite{PhysRevLett.67.661}, quantum
teleportation \cite{PhysRevLett.70.1895} or measurement based quantum
computation \cite{PhysRevLett.86.5188}. Together with the rapid experimental
progress on quantum control, this lead to a rapidly growing interest in
entanglement theory and many experiments nowadays aim at the generation
of entanglement.

Indeed, in the last years an enormous progress on the generation of
entanglement has been achieved. For instance, six or eight ions have
been entangled \cite{leibfriedsixghz,haeffner-2005-438}, photons
have been used to demonstrate entanglement between six particles
or ten qubits \cite{lu-2007-3,gao-2008} and in diamond nuclear and
electronic spins have been entangled \cite{wrachtrupscience}. These
systems allow for individual addressing of the parties. In other
types of systems only collective measurements are possible. Through
spin squeezing, entanglement of $10^7$ atoms were created in cold
atomic clouds \cite{PhysRevLett.83.1319} and large scale entangling
operations were realized in optical lattices of $10^5$ two-state
atoms \cite{mandel2003ccm}. As the underlying
techniques of quantum control improve continuously, it can be
expected that in the near future even larger systems can be
entangled.

In any of these experiments, typical questions arise: How can one
be sure that entanglement was indeed produced? How can one detect
the presence of entanglement? Can we quantify the entanglement in
the experiment?  These questions are difficult to answer and many
possible ways to tackle this problem have been proposed. These range
from Bell inequalities, entanglement witnesses and spin squeezing
inequalities to entropic inequalities, the measurement of nonlinear
properties of the quantum state and the approximation of positive maps.

In general, one could imagine several desirable properties for an
entanglement verification procedure. First, a crucial requirement is
that the scheme under consideration must be easy to implement.
Depending on the type of experiment, some measurements are easy to
carry out, and some are not. In experiments using photons, the count
rates are often not very high, making it impossible to characterize
the experimental state completely. Then, one has to find the
measurements that allow to conclude much about the entanglement
content of the given state. In addition, the scheme should be robust
against noise, and it should also detect weakly entangled states.

Second, most experiments nowadays aim at the generation of
entanglement between more than two particles and for this case
different entanglement classes exist. Therefore, an entanglement
detection scheme has to be capable of distinguishing between these
different classes. In an $N$-qubit  experiment it is not sufficient
to say that some qubits are entangled, one has to prove that all of the  $N$
qubits are entangled.

Third, the conclusion that the experimentally generated state was entangled
should not depend on some assumptions about the state. Most experiments aim
at the generation of a special state, and clearly one can use this to design
appropriate measurements. But the final conclusion that the state was 
entangled,
must not rely on assumptions concerning a special form of the 
state or its purity.

In this paper, we review different methods for the characterization
of entanglement in experiments. This includes the mere detection of
entanglement, but also its experimental quantification, the
estimation of the state fidelity and the characterization whether a
prepared state is useful for some task or not. We also explain
several experiments,  where these methods have been used. Due to
their higher experimental relevance, we especially explain methods
that can be used for the characterization of entanglement between
more than two particles. A special emphasis is given on the method
of entanglement witnesses, as they are the most frequently used tool
nowadays, moreover, they have also applications in other areas of
quantum information theory.

Throughout this paper we restrict our attention to the case of
discrete systems; mainly we consider the case of qubits. It should
be stressed that entanglement in continuous variable systems like
harmonic oscillators or light modes is also under intensive
research. However, as the underlying theory of Gaussian states is
significantly different from the case of qubits, we have not
addressed these systems here. Moreover, for the entanglement theory
of continuous variable systems already several good reviews exist
\cite{braunstein:513,eisert-2003-1,adesso-2007-40,wang-2007-448}.

As entanglement is a central topic in quantum information theory,
there are some other excellent review articles that are related to
some questions addressed here. For the case of bipartite
entanglement, a recent exhaustive review was written by the
Horodecki family \cite{horodecki-2007} and entanglement measures
have been reviewed in detail by Virmani and Plenio
\cite{plenio-2005-1}. Some older proposals for entanglement
detection schemes have been summarized by Terhal
\cite{terhal-2002-287} and a survey about conceptual differences and
possible problems of entanglement verification schemes has been
given by van Enk, L\"utkenhaus and Kimble \cite{vanenk-2007-75}.
Moreover, several other overview articles concerning bipartite and
multipartite entanglement
\cite{horodecki-2001,bruss-2002-43,bruss-2002-49,eisert-2006,amico:517},
entanglement measures \cite{mintertreview,horoqicreview} or Bell
inequalities \cite{PeresConj,wernerwolfbell,genovese-2005-413} have
been published. Concerning the experimental techniques, photon
experiments have been reviewed 
in Refs.~\cite{pan-2008,kok:135,tittel-2001} and ion trap experiments in
Refs.~\cite{RevModPhys.73.565,RevModPhys.75.281,blattnature,eschnervarenna,haeffner-2008}.

This review is organized as follows: The Sections 2 - 4 give an 
overview about theoretical aspects of entanglement, while the 
Sections 5 - 8 explain different entanglement detection methods 
in detail. 

In Section 2, we give an introduction into entanglement theory 
of bipartite systems. We introduce the notions of entanglement 
and separability, explain
several separability criteria and the phenomenon of bound
entanglement. We also introduce entanglement witnesses as a
theoretical concept and explain how to construct and optimize them.
In Section 3, we discuss entanglement between more than two
particles. We explain the different classes of multipartite
entanglement and introduce the most relevant families of
multiparticle entangled states, such as GHZ states, cluster states
or Dicke states. In Section 4, we introduce entanglement measures.
We explain their basic properties and give the definitions of the
most popular measures.

In Section 5, we review Bell inequalities as the oldest tool to
verify entanglement. We explain how Bell inequalities bound the set
of correlations originating from local hidden variable models 
and how the violation of a
Bell inequality implies the quantum phenomenon of entanglement. We
also explain parametric down-conversion as a process to generate
entangled photons, leading to a violation of Bell inequalities and
also an experiment, where this process has been used to demonstrate
violation of a Bell inequality under locality conditions.

In Section 6, we review entanglement witnesses as a tool for the
detection of entanglement. We show how entanglement witnesses can be
implemented with simple local measurements and that they are
especially useful for the detection of multipartite entanglement. We
also explain several experiments, where witnesses have been used in
various experimental situations. Finally, we explain how
entanglement witnesses can be useful in other parts of quantum
information theory, such as quantum cryptography.

In Section 7, we review other entanglement detection schemes that
have been implemented or proposed. This includes entropic
inequalities, estimates of the concurrence and positive maps via
measurements on several copies, variance based criteria, and
nonlinear entanglement witnesses.

Finally, in Section 8 we give an overview over entanglement
detection with collective observables. This concerns spin squeezing
inequalities and optical lattices,  but also entanglement
verification schemes in spin models, where global observables such
as the magnetic susceptibility of the energy can be used to derive
statements about the entanglement in the system.


\section{Bipartite entanglement}

In this Section, we will explain the basic notions of bipartite
entanglement. As many experiments nowadays aim at the generation of
multiparticle entangled states, we will explain the  theory of
multipartite entanglement in the following section. However, the
study of bipartite entangled states will already enable us to
introduce the central concepts of entanglement detection.

\subsection{Entanglement of pure states}
\label{subsec_purestates}

Let us assume that we are given two quantum systems. The first one
is owned by one physicist, called Alice, and the second one by
another one, called Bob. The physical states of Alice's system may
be described by states in a Hilbert space $\HH_A$ of dimension
$d_A$, and in Bob's system in a Hilbert space $\HH_B$ of dimension
$d_B.$ The composite system of both parties is then described by
vectors in the tensor-product of the two spaces $\HH =
\HH_A\otimes\HH_B.$ Thus, any vector in $\HH_A \otimes \HH_B$ can be
written as
\begin{equation}
\ket{\psi} = \sum_{i,j=1}^{d_A,d_B} c_{ij} \ket{a_i} \otimes \ket{b_j}
\in \HH_A\otimes\HH_B,
\label{be1}
\end{equation}
with a complex $d_A \times d_B$ matrix $C=(c_{ij}).$ To keep the
notation simple, we often write tensor products of vectors as
$\ket{a} \otimes \ket{b} \equiv \ket{a}\ket{b} \equiv \ket{ab}.$ Now
one can define separability and entanglement for pure states.

{\bf Definition 1 (Entanglement for pure states).}
A pure state $\ket{\psi} \in \HH$ is called a {\it product state}
or {\it separable} if we can find states $\ket{\phi^A}\in \HH_A$
and $\ket{\phi^B}\in \HH_B$ such that
\begin{equation}
\ket{\psi}=\ket{\phi^A}\otimes \ket{\phi^B}
\label{be2}
\end{equation}
holds. Otherwise the state $\ket{\psi}$ is called {\it entangled.}

Physically, the definition of product states means that
the state is uncorrelated. A product state can thus easily
be prepared in a local way: Alice produces the state
$\ket{\phi^A}$ and Bob produces independently the state
$\ket{\phi^B}.$ If Alice measures any observable $A$ and
Bob measures $B,$ then the probabilities of the different
outcomes factorize. Thus, the measurement outcomes for
Alice do not depend on the outcomes on Bob's side.

Before proceeding to the definition of entanglement for
mixed states we shall mention a very useful tool in the
description of entanglement for bipartite systems. This
is the so-called {\it Schmidt decomposition.}

{\bf Lemma 2 (Schmidt decomposition).} Let
$\ket{\psi} = \sum_{i,j=1}^{d_A,d_B} c_{ij} \ket{a_i b_j}
\in \HH_A\otimes\HH_B$ be a vector in the tensor product
of two Hilbert spaces. Then there exists an orthonormal
basis $\ket{\alpha_i}$ of $\HH_A$ and an orthonormal
basis $\ket{\beta_j}$ of $\HH_B$ such that
\begin{equation}
\ket{\psi}=\sum_{k=1}^R \lambda_k \ket{\alpha_k \beta_k}
\label{be3}
\end{equation}
holds, with positive real coefficients $\lambda_k.$ The $\lambda_k$
are uniquely determined as the square roots of the eigenvalues of
the matrix $CC^\dagger,$ where $C=(c_{ij})$ is the matrix formed by
the coefficients in Eq.~(\ref{be1}). The number $R \leq \min\{d_A,
d_B\}$ is called the Schmidt rank of $\ket{\psi}.$  If the
$\lambda_k$ are pairwise different, then also the $\ket{\alpha_k}$
and $\ket{\beta_k}$ are unique up to a phase.

The proof of this statement can be found in many textbooks
\cite{peresbook, brussbook, nielsenbook, vedralbook}. Note that pure
product states correspond to states of Schmidt rank one.
Therefore, the Schmidt decomposition
can be used to decide whether a given pure state is entangled
or separable.

\subsection{Entanglement of mixed states}

In a more general situation one does not know the exact
state of a quantum system. It is only known that it is,
with some probabilities $p_i,$ in one of some states
$\ket{\phi_i}\in \HH.$ This situation is described by
a density matrix
\begin{equation}
\varrho=\sum_i p_i \ketbra{\phi_i}, \;\;\;
\mbox{ with } \;\;\;  \sum_i p_i = 1 \;\;\;
\mbox{ and } \;\;\; p_i \geq 0.
\label{ms1}
\end{equation}
In a given basis this density matrix or state is
represented by a complex matrix. This matrix is
positive semidefinite\footnote{A hermitian matrix $M$ is called positive semidefinite
($M \geq 0$), iff the eigenvalues are non negative. Equivalently,
$M \geq 0$ if $\bra{\psi}M \ket{\psi}\geq 0$ for all $\ket{\psi}.$}
and hermitian, since all the operators $\ketbra{\phi_i}$ are
positive and hermitian. Due to the condition on the $p_i$
also $Tr(\vr)=1$ holds. Conversely, any positive semidefinite
matrix of trace one can be interpreted as a density matrix of
some state. This leads to a geometrical picture of the set of
all states as a convex set. This means that given two states
$\vr_1$ and $\vr_2,$ their convex combination
$\vr = \alpha \vr_1 + (1-\alpha) \vr_2$ with $\alpha \in [0;1]$
is again a state. This property holds also for combinations of
more than two states. Given some $p_i \geq 0$ with $\sum_i p_i = 1$
then the convex combination $\sum_i p_i \vr_i$ of some states
is again a state. We call coefficients $p_i\geq 0$ with the
property $\sum_i p_i = 1$ often {\it convex weights.}

Now one can define separability and entanglement for mixed states
according to Ref.~\cite{werner89}. The idea is the same as for
the pure state case: A state is separable, if it can be produced
locally, otherwise it is entangled.

{\bf Definition 3 (Entanglement for mixed states).}
Let $\vr$ be a density matrix for a composite system. We say
that $\vr$ is a {\it product state} if there exist states
$\vr^A$ for Alice and $\vr^B$ for Bob such that
\begin{equation}
\vr=\vr^A \otimes \vr^B.
\label{ms2}
\end{equation}
The state is called {\it separable}, if there are convex weights $p_i$
and product states $\vr^A_i \otimes \vr^B_i$ such that
\begin{equation}
\vr=\sum_i p_i \vr^A_i \otimes \vr^B_i
\label{ms3}
\end{equation}
holds. Otherwise the state is called {\it entangled.}

\begin{figure}[t]
\centerline{
\includegraphics[width=0.9\columnwidth]{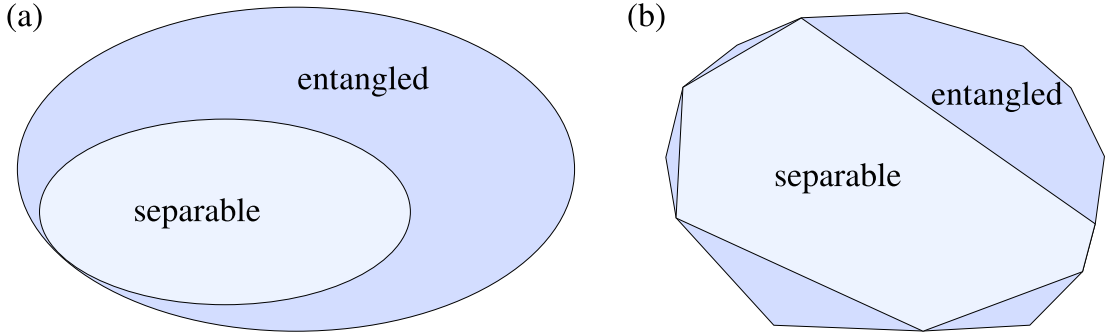}
}
\caption{(a) Schematic picture of the set of all states as a convex
set and the set of separable states as a convex subset. (b) Different 
schematic picture of the same sets. Here, it is stressed that the 
extremal points of the separable states (the pure product states),
are also extremal points of the set of all states,
hence they are located on the border of the total set.
\label{eierbild1}
}
\end{figure}

Physically, this definition discriminates between three scenarios.
First, a product state is an uncorrelated state, where Alice and Bob
own each a separate state. For non-product states there are two
different kinds of correlations. Separable states are classically
correlated. This means that for the production of a separable state
only local operations and classical communication (LOCC) are
necessary. Alice and Bob can, by classical communication, share a
random number generator that produces the outcomes $i$ with
probabilities $p_i.$ For each of the outcomes, they can agree to
produce the state $\vr^A_i \otimes \vr^B_i$ locally. By this
procedure they  produce the state $\vr=\sum_i p_i \vr^A_i \otimes
\vr^B_i.$ This procedure is not specific for quantum theory, which
justifies the notion of {\it classical} correlations.  Otherwise, if
a state is entangled, the correlations cannot originate from the
classical procedure described above. In this sense entangled states
are a typical feature of quantum mechanics.

For our later discussion it is very important to note that the 
set of separable states is a convex set. This is clear from the
definition of separability, obviously a convex combination of 
two separable states is again separable. Furthermore, the definition 
of separability implies that any separable state can be written as 
a convex combination of pure product states. Thus, the set of
separable states is the so called convex hull of the pure product
states. Further, any separable state can be written as a convex 
combination of maximally $d_A^2 d_B^2$ pure product states.  This 
follows from Carath\'eodory's theorem, which states that any element 
of a $d$-dimensional convex set can be written as a convex combination 
of $d+1$ extremal points of this set \cite{carate}. For the special case 
of two qubits, however, this bound can be improved and  any separable 
state can be written as a convex combination of four product states only 
\cite{wootters98,sanpera-1998-58}. The set of separable states is also shown in 
Fig.~\ref{eierbild1}.\footnote{A quantitative study on the shape of these
sets in space where the coordinates are the density matrix elements
is given in Ref.~\cite{verstraete-2002-49}.}

Given the definition of entanglement and separability, it is very
natural to ask whether a given density matrix is separable or
entangled. This is the so-called {\it separability problem.} There
are several criteria known that imply separability or entanglement
of a state. However, up to now, no general solution for the
separability problem is known.

\subsection{Separability criteria}
\label{criteria}

In this Section, we will present some criteria for bipartite
entanglement. As it is not our aim to discuss all of them in
detail, we focus our discussion on the most important ones.

\subsubsection{The PPT criterion}
Let us start with the criterion of the {\it partial transposition.}
In order to formulate this, note that we can expand any density
matrix of a composite quantum system in a chosen product basis as
\begin{equation}
\vr = \sum_{i,j}^N \sum_{k,l}^M \vr_{ij,kl}
\ket{i}\bra{j}\otimes\ket{k}\bra{l}.
\end{equation}
Given this decomposition, one defines the partial transposition of
$\vr$ as the transposition with respect to one subsystem. Thus,
there are two partial transpositions: The partial transposition with
respect to Alice is given by
\begin{equation}
\vr^{T_A} = \sum_{i,j}^N \sum_{k,l}^M \vr_{ji,kl}
\ket{i}\bra{j}\otimes\ket{k}\bra{l}
\end{equation}
and similarly we can define $\vr^{T_B}$ by exchanging $k$ and $l$
instead of the $i$ and $j.$ Note that the partial transposition
is related to the usual transposition by $\vr^T=(\vr^{T_A})^{T_B}$
and thus $\vr^{T_B}=(\vr^{T_A})^{T}.$

It is worth mentioning that the partial transposition depends
on the product basis in which it is performed. But one can show that
its spectrum does not depend on the basis.\footnote{Note that this
also holds for the full transposition.} We say a density matrix
$\vr$ has a {\it positive partial transpose} (or: the matrix is PPT)
if its partial transposition has no negative eigenvalues, {i.e.},
it is positive semidefinite: \be \vr^{T_A}\geq 0 \Leftrightarrow
\vr^{T_B}\geq 0. \ee If a matrix is not PPT, we call it NPT. Now we
can formulate the PPT criterion (also called Peres-Horodecki
criterion), originally introduced in Ref.~\cite{peresppt}.

{\bf Theorem 4 (PPT Criterion).} Let $\vr$ be a bipartite
separable state. Then $\vr$ is PPT.

{\it Proof.} This  fact follows directly from the definition
of separability in Eq.~(\ref{ms3}) since for a separable
$\vr=\sum_k p_k \vr^A_k\otimes \vr^B_k$ we have
$\vr^{T_A}=\sum_k p_k (\vr^A_k)^T \otimes \vr^B_k  =
\sum_k p_k \tilde{\vr}^A_k \otimes \vr^B_k \geq 0.$
\qqed

This theorem provides a very strong criterion for the detection of
entanglement. For a given density matrix one can easily calculate
the partial transpose and compute its spectrum. If one finds negative
eigenvalues one can conclude that the state is entangled. Given this
result, the question arises if this criterion is also sufficient for
separability, {i.e.},  whether $\vr^{T_A}\geq 0$ implies
separability. As it was shown already shortly after the discovery
of the PPT criterion, this is the case only in low dimensional
systems:

{\bf Theorem 5 (Horodecki Theorem).}
If $\vr$ is a state in a $ 2 \times 2$ or $ 2 \times 3$
system, then $\vr^{T_A}\geq 0$ implies that $\vr$ is
separable. In other dimensions this is not the case.

{\it Proof.} For the proof of the $ 2 \times 2$ or $ 2 \times 3$
case, see Ref.~\cite{horoppt}. For the first counterexample in a $ 2
\times 4$- or $3 \times 3$-system see Ref.~\cite{horobound}. We will
give an example in the context of bound entanglement later in
Section \ref{be}. \qqed

Although the PPT criterion does not constitute a necessary and
sufficient criterion, it is the most popular criterion. This is,
beside its simplicity, due to several reasons. First, the fact that
it provides a complete characterization of entanglement for
two-qubit systems makes it very appealing, since  two-qubit systems
are the most studied bipartite systems. Second, it has been shown
that the amount of violation of the PPT condition can be used to
{\it quantify} entanglement \cite{zyczkowski-1998-58, vidal-2002-65}
(see also Section \ref{negativitymeasure}). As we will see in Section 
\ref{distillabilityversusentanglement}, however,
the PPT criterion is of limited use for the investigation of
multipartite entanglement.

\subsubsection{The CCNR criterion}
Since the PPT criterion does not detect all states, the question
arises, how one can prove that a state is entangled, if the PPT
criterion fails. For this  problem, many criteria have been
proposed. Here, we want to explain the {\it computable cross norm}
or {\it realignment} criterion  (CCNR criterion) \cite{rudolph-2002,
chen-2003-3}, since it is simple and strong.

In order to formulate the CCNR criterion, we utilize the Schmidt
decomposition in operator space. For a density matrix $\vr,$ such a
decomposition is
 \be
 \vr=\sum_k \lambda_k G^A_k \otimes G^B_k,
 \label{rhodecompose}
 \ee
where the $\lambda_k \geq 0$ and $G^A_k$ and $G^B_k$ are orthonormal
bases of the observable spaces of $\HH_A$ and $\HH_B.$ Such a basis
consists of $d^2$ hermitian observables which have to fulfill
$
Tr(G^A_k G^A_l) = Tr(G^B_k G^B_l)=\delta_{kl}.
$
Such observables are often called {\it local orthogonal
observables} (LOOs) \cite{yu-150504}. For instance, for qubits
the (appropriately normalized) Pauli matrices together with
the identity form a set of LOOs.

Eq.~(\ref{rhodecompose}) is nothing but the Schmidt decomposition
from Sec.~\ref{subsec_purestates}, but applied to the space of
observables.\footnote{Note that one can also define a Schmidt rank
for density matrices \cite{terhal-2000-61}, this is, however, not
related to Eq.~(\ref{rhodecompose}).} Therefore, the $\lambda_k$ are
(up to a permutation) unique and if the  $\lambda_k$ are pairwise
different, the $G^A_k$ and $G^B_k$ are also unique (up to a sign).
The $\lambda_k$ can be computed as in the Schmidt decomposition:
First, one decomposes $\vr=\sum_{kl} \mu_{kl} \tilde{G}^A_k \otimes
\tilde{G}^B_l$ with arbitrary LOOs $\tilde{G}^A_k$ and
$\tilde{G}^B_l,$ then, by performing the singular value
decomposition of $\mu_{kl}$ one arrives at Eq.~(\ref{rhodecompose}),
the $\lambda_k$ are the roots of the eigenvalues of the matrix $\mu
\mu^\dagger.$ One can formulate:

{\bf Theorem 6 (CCNR criterion).}
If the state $\vr$ is separable, then the sum of all $\lambda_k$
in Eq.~(\ref{rhodecompose}) is smaller than one:
\be
\sum_k \lambda_k \leq 1.
\ee
Hence, if $\sum_k \lambda_k > 1$ the state must be entangled.

{\it Proof.} For a pure product state, $\vr=\ketbra{a}\otimes
\ketbra{b}$ the Schmidt decomposition is directly given, and the
statement is trivial. For the general case $\vr=\sum_k p_k
\ketbra{a_k}\otimes \ketbra{b_k}$, note that the sum over the
Schmidt coefficients defines a norm on the set of density matrices.
Then we have due to the triangle inequality $\Vert \vr \Vert \leq
\sum_k p_k \Vert \ketbra{a_k}\otimes \ketbra{b_k}\Vert = 1.$ 
Note that the question, whether other functions of the Schmidt 
coefficients can be used for entanglement detection, has been 
studied in Ref.~\cite{2008arXiv0812.4167A}.
\qqed

Let us discuss this criterion. First, the two different names come
from the fact that this criterion has been discovered in two
different forms. It has been first found while investigating cross
norms of density matrices \cite{rudolph-2000-33}, the above
formulation was taken from \cite{rudolph-2002}. Independently, it
has been formulated in a way that one realigns the entries of a
density matrix and calculates then the trace norm of this new matrix
\cite{chen-2003-3}. This second formulation is somehow more
complicated, it allows, however, to see formal similarities with the
PPT criterion \cite{horodecki-2006-13} or to generalize it into
different directions
\cite{chen-2002-306,wocjan-2005-12,clarisse-2006-6}.

The remarkable fact is that the CCNR criterion allows to prove the
entanglement for many states  where the PPT criterion fails.
Combined with its simplicity, this makes it a useful tool for the
analysis of entanglement. However, it does not detect all entangled
states of two-qubits \cite{rudolph-2003-67}. Therefore, one may view
it as complementary to the PPT criterion.

\subsubsection{Other criteria}
Beside the PPT and CCNR criterion, there are many other approaches
to derive separability criteria. Not all of them will be needed for
the future discussion. However, in order to give the reader a
coherent overview we would like to give a small list:

\begin{enumerate}

\item The {\it range criterion} was one of the first criteria
for the detection of states for which the PPT criterion fails
\cite{horobound}. It states that if a state $\varrho$ is separable,
then there is a set of product vectors $\ket{a_i b_i}$ such that the
set $\{\ket{a_i b_i}\}$ spans the range of $\vr$ as well as  the set
$\{\ket{a^*_i b_i}\}$\footnote{The symbol $\ket{a^*}$ denotes the
vector resulting when all coefficients of $\ket{a}$ in a certain
basis are complex conjugated. Note that $\ketbra{a^*}=\ketbra{a}^T.$
} spans the range of $\vr^{T_A}.$ This condition allowed to discover
many states which are entangled, but not detected by the PPT
criterion \cite{horobound, bennett-1999-82, bruss-2000-61}. However,
it cannot be used if some state is affected by noise: Then, the
density matrix and its partial transpose  will usually have full
rank, hence  the condition in the range criterion is automatically
fulfilled.

\item
The PPT criterion is an example of a criterion using a
{\it positive, but not completely positive map.} These
objects are defined
as follows. Let $\HH_B$ and $\HH_C$ be Hilbert spaces and let $\BB(\HH_i)$
denote the linear operators on it.
A linear map $\Lambda: \BB(\HH_B) \rightarrow \BB(\HH_C)$
is called {positive} if it maps hermitian operators onto hermitian
operators, fulfilling $\Lambda(X^\dagger)=\Lambda(X)^\dagger$ and it
preserves the positivity, i.e., if $X \geq 0$ then $\Lambda(X) \geq 0.$
Note that the second condition implies that it maps valid density matrices
onto density matrices, up to normalization. A positive map $\Lambda$ is
called {\it completely} positive when for an arbitrary Hilbert space $\HH_A$ the map
$\II_A \otimes \Lambda$
is positive, otherwise, $\Lambda$ is positive, but not
completely positive. Here, $\II_A$ denotes the identity on $\BB(\HH_A).$
For example, the transposition map $T$ is positive, but not completely
positive: while $X \geq 0$ implies $X^T \geq 0,$ the {\it partial}
transposition does not necessarily preserve the positivity of the
state.

{From} this, other entanglement criteria similar to the PPT criterion
can be formulated from other positive, but not completely positive
maps. For any separable state $\vr$ and any positive map $\Lambda$
we have
\be
(\II_A \otimes \Lambda) (\vr) \geq 0.
\label{pmeq}
\ee
Furthermore, it has been shown in Ref.~\cite{horoppt}
that a state $\vr$ is separable if and only if for all positive
maps $\Lambda$ the relation in Eq.~(\ref{pmeq}) holds. In this sense,
the separability problem is equivalent to the classification
of all positive maps. Of course, in order to develop a
separability criterion, in Eq.~(\ref{pmeq}) only the positive,
but not completely positive maps are of interest.

This problem was considered in the mathematical literature for a
longer time \cite{stormer-63, woronowicz-76, choi-75, maurer-76},
however, it has not been solved yet. From the perspective of
quantum information theory, the classification of positive maps was
under intensive research and has led to many new positive, but not
completely positive maps, resulting in strong separability tests
\cite{terhal-2000-323, breuer-2006-97, chruscinski-2006,
piani-2006-73,chruscinski-2009}.

Another example of a positive, but not completely positive map is
the {\it reduction map} \cite{PhysRevA.59.4206}. This is on one
system defined as 
\be \Lambda^R (X) = Tr(X)\cdot \eins - X.
\label{reductionequation} 
\ee Consequently a separable state has to
fulfill $\II_A \otimes \Lambda^R (\vr) = \vr_A \otimes \eins - \vr
\geq 0.$ This is the {\it reduction criterion for separability.} In
Ref. \cite{PhysRevA.59.4206}, it has been shown that the reduction
map is a so-called decomposable map, that is, it can be written as
\be \Lambda^R  = P_1 + P_2 \circ T, \ee where $P_1, P_2$ are
completely positive maps, and $T$ is the transposition. This
implies, that $\II_A \otimes \Lambda^R$ can never detect
entanglement, unless $\II_A \otimes T$ detects it.\footnote{In fact,
all positive maps with ${\rm dim}\HH_B =2$ and ${\rm dim}\HH_B \leq
3$ are decomposable \cite{woronowicz-76, maurer-76}, from which it
follows, that the PPT criterion is necessary and sufficient for low
dimensions.} Therefore, the reduction criterion is weaker than the
PPT criterion. Nevertheless, the reduction criterion is interesting
for several reasons: For two qubits, it also provides a necessary
and sufficient criterion for separability \cite{PhysRevA.59.4206},
and it is further closely connected to the process of distillation
(see Section \ref{be}) and certain entanglement witnesses (see
Section \ref{walbornexp}). Furthermore, one can extend the reduction map
in the following way \cite{breuer-2006-97,0305-4470-39-45-020}: If $U$
is a unitary matrix with $U^T= -U,$ then the map 
$\Lambda(X)=Tr(X)\cdot \eins - X - U X^T U^\dagger$ is a positive, but not 
completely positive map. Such a unitary transformation can only exist in even 
dimensions. But then, the map clearly improves the reduction criterion
and can detect states that are not detected by the PPT criterion.

For our purposes, the theory of positive maps will become important for
the following reason. We will see that entanglement witnesses are good
tools for the experimental detection of entanglement. Any entanglement
witness, however, gives rise to a positive, but not completely positive
map via the Choi-Jamio{\l}kowski isomorphism, as will be explained in Section
\ref{jamiolsection}. Furthermore, the theory of positive maps will later
enable us to construct nonlinear witnesses and to derive further entanglement
tests (see Sections \ref{sectionnonlinearwitnesses} and \ref{sectionmaps}).

\item
The {\it majorization criterion} relates the eigenvalues of the
total state with reduced states \cite{nielsen-2001-86}. For a
general state $\vr$ one takes $\vr_A= Tr_B(\vr)$ as the reduced
state with respect to Alice and denotes by $\PP=(p_1,p_2,...)$ the
decreasingly ordered eigenvalues of $\vr$ and by $\QQ=(q_1,q_2,...)$
the decreasingly ordered eigenvalues of  $\vr_A.$ The majorization
criterion states that if $\vr$ is separable, then 
\be \sum_{i=1}^k
p_i \leq \sum_{i=1}^k q_i \label{majorizationequation} 
\ee holds for
all $k.$ The same inequality holds, when $\vr_A$ is replaced by the
reduced density matrix of the second system $\vr_B= Tr_A(\vr).$
Physically, this criterion means that for a separable state $\vr$ is
more disordered than $\vr_A,$ and this criterion results also in
inequalities for the entropies of $\vr$ and $\vr_A$ 
\cite{cerf-1997-79,PhysRevA.54.1838,vollbrecht-2002,abe-1999-60}. 
It was shown in Ref. \cite{PhysRevLett.91.057902}, however, that 
the majorization criterion follows from the reduction criterion 
hence any state which can be detected by the majorization criterion, 
can also be detected by the reduction criterion and, consequently, 
by the PPT criterion. Similarly, one can derive criteria 
for entropies  from other positive, but not completely positive 
maps, leading to a generalization of the majorization criterion 
\cite{augusiak-2008-77A,augusiak-2008-77B,Augusiak:arXiv0811.3604}.

\item
There are further {\it algorithmic approaches} to the separability
problem. These methods formulate separability in terms of (convex)
optimization problems or semidefinite programs, in order to derive
numerical algorithms for separability testing.

Let us explain the method of the {\it symmetric extensions}
\cite{doherty-2002-88,wernersym1}, since this is one of the most
powerful separability criteria. The idea of this algorithm comes
from the following  observation. If a state $\vr=\sum_k p_k
\ketbra{a_k}\otimes \ketbra{b_k}$, is separable, we can find a
symmetric extension to three parties ($A,B$ and $A_1$) as \be
\vr^{(ABA_1)}=\sum_k p_k \ketbra{a_k}\otimes
\ketbra{b_k}\otimes\ketbra{a_k}. \ee The state $\vr^{(ABA_1)}$ has
now the following properties: (a) it is PPT with respect to each
partition, (b) the reduced state on the first two parties is given
by $\vr$ and (c) it is symmetric under the exchange of the parties
$A, A_1.$ Similarly, one can define for a separable symmetric
extensions to more parties $A,B, A_1, ...,A_k.$

The key observation in Ref.~\cite{doherty-2002-88} is that the
question ``Is there a symmetric extension with the properties (a) -
(c)?'' can be directly formulated as a feasibility problem in
semidefinite programming. Semidefinite
programs are a family of optimization problems that can not only be
solved efficiently, but under weak conditions the global optimality
of the found solution can also be proven (see Ref.~\cite{vandenberghe-96}
for a review). Practically this means
that the above mentioned feasibility problem can be directly tackled
with standard numerical packages, and if no extension is found, the
algorithm  can also prove that no extension exists, consequently the
state must be entangled.

In fact, the  method of symmetric extensions delivers a hierarchy of
separability criteria, the criterion looking for the extension to
$k+1$ parties is always stronger than the criterion with the
extension to $k$ parties. It was then shown in
Ref.~\cite{doherty-2004-69} (see also Ref.~\cite{wernersym1}) that
this hierarchy is complete in the sense that any entangled state is
detected in some step of the hierarchy. In addition, due to symmetry
requirements, the number of parameters in the semidefinite program
increases only polynomially in the number of extensions. Still,
however, the numerical effort is considerable: while the first step
of the hierarchy is, for system up to $5 \times 5$ feasible on a
standard computer\footnote{Numerical code for this first step is
available at {\tt
http://www.iqi.caltech.edu/documents/spedalieri/pptsetest1.m}. },
the higher order steps are difficult to  perform.

Beside the above mentioned algorithms, there are several other
algorithmic approaches. First, the method of symmetric extensions
was extended to the multipartite case \cite{doherty-2005-71}. Then,
in Ref.~\cite{eisert-2004-70} it was shown that on can view the
separability problem as an optimization problem with polynomial
constraints, which can, in principle, be tackled with a hierarchy 
of semidefinite programs \cite{lasserre-2001}. Other algorithms 
for separability testing have been proposed by making use of 
convex geometry \cite{ioannou-2004-70,zapatrin-2005,hulpke-2005-38,
ulmwolf-2006,brandao-2004-93,brandao-2004-70,perezgarcia-2004-330}, 
see also Ref.~\cite{ioannou-2007-7} for an overview. In addition, it
should be noted that semidefinite programs can also be used to
tackle other problems in quantum information theory, such as the
existence of local hidden variable models \cite{terhal-2003-90},
distillability of states \cite{vianna-2006-74} and the estimation of
localizable information \cite{synak-2005-46}.

Finally, these algorithms pose the question concerning the algorithmic 
complexity of the separability problem, that is, the question how much 
time is needed to decide whether a given state is entangled or not. It 
has been shown that the separability problem, considered as a weak membership 
problem, is an NP hard problem \cite{DBLP:conf/stoc/2003}. That is, deciding 
separability with an accuracy of $\varepsilon$ in an $d\times d$ system requires 
an effort exponentially in $d$, if the inverse accuracy $1/\varepsilon$ 
should increase exponentially in $d.$ \cite{ioannou-2007-7,
DBLP:conf/stoc/2003}. As recently shown, this result also holds if the 
if the inverse accuracy $1/\varepsilon$ is required to increase only 
polynomially \cite{gharibian-2008}.
It should be noted, however, that this statement
does not exclude the possibility of a simple solution for any fixed
dimension. In addition, as this statement concerns only the
computational complexity, it does not exclude possible physical
insights into the separability problem.

\item
Another approach to the separability problem makes use of criteria
based on {\it covariance matrices} (CMs) \cite{hofmann-2003-68,
guehne-2004-92, guhne-2007-99, guhne-2006-74}. Here, one considers
for some observables $M_k$ the covariance matrix $\gamma$ of these
observables in some state, given by its entries \be
\gamma_{ij}=\mean{M_i M_j}- \mean{M_i}\mean{M_j}. \label{gammadef}
\ee Then, one tries to formulate separability criteria from this
representation. The motivation for this approach comes from the
theory of infinite-dimensional systems. There, an important family
of states consists of Gaussian states, which are described by such a
CM, using the canonical conjugate variables $X$ and $P$ (for reviews
on this subject see Refs.~\cite{braunstein:513,
eisert-2003-1,adesso-2007-40,wang-2007-448}). Moreover, for these
Gaussian states the separability problem has been solved
\cite{PhysRevLett.86.3658,giedke-2001-87,hyllus-2006-8}.

With the help of these covariance matrices one can formulate
a general  {\it covariance matrix criterion} (CMC) \cite{guhne-2007-99,
gittsovich-2008}. For this, one chooses the observables in
Eq.~(\ref{gammadef}) to be the set $\{G_i^A \otimes \eins, \eins \otimes G^B_j\}$
with local orthogonal observables $G_i^A$ and $G_i^B.$ Then, $\gamma$ has a
$2\times 2$ block structure,
and if $\vr$ is separable, one can find matrices
$\kappa_{A/B} = \sum_k p_k \gamma (\ket{\psi^{A/B}_k})$ which are convex combinations
of CMs of pure states on the subsystems, such that
\be
\gamma =
\left[
\begin{array}{c c}
A & C \\
C^T & B
\end{array}
\right]
\geq
\left[
\begin{array}{c c}
\kappa_A & 0 \\
0 & \kappa_B
\end{array}
\right]. \label{cmccriterion}
\ee
On the one hand, this criterion is very strong, as is detects many bound 
entangled states, and, with the help of filtering operations 
(see Sections \ref{jamiolsection} and  \ref{sectionionexp}), 
it can  detect all entangled states for two qubits, just as the PPT 
criterion.\footnote{For general systems, however, there are NPT states which
are not detected by the CMC and filtering; moreover, there are entangled
PPT states which not detected by CMC.}
Moreover, the CCNR criterion can be shown to be a corollary of the CMC. For 
instance, from the CMC it follows that for separable states 
\be 
\Vert C
\Vert_1 \leq \sqrt{[1-Tr(\vr^2_A)][1-Tr(\vr^2_B)]}, 
\ee 
where $\Vert C \Vert_1$ is the trace norm of the matrix 
$C_{ij}= \mean{G_i^A\otimes G_j^B}-\mean{G_i^A} \mean{G_j^B},$ 
which is a systematic improvement of the CCNR criterion 
\cite{zhang-2007,gittsovich-2008,2008arXiv0812.4167A}. 
In addition, the CMC can be related to other known separability criteria 
\cite{zhang-2007-76,devicente-2007-7}.

It is not clear, how to use the CMC for entanglement detection 
in the multipartite case. The closely connected criterion of 
the {\it local uncertainty relations} (LURs) \cite{hofmann-2003-68} 
is, however, experimentally relevant and will be discussed
in Section \ref{sectionlurs}.

\item A similar approach uses the {\it expectation value matrix}
(EVM) of a quantum state to investigate its entanglement properties
\cite{vogelprl,korbicz:022318,moroder-2008-evm,miranowicz-2006-evm}.
For a set of operators $\{M_k\}$ one defines the positive semidefinite
EVM $\chi$ as
\be
\chi_{ij} = \mean{M_i^\dagger M_j}.
\label{evmdef}
\ee
Typically, one chooses the observables to be of the type
$\{M_k\}= \{A_k \otimes B_l\}.$ Then it is straightforward
to see that for a separable state $\vr=\sum_k p_k \vr^A_k \otimes \vr^B_k$
one has $\chi(\vr) = \sum_k p_k \chi(\vr^A_k) \otimes \chi(\vr^B_k),$
hence $\chi(\vr)$ is also separable and its separability can be
investigated by criteria such as the PPT criterion. Especially, if
$B_l^T=B_l^\dagger$ one has $\chi^{T_B} = \chi(\vr^{T_B}).$ This
has two interesting consequences: First, a violation of the PPT
criterion for $\chi$ may be indicated by the negativity of some
subdeterminants of $\chi$, leading to nonlinear or variance-like
criteria for separability \cite{vogelprl, moroder-2008}. Second, if
the PPT criterion for $\chi$ is violated, there is a vector $\ket{x}=(x_i)$
such that  $\bra{x}\chi^{T_B}\ket{x}<0$ and hence a positive operator
$P= X^\dagger X$ with $X = \sum_{i} x_i M_i$ such that
$Tr(\vr P^{T_B})<  0$ while $Tr(\vr P^{T_B}) = \bra{x}\chi^{T_B}\ket{x} \geq  0$
for all PPT states. The observable $P^{T_B}$ will later be called an entanglement
witness (see Section \ref{ewsection}) and therefore the EVM allows to investigate
for which states a violation of the PPT criterion can be proved with a given,
restricted set of observables $\{M_k\}= \{A_k \otimes B_l\}$ \cite{moroder-2008-evm}.
Furthermore, the EVM can be used to investigate entanglement and separability for
qubit-mode systems \cite{haseler-2008-77,rigas-2006-73}.

\item A different line of developing separability criteria uses
{\it linear contractions} and {\it permutations}
\cite{horodecki-2006-13}. The  main idea behind this approach is the
following: given a density matrix $\vr =\sum_{ij,kl} \vr_{ijkl}
\ket{i}\bra{j}\otimes\ket{k}\bra{l}$ expanded in a product basis,
all information concerning the state is encoded in the coefficients 
$\vr_{ijkl},$ which build a $(d_A \times d_A) \times (d_B
\times d_B)$ tensor. For pure states, it is normalized,
$\Vert\vr_{ijkl}\Vert_1=1$, where $\Vert X \Vert_1 =Tr(\sqrt{X
X^\dagger}) $ denotes the trace norm, and $\vr_{ijkl}$ is considered
as an $(d_A^2)\times (d_B^2)$ matrix. It is easy to see that for a
product state, any permutation $\pi(ijkl)$ of the indices $i,j,k,l$ 
yields also a tensor with $\Vert\vr_{\pi(ijkl)}\Vert_1=1,$ where, depending 
on the permutation, $\vr_{\pi(ijkl)}$ might be considered as a 
$(d_A d_B) \times (d_A d_B)$ matrix.
Since the 
trace norm obeys the triangle inequality, this means that for any 
separable state $\Vert\vr_{\pi(ijkl)}\Vert_1 \leq \sum_\alpha p_\alpha
\Vert\vr^{(\alpha)}_{\pi(ijkl)}\Vert_1 = 1$, where
$\vr=\sum_\alpha p_\alpha \vr^{(\alpha)}$ It turns out, that  in the
end for two parties only two permutations yield independent
separability criteria, resulting in the conditions
\begin{equation}
\Vert\vr_{(ij lk)}\Vert_1 \leq 1 \mbox{ and } \Vert\vr_{(ik
jl)}\Vert_1 \leq 1.
\label{permcrit}
\end{equation}
It can be directly seen that the first condition is nothing but the
PPT criterion, while the second one corresponds to the CCNR
criterion. Thus, this approach delivers a way  to formulate these
two criteria in a unified way. Further, it can be extended to the
multipartite case, where many new separability criteria arise
\cite{wocjan-2005-12,clarisse-2006-6}, see also 
Section \ref{section:multipartitesepcrit}.

\end{enumerate}
Other approaches to detect entanglement use special observables, they are called
either Bell inequalities or entanglement witnesses. Bell inequalities will be
discussed in Section \ref{sectionbellinequalities} and entanglement witnesses
in Section \ref{ewsection}. Before doing that, we would like
to discuss the phenomenon of bound entanglement.

\subsection{Bound entanglement}
\label{be}
For many tasks in quantum information theory, like teleportation or
cryptography one ideally needs maximally entangled two-qubit states,
{i.e.}, singlet states. In the real world, however, noise is
unavoidable, thus only mixed states are available.  So one has to
deal with the question, if and how  one can create a singlet state
out of some given mixed state. This leads to the problem of the
so-called {\it distillation of entanglement.} It can be posed as
follows: Assume that there is  an arbitrary, but finite, number of
copies of an entangled quantum state $\vr$ distributed between Alice
and Bob. Can they perform local operations on the states, assisted
by classical communication, such that at the end they share a
singlet state, \be \underbrace{\vr \otimes \vr \otimes ... \otimes
\vr }_{\mbox{ k copies}} \stackrel{LOCC}{\longrightarrow}
\ket{\psi^-}=\frac{1}{\sqrt{2}}(\ket{01}-\ket{10}) ? \ee If this is
the case, the state $\vr$ is called distillable, otherwise we call
it undistillable or {\it bound entangled.}

It is not clear from the beginning that one can distill quantum
states at all. However, the first distillation protocols were
derived in Ref.~\cite{bennett-1996-76, PhysRevLett.77.2818}, showing
that it is in principle possible. The question whether a given state
is distillable is, of course, difficult to decide, since there is no
restriction to a specific kind  of local operations, as well as to
the number of copies. The failure of a special protocol is not
enough to conclude that a state is undistillable. However, some
simple criteria are known:

{\bf Theorem 7.} If a bipartite state is PPT, then the state is
undistillable. If a state violates the reduction criterion (e.g.,
due to a violation of the majorization criterion) then the state is
distillable.

{\it Proof.} For the sufficient condition to be undistillable, see
Ref.~\cite{horodecki-1998-80}. For the sufficient condition to be
distillable, see Refs.~\cite{PhysRevA.59.4206,
PhysRevLett.91.057902}. $\qqed$

By definition, bound entangled states require some entanglement
for its creation, however, afterwards the entanglement cannot be
distilled back. Naturally, such states are interesting, and the
above Theorem implies that the search for PPT entangled states is of
great interest in studying bound entanglement. The first example of
a PPT entangled state was given in Ref.~\cite{horobound} and in the
meantime many other examples have been found \cite{bennett-1999-82,
bruss-2000-61, piani-2007-75,piani-2006-73}.

In order to demonstrate
how such states may be constructed, let us  discuss a simple example
from Ref.~\cite{bennett-1999-82}.
In a $3\times 3$-system, one can consider the vectors
\begin{align}
\ket{\psi_0} &= \frac{1}{\sqrt{2}}\ket{0}(\ket{0}-\ket{1}),
\;\;\;
\ket{\psi_1}=\frac{1}{\sqrt{2}}(\ket{0}-\ket{1})\ket{2},
\;\;\;
\ket{\psi_2}=\frac{1}{\sqrt{2}}\ket{2}(\ket{1}-\ket{2}),
\;\;\;
\nonumber\\
\ket{\psi_3}&=\frac{1}{\sqrt{2}}(\ket{1}-\ket{2})\ket{0},
\;\;\;
\ket{\psi_4}=\frac{1}{3}(\ket{0}+\ket{1}+\ket{2})(\ket{0}+\ket{1}+\ket{2}).
\end{align}
These five product vectors form a so-called {\it unextendible
product basis:} they are all pairwise orthogonal and there is not
another product vector orthogonal to all of them, as can be seen by
direct inspection. Therefore, the mixed state \be \varrho_{\rm
BE}=\frac{1}{4}(\eins-\sum_{i=0}^4\ket{\psi_i}\bra{\psi_i}) \ee has
no product vector in its range, and must be entangled due to the
range criterion. Since $\varrho_{\rm BE}=\varrho_{\rm BE}^{T_B}$ it
is also PPT, and thus bound entangled. This state has also the minimal
possible rank, since the minimal rank of a PPT entangled state in
a $d_A \times d_B$-system is $\max\{d_A,d_B\}+1$ \cite{horodecki-2000-62bound}.

Beside the mere existence, there are several results and questions,
which make the phenomenon of bound entanglement interesting:
\begin{enumerate}

\item At first sight, entangled states that cannot be distilled
seem to be not very useful, since they cannot be used for certain tasks
like quantum key distribution via the Ekert protocol \cite{PhysRevLett.67.661}.
It has been proved, however, that some bound entangled states can indeed be
used for secure quantum key distribution \cite{horodecki-2005-94, horodecki-2005},
showing that entanglement distillation and secure key distillation are not equivalent.

\item Bound entangled states require entanglement for their
creation, but the entanglement can then not be distilled again.
Similarly, one may ask, whether there are classical probability
distributions between two or several parties that require secret
communication for their creation,  but they do not allow generating
a secret key \cite{gisin00linking}. It has been shown for the
tripartite case, that such {\it bound information} exists
\cite{acin-2004-92}. Interestingly, the proof is based on a bound
entangled tripartite quantum state, from which the classical
probability distribution is derived.

\item It has already been conjectured by Peres in 1999, that bound entangled
states admit a local hidden variable model and thus do not violate any Bell
inequality \cite{peres-1999-29}. We will discuss this open question in more
detail in Section \ref{sectionbellconsequences}.

\item Since PPT entangled states are bound entangled, the question arises,
whether NPT states are always distillable. It has been conjectured that
this is not the case \cite{dur-2000-61, divincenzo-2000-61}. Examples
of states have been found, which cannot be distilled, if only few copies
are available \cite{dur-2000-61, divincenzo-2000-61}. However, despite of
several partial results \cite{pankowski-2007} the question
remains open. For $2\times d_B$-systems, any NPT state is 
distillable \cite{dur-2000-61}.

\item The given example of a bound entangled state may seem artificial. An
interesting question is whether bound entangled states occur under
natural conditions, e.g., as thermal states in spin models. For
multipartite bound entanglement, this is indeed the case
\cite{optimalspsq,ferraro-2008-100}.

\end{enumerate}
To conclude, all these results indicate that bound entangled states
are an interesting object of study. Their construction, characterization
and detection is a challenging and important task in entanglement theory.

\subsection{Entanglement witnesses}
\label{ewsection}

Let us now come to a completely different type of separability
criteria. The criteria from above have all something in common:
at first sight they all assume that the density matrix is already
known. They all require applying certain operations to a density
matrix, to decide whether the state is entangled or
not\footnote{Note
that in principle one can decide whether a state violates the PPT
criterion or not without knowing the state completely. For schemes
to do that see Section \ref{sectionmaps}}.
There is, however, a necessary and sufficient entanglement criterion
in terms of directly measurable  observables. These are the so called
entanglement witnesses \cite{horoppt,terhal-2000-271, PhysRevA.62.052310,
bruss-2002-49}, which we introduce now.

{\bf Definition 8.} An observable $\WW$ is called an {\it entanglement
witness} (or witness for short), if
\bea
Tr(\WW \vr_s) & \geq & 0 \mbox{  for all separable }\vr_s,
\nonumber
\\
Tr(\WW \vr_e) & < & 0 \mbox{  for at least one entangled }\vr_e
\label{ew1}
\eea
holds. Thus, if one measures $Tr(\WW \vr)<0$ one knows for sure
that the state $\vr$ is entangled.  We call a state
with $Tr(\WW \vr)<0$ to be detected by $\WW.$

The fact that entanglement witnesses are directly measurable
quantities makes them a very useful tool for the analysis
of entanglement in experiment. As we will see in the further
course of this review, entanglement witnesses are one of the
main methods to detect entanglement experimentally.

For the further understanding, it is crucial to note that entanglement
witnesses have a clear geometrical meaning. The expectation value
of an observable depends linearly on the state. Thus, the set of states
where $Tr(\WW \vr)=0$ holds is a hyperplane in the set of all
states, cutting this set into two parts. In the part with $Tr(\WW \vr)>0$
lies the set of all separable states, the other part [with $Tr(\WW \vr)<0$]
is the set of states detected by $\WW.$ This scheme is shown in Figure
\ref{witnessfigure}.
{From} this geometrical interpretation it follows that all entangled
states can be detected by witnesses:

\begin{figure}[]
\centerline{\includegraphics[width=0.5\columnwidth]{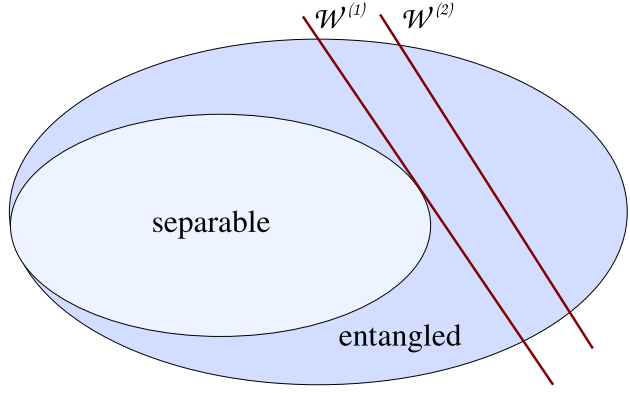}
}
\caption{
Schematic picture of the set of all states and the set of
separable states as nested convex sets and two witnesses,
$\WW^{(1)}$ and $\WW^{(2)}.$ The red lines represent the hyperplanes,
where $Tr(\WW\vr)=0.$ The witness $\WW^{(1)}$ is finer than the
witness $\WW^{(2)}$ [see also Eq.~(\ref{witnessfinerdefinition})].
}
\label{witnessfigure}
\end{figure}

{\bf Theorem 9 (Completeness of witnesses).}
For each entangled state  $\vr_e$ there exists
an entanglement witness detecting it.

{\it Proof.} This theorem was proved in Ref.~\cite{horoppt}. The
idea for the proof relies on the fact that the set of separable
states is convex and closed. Thus, for a point outside this set,
there is a hyperplane separating this point from the convex set. For
finite-dimensional spaces with a scalar product this is easy to see,
but the statement holds also for infinite-dimensional Banach spaces,
where the proof requires a corollary of the Hahn-Banach-theorem
\cite{scharlaubook}. $\qqed$

Although this theorem ensures that any entangled state can
in principle be detected with an entanglement witness, the task
remains to construct witnesses. This is not an easy problem, since
solving this problem would also solve the separability problem.
Indeed, a large part of this review is concerned with the
proper construction and evaluation of witnesses for the multipartite
case.

\subsubsection{Construction of witnesses}
\label{Sec_ConstWit}

Let us discuss in three simple examples, how witnesses for bipartite
entanglement can be constructed. Typically, if a state violates some
criterion for separability, an entanglement witness detecting the state
can be  written down.

As the first example, let us take a state $\vr_e$ which is NPT. Then
there is a negative eigenvalue $\lambda_-<0$ of $\vr_e^{T_A}$ and a
corresponding eigenvector $\ket{\eta}.$ Now \be
\WW=\ketbra{\eta}^{T_A} \label{pptwitness} \ee is a witness
detecting $\vr_e$: Due to the fact that for general matrices $X,Y$
the relation $Tr(X Y^{T_A})=Tr(X^{T_A} Y)$ holds\footnote{This is
obvious for the case $X=A\otimes B$ and $Y=C\otimes D$, for the
general case it follows from linearity.}, we have
$Tr(\WW\vr_e)=Tr(\ketbra{\eta}^{T_A}\vr_e)=
Tr(\ketbra{\eta}\vr_e^{T_A}) =\lambda_- < 0$ and $Tr(\WW\vr_s) =
Tr(\ketbra{\eta}\vr_s^{T_A}) \geq 0$ for all separable $\vr_s$,
since they are PPT.

More generally, if $\vr_e$ is entangled and detected by some
positive map $\Lambda,$ then $\II \otimes \Lambda (\vr_e)$ has
again a negative eigenvalue  $\lambda_-<0$ for the eigenvector
$\ket{\eta}.$ Then, $\WW = \II \otimes \Lambda^+ (\ketbra{\eta})$
is a witness detecting $\vr_e.$ Here, $\Lambda^+$ denotes the
adjoint map, i.e., the map fulfilling $Tr[\Lambda^+(X)Y]=Tr[X \Lambda(Y)]$ for
all $X,Y$.

As a second example, let us consider the case when $\vr_e$ violates
the CCNR criterion. Then, by definition, we have in the Schmidt
decomposition in Eq.~(\ref{rhodecompose}) $\sum_k \lambda_k > 1.$ A
witness is now given by Ref.~\cite{guhne-2006-74, yu-150504} \be \WW
= \eins - \sum_k G^A_k \otimes G^B_k, \label{ccnrwitness} \ee where
the $G^{A/B}_k$ are the observables from the Schmidt decomposition.
Clearly, for $\vr_e$ we have $Tr(\WW \vr_e)= 1-\sum_k \lambda_k <0.$
A general separable state $\vr_s=\sum_{kl}\mu_{kl} G^A_k \otimes
G^B_l$ can be expressed in the basis of $G^{A/B}_k.$ Then,
$Tr(\WW\vr_s)= 1- \sum_k \mu_{kk} \geq 0,$ where the last estimate
comes from the CCNR criterion and the fact, that a trace of a matrix
is always smaller than the sum of its singular values
\cite{hornjohnsonbook}. The witness in Eq. (\ref{ccnrwitness}) can
also be derived from the realignment map \cite{chen-2004-69,
guhne-2006-74}.

As a third example, witnesses can also be constructed from the
consideration that states close to an entangled state must also be
entangled. Therefore, one may try for a given entangled pure state
$\ket{\psi}$ to write down a projector-based witness like \be \WW =
\alpha \eins - \ketbra{\psi}. \label{projwitdef} \ee This witness
can be interpreted as follows: The quantity $Tr(\vr\ketbra{\psi}) =
\bra{\psi}\vr\ket{\psi}$ is the fidelity of the state $\ket{\psi}$
in the mixed state $\vr,$ and if this fidelity exceeds a critical
value $\alpha,$ then the expectation value of the witness is
negative and the state $\vr$ must be entangled.

The problem remains to calculate the smallest possible value for
$\alpha,$ in order to ensure that $\WW$ is still positive on all
separable states. This can be done by taking \be \alpha = \max_{\vr
\;\;{\rm is\;\;separable}} Tr(\vr\ketbra{\psi}) = \max_{\tiny
\ket{\phi}=\ket{a}\otimes \ket{b}} |\braket{\psi}{\phi}|^2.
\label{projwitdef2} \ee Here, we used that a linear function takes
its maximum on a convex set in one of the extremal points, and for
the convex set of the separable states these extremal points are
just the pure product states. Then, as shown in
Ref.~\cite{bourennane:087902} the maximum can be directly computed
and is given by the square of the maximal Schmidt coefficient of
$\ket{\psi}.$ Finally it is worth mentioning that the witness $\WW$
in Eq.~(\ref{projwitdef}) does only detect states which are NPT
therefore the construction is less general than the one in
Eq.~(\ref{pptwitness}).\footnote{This follows from the fact that the
witness $\WW=\eins/d-\ketbra{\psi}$, where $\ket{\psi}$ is maximally
entangled, does not detect NPT states \cite{sanpera-2001-63},
combined with an argument as in Theorem 3 of
Ref.~\cite{PhysRevLett.95.120405}.} It has the advantage, however,
that it can be generalized to the multipartite case, as we will see
later.

Finally, let us mention other ways of constructing entanglement
witnesses. Most of them are connected to special separability
criteria.

\begin{enumerate}

\item As already mentioned, any positive, but not completely positive
map gives rise to a construction of entanglement witnesses. As we will
see soon, also a connection in the other direction holds.

\item Many bound entangled states violate the range criterion in a
way that for any product vector $\ket{a,b}\in R(\vr)$ in the range
of $\vr$ we have already $\ket{a,b^*}\notin R(\vr^{T_B}).$ States with
such an extremal violation are called {\it edge states}, and if  an
edge state $\vr$ is PPT, then it lies on the boundary between the PPT
and the NPT states. For such a state, one can write down a witness of
the type
\be
\WW = \alpha \eins - P - Q^{T_B}
\ee
where $P$ ($Q$) denotes the projector onto the range of
$\vr$ ($\vr^{T_B}$) and $\alpha$ is defined as
$\alpha = \max_{\tiny \ket{\phi} =
\ket{a,b}} \bra{\phi}(P - Q^{T_B})\ket{\phi} < 2$ \cite{PhysRevA.62.052310}.
This construction guarantees that $\WW$ is a valid witness, and
$\vr$ is clearly detected. The parameter $\alpha$ can either be
estimated analytically \cite{terhal-2000-323}, or calculated numerically
\cite{guehne-2003-50, hyllus-2004-70, eisert-2004-70}.

\item The semidefinite program in the criterion of the symmetric
extensions \cite{doherty-2002-88} gives as a byproduct an entanglement
witness, if no symmetric extension was found.

\item 
Entanglement witnesses can also be constructed from geometrical 
considerations: If $\vr_e$ is an entangled state, and $\sigma$ is 
the closes separable state to $\vr_e$ [in Hilbert-Schmidt norm, 
$\Vert X \Vert := \sqrt{Tr(X^\dagger X)}$], then
\be
\WW = \frac{1}{\mathcal{N}}
\big(\sigma - \vr_e + Tr[\sigma(\vr_e - \sigma)]\cdot \eins \big)
\ee
with $\mathcal{N} = \Vert \vr_e - \sigma \Vert$ is an entanglement 
witness, since the set of states with $Tr(\WW \vr) = 0$ is the 
hyperplane orthogonal to the line from $\vr_e$ to $\sigma$
\cite{pittenger-2002-346,bertlmann-2002-66}. The normalization
is chosen in such a way that its expectation value for $\vr_e$ 
equals (up to the sign) the distance to the separable states 
\cite{bertlmann-2002-66,bertlmann-2005-72}. Such geometric 
entanglement witnesses have been applied to different 
situations \cite{bertlmann-2007,jafarizadeh-2008,krammer-2008,bertlmann-2009}.

\item The criteria based on variances and covariance matrices also
result in certain types of entanglement witnesses. These witnesses,
however, are nonlinear, we will discuss them in Section \ref{sectionvariances}.
Further, special constructions of witnesses have derived from
estimations of entanglement measures \cite{mintert:052302} (see also Section 
\ref{walbornexp}). Also, from a given witness detecting some state further 
witnesses can be constructed \cite{chruscinski-2007}.

\end{enumerate}

\subsubsection{Optimization of witnesses}
\label{sectionoptimization}

Every entanglement witness can, by definition, detect
some entangled states. However, as can directly be seen in Fig.~\ref{witnessfigure},
some witnesses are better in this task than others. In this sense, one
can {\it optimize} entanglement witnesses \cite{PhysRevA.62.052310}. First,
one calls a witness $\WW^{(1)}$ {\it finer} than another witness $\WW^{(2)}$ if
$\WW^{(1)}$  detects all the states detected by $\WW^{(2)}$ and
also some states in addition. This is the case, if
\be
\WW^{(2)} = \WW^{(1)} + P,
\label{witnessfinerdefinition}
\ee
where $P$ is a positive operator. Consequently, for any state
$Tr(\vr\WW^{(1)}) \leq Tr(\vr\WW^{(2)})$ holds.

Furthermore, a witness $\WW$ is called {\it optimal}, if there is no
other witness, which is finer than $\WW.$ This implies that for any
positive operator $P$ the observable $X=\WW-P$ is not a witness
anymore. From  this it can be easily seen that $\WW$ is optimal, if
and only if the product vectors $\ket{\phi_i}=\ket{a_i b_i}$ with
$\bra{\phi_i}\WW\ket{\phi_i}=0$ span the whole space. A necessary
condition for a witness to be optimal is that it ``touches'' the
set of separable states (see Fig.~\ref{witnessfigure}), i.e., there
must be a separable $\vr$ with $Tr(\vr\WW)=0.$ This is, however, not
a sufficient condition, witnesses which fulfill this condition are
sometimes also called weakly optimal.

The optimization of a given witness can be done as follows
\cite{PhysRevA.62.052310}: If one finds a positive operator
$P,$ which fulfills $\bra{a_i b_i} P \ket{a_i, b_i}=0$
for all product vectors with $\bra{a_i b_i}\WW\ket{a_i, b_i}=0,$
then $\tilde{\WW}= \WW - \lambda P$ is an entanglement witness
finer than $\WW$ if
\be
\lambda \leq \lambda_0 := \inf_{a} \min_{\rm eigenvalues}
\big[\frac{1}{\sqrt{P_a}} \WW_a \frac{1}{\sqrt{P_a}}\big]
\label{opticheck}
\ee
with the notation $X_a= \bra {a}X\ket{a} \in \BB(\HH_B)$ for
$\ket{a}\in \HH_A$ and $X \in \{\WW, P\}.$ This can be iterated, 
until the product vectors with $\bra{a_i b_i}\WW\ket{a_i b_i}=0$ 
span the complete space. 
However, the optimization in Eq.~(\ref{opticheck}) is not 
straightforward, as it is equivalent to checking whether
$\inf_{\tiny \ket{a}\ket{b}} \bra{a,b}[\WW-\lambda P]\ket{a,b} 
\geq 0.$ In fact, such an optimization for general observables 
is equivalent to the separability problem \cite{sperling-2008}, 
and it can be shown that the question whether a given state is
separable or not is equivalent to the question whether some 
witness in a higher dimensional space is weakly optimal or not 
\cite{badziag-2007-single}.

\subsubsection{The Choi-Jamio{\l}kowski isomorphism}
\label{jamiolsection}
Now we explain a close connection between witnesses and positive
maps. This connection is given by an isomorphism between quantum
states on a space of matrices, and maps between such spaces. In the
literature, this isomorphism is known as the Choi-Jamio{\l}kowski
isomorphism
\cite{pillis-67,jamiolkowski-72,choi-75,choi-82}.\footnote{To our
knowledge, an isomorphism of this type was first investigated by J.
de Pillis \cite{pillis-67}, who used a map as in
Eq.~(\ref{pillisversion}) without the transposition on the first
subsystem and proved that for this map $\varepsilon$ is positive if
$E$ is positive semidefinite. In Ref.~\cite{jamiolkowski-72}, it was
proved for the same map that $\varepsilon$ is positive, iff $E$ is
positive on all product states, i.e., it is an entanglement witness,
and the explained version and the proof of main properties in
Theorem 10 are given in Refs.~\cite{choi-75,choi-82}. See also
Ref.~\cite{ranade-2007}.}

According to this, an operator $E$ in $\BB(\HH_B) \otimes
\BB(\HH_C)$ corresponds to a map $ \varepsilon$ that maps states
from $\BB(\HH_B)$ to an element of $\BB(\HH_C).$ This map is given
by \be \varepsilon (\vr) = Tr_B (E \vr^T \otimes \eins_C).
\label{bernhardkohl} \ee The inverse relation reads \be E =
(\II_{B'} \otimes \varepsilon)(\ketbra{\phi^+}) \label{nl15}, \ee
where $\HH_{B'}\cong \HH_{B}$ and $\ket{\phi^+}= \sum_i \ket{ii}$ is
a non-normalized maximally entangled state on $\HH_{B'} \otimes
\HH_B.$ Using the fact that $\ketbra{\phi^+} = \sum_k (G^{B'}_k)^T
\otimes G^{B}_k$ (see, e.g., Ref.~\cite{guhne-2006-74}) for any
orthogonal basis $G^{B'/B}_k$, this may be expressed as \be E=\sum_k
(G^{B'}_k)^T \otimes \varepsilon(G^{B}_k). \label{pillisversion} \ee
from which it is also easy to see that Eq.~(\ref{bernhardkohl}) is
the inverse of Eq.~(\ref{nl15}). Physically, these relations can be
interpreted as follows. If the map $\varepsilon$ can be realized,
then the state $E$ can be prepared by applying the map on one part
of a maximally entangled state. On the other hand (due to the
relation $Tr_{B'B}[X_{B'} \otimes Y_B \ketbra{\phi^+}] =
Tr_{B'B}[(X_{B'})^T \otimes Y_B \ketbra{\phi^+}^{T_{B'}}]
=Tr_B[X^TY]$) one can view the map $\varepsilon$ as generated via
teleportation with the state $E.$

The Choi-Jamio{\l}kowski isomorphism is useful for several purposes.
For instance, it can be used to define a distance measure for maps
as quantum operations by looking at the distance of the
corresponding states. This distance has then several natural
properties \cite{gilchrist:062310}. It can also serve to investigate
the entanglement generating capacity of positive maps on a bipartite
system \cite{cirac-2001-86}. For the theory of witnesses, the
following theorem concerning the relation between the map
$\varepsilon$ and the state $E$ is relevant (for a proof see
Refs.~\cite{choi-75, choi-82, lewenscript}):

{\bf Theorem 10 (Properties of the Choi-Jamio{\l}kowski isomorphism).}
The  Choi-Jamio{\l}kowski isomorphism has the following properties:
\\
(i) The map $\varepsilon$ is completely positive, iff $E$ is a
positive semidefinite operator.
\\
(ii) The map $\varepsilon$ is positive but not completely positive, iff $E$ is an
entanglement witness
\\
(iii) The map $\varepsilon$ is a decomposable map, iff $E$ is a decomposable
entanglement witness.

Concerning the last point, note that an entanglement witness is called
decomposable, if it can be written as $\WW=P_1+P_2^{T_B},$ where
$P_1,P_2$ are positive operators \cite{PhysRevA.62.052310}.
Clearly, such a witness cannot detect any PPT entangled state.

Due to this isomorphism, any witness, which detects some PPT entangled state
gives rise to a positive, but not completely positive map, which is non
decomposable. In this way, entanglement witnesses can be used to construct new
and interesting positive maps.

The question arises, whether the map $\varepsilon (\WW)$ detects more states than
the witness $\WW.$ Indeed, one can directly see that if
$\Lambda: \BB(\HH_B) \rightarrow \BB(\HH_{A'})$ is a map detecting the state
$\vr_{AB},$ then it detects all the states
$(F_A \otimes \eins_B) \vr_{AB} (F_A^\dagger \otimes \eins_B),$ where 
$F_A$ is an arbitrary invertible matrix \cite{lewenscript}. On the other 
hand, if $\WW_{AB}$ is a witness with $Tr(\WW_{AB}\vr_{AB})<0,$ then the adjoint map
$[\varepsilon(\WW)]^+:\BB(\HH_B) \rightarrow \BB(\HH_{A'})$ detects the state,
from which it follows that the map $\varepsilon^+$ detects all states which
are detected by the witness
\be 
\tilde{\WW} = [(F_A)^{-1} \otimes \eins_B] \WW [(F_A^\dagger)^{-1}
\otimes \eins_B].
\label{filterwitness}
\ee
This is a larger set of states, for examples see 
Refs.~\cite{PhysRevA.63.044304,bodoky-2008}. 
The operator $F_A$ corresponds to local filters \cite{gisin-1996,
verstraete-2003-68,leinaas:012313}. Such transformations do not change 
the fact whether a state is entangled or not, hence they may be used to 
simplify entanglement detection. For instance, it has been shown that 
any state $\vr$ of full rank can be simplified via filters to 
(assuming $d_A \leq d_B$)
\be
\vr \mapsto \tilde{\vr} 
= (F_A \otimes F_B) \vr (F_A \otimes F_B)^\dagger 
= \frac{\eins}{d_A d_B} + \sum_{k=1}^{d_A^2-1}\xi_i G^{A}_k \otimes G^B_k,
\ee
where the $G^{A/B}_k$ are traceless orthogonal observables
\cite{verstraete-2003-68,leinaas:012313}. Such normal forms 
under filtering operations have been used to improve given 
separability criteria \cite{guhne-2007-99,gittsovich-2008} 
and we will see in Section \ref{sectionionexp} 
how filters can be used to 
improve witnesses systematically.

Finally, another interesting consequence from the Choi-Jamio{\l}kowski
isomorphism is that any witness can be written in the form
Eq.~(\ref{nl15}). This form is similar to Eq.~(\ref{pptwitness}),
but with a different positive map. This idea will be later used for
the construction of nonlinear witnesses.

\section{Multipartite entanglement}

In this Section we discuss the structure of entanglement
when more than two parties are involved. It will
turn out that this structure is much richer
than the structure of entanglement in the bipartite
case. Especially, it will turn out that several
inequivalent classes of  entanglement exist.
We first discuss different notions of entanglement
and separability for the case of three qubits.
Then, we discuss the case of general multipartite
systems. After that, we introduce the main families of
multipartite entangled states, such as GHZ states, Dicke
states and graph states, which are interesting because of
their entanglement properties or their usefulness for
applications. Finally, we discuss the construction
of entanglement witnesses for the multipartite case.

\subsection{Entanglement of three qubits}

\subsubsection{Pure states}
Let us first consider pure three-qubit states. There are two different types
of separability: the {\it fully separable} states that can be written as
\begin{equation}
\Ket{\phi^{\rm{fs}}}_{A|B|C}=
\Ket{\alpha}_A\otimes\Ket{\beta}_B\otimes\Ket{\gamma}_C,
\end{equation}
and the {\it biseparable} states that can be written as a product
state in the bipartite system. A biseparable state can be created,
if two of the three qubits are grouped together to one party. There
are three possibilities of grouping two qubits together, hence there
are three classes of biseparable states. One example is
\begin{equation}
\Ket{\phi^{\rm{bs}}}_{A|BC}=
\Ket{\alpha}_A\otimes\Ket{\delta}_{BC}.
\end{equation}
The other possibilities read $\Ket{\phi^{\rm bs}}_{B|AC}=
\Ket{\beta}_B\otimes\Ket{\delta}_{AC}$ and $\Ket{\phi^{\rm
bs}}_{C|AB}=\Ket{\gamma}_C\otimes\Ket{\delta}_{AB}.$ Here,
$\ket{\delta}$ denotes a two-party state that might be entangled.
Finally, a pure state is called {\it genuine tripartite entangled}
if it is neither fully separable nor biseparable. Examples of such
states are the GHZ state \cite{PhysRevD.35.3066,greenberger-1989}
\begin{equation}
\Ket{GHZ_3}=\frac{1}{\sqrt{2}}(\Ket{000}+\Ket{111}), \label{GHZ3}
\end{equation}
and the so-called W state \cite{zeilinger-1992,PhysRevA.62.062314}
\begin{equation}
\Ket{W_3}=\frac{1}{\sqrt{3}}(\Ket{100}+\Ket{010}+\Ket{001}.
\label{W3}
\end{equation}

{From} a physical point of view, the generation of fully separable
or biseparable states does not require interaction of all parties.
Only for the creation of genuine tripartite entangled state all
three parties have to interact.\footnote{The notion of genuine
multipartite entanglement was (in the context of Bell inequalities)
already anticipated in Ref.~\cite{PhysRevD.35.3066}, see also
Section~\ref{bellgenuine}.}

The  genuine entangled three-qubit states can, however, be further
divided into two inequivalent classes in the following way: Given
two three-qubit states, $\Ket{\phi}$ and $\Ket{\psi},$ one can ask
whether it is possible to transform  a single copy of $\Ket{\phi}$
into $\Ket{\psi}$ with local operations and classical communication,
without requiring that this can be done with certainty. These
operations are called stochastic local operations and classical
communication (SLOCC). It turns out \cite{PhysRevA.62.062314} that $\Ket{\phi}$
can be transformed into $\Ket{\psi}$ iff there exist invertible
operators $A,B,C,$ acting on the space of one qubit with
\begin{equation}
\Ket{\psi}=A \otimes B \otimes C \Ket{\phi}.
\end{equation}
Since the operators  $A,B,C$ are invertible, this defines an
equivalence relation with a clear physical meaning.

Surprisingly, it was proved in Ref.~\cite{PhysRevA.62.062314} that
there are two different equivalence classes of genuine tripartite
entangled states, which cannot be transformed into another by SLOCC.
One class, the class of GHZ states is represented by the GHZ state $
\Ket{GHZ_3}$ defined in Eq.~(\ref{GHZ3}). The other class, the class
of W states can be transformed via SLOCC into $\Ket{W_3}$ given in
Eq.~(\ref{W3}). In this sense there are two different classes of
tripartite entanglement. There are much more pure GHZ class states
than W class states: By local unitary operations one can transform
any pure three-qubit state into\footnote{This is a generalization of
the Schmidt decomposition to three qubits, see
Refs.~\cite{verstraete-2003-68,carteret-2000-41,sudbery-2001-34,acin-2001-34,kempe-1999-60,tamaryan-2008}
for other results on this problem.} \be \ket{\psi}= \lambda_0
\ket{000} + \lambda_1 e^{i\theta} \ket{100} + \lambda_2 \ket{101} +
\lambda_3 \ket{110} + \lambda_4 \ket{111}, \label{acinform} \ee
where $\lambda_i \geq 0, \sum_i \lambda_i^2=1$ and $\theta \in
[0;\pi],$  see Ref.~\cite{acin-2000-85}. Thus, six real parameters
are necessary to characterize the nonlocal properties of a pure
state. For the W class states, however, $\theta=\lambda_4=0$ holds,
which shows that they are a set of measure zero in the set of all
pure states.

Physically, there are also differences between the two
classes: On the one hand, the GHZ state is
maximally entangled and a generalization of the Bell states
of two qubits. For instance, for the most known Bell
inequalities the violation is maximal for GHZ states
\cite{scarani-2001-34}. On the other hand, the entanglement of the
W state is more robust against particle losses: If one
particle is lost in the GHZ state, the state
$\vr_{AB}=Tr_{C}(\ketbra{GHZ_3})$ is separable, for the W state
the resulting reduced density matrix $\vr_{AB}=Tr_C(\ketbra{W})$
is entangled. Indeed, it can be shown that the W state is the state
with
the maximal possible bipartite entanglement in the reduced two-qubit
states
\cite{PhysRevA.62.062314,PhysRevA.62.050302}.

\subsubsection{Mixed states}
The classification of mixed states according to
Refs.~\cite{PhysRevLett.87.040401, dur-1999-83} is similar to the
definition of bipartite entanglement via convex combinations. We
define a mixed state $\varrho^{\rm{fs}}$ as fully separable if
$\varrho^{\rm{fs}}$ can be written as a convex combination of fully
separable pure states, {i.e.,} if there are convex weights $p_i$ and
fully separable states $\ket{\phi^{\rm{fs}}_{i}}$ such that we can
write \be \varrho^{\rm{fs}}=\sum_i p_i \ketbra{\phi^{\rm{fs}}_{i}}.
\ee A state $\varrho^{\rm{bs}}$ that is not fully separable is
called biseparable if it can be written as a convex combination of
biseparable pure states: \be \varrho^{\rm{bs}}=\sum_i p_i
\ketbra{\phi^{\rm{bs}}_{i}}. \ee Note that the biseparable states
$\ket{\phi^{\rm{bs}}_{i}}$ might be biseparable with respect to
different partitions. We will see in Section
\ref{distillabilityversusentanglement} why it is reasonable
to exclude mixtures of biseparable states with respect to different
partitions from the genuine multipartite entangled states.
Of course, one may define three classes of biseparable mixed
states that are biseparable with respect to a fixed partition.

Finally, $\varrho$ is fully entangled if it is neither biseparable
nor fully separable. There are again two classes of fully entangled
mixed states: A fully entangled state belongs to the W class if it
can be written  as a convex combination of W-type pure states \be
\varrho^{\rm{w}}=\sum_i p_i \ketbra{\phi^{\rm{w}}_{i}}, \ee
otherwise it belongs to the GHZ class. One can show that the W class
forms a convex set inside the GHZ class. Also, in contrast to the
case of pure states, the set of mixed W class states is not of
measure zero compared to the GHZ class \cite{PhysRevLett.87.040401}.

\begin{figure}[]
\centerline{\includegraphics[width=0.55\columnwidth]{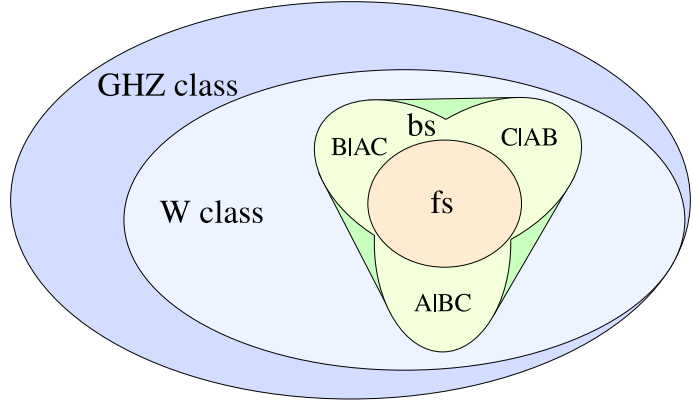}}
\caption{Schematic picture of the structure of mixed states for
three qubits. The convex set of all fully separable states (fs) is
a subset of the set of all biseparable states (bs). The biseparable
states are the convex combinations of the biseparable states with
respect to fixed partitions, sketched by the three different
leafs. Outside are the genuine tripartite entangled states, the W class and
the GHZ class.
\label{3qubitschema}
}
\end{figure}

Again, this classification can be represented in a schematic picture
of nested convex sets (see Fig.~\ref{3qubitschema}). Here, it is
important to note that this picture is, of course, only a schematic
picture, which does not take all properties into account. For
instance, it has been shown that there exist states that are
biseparable with respect to each fixed partition, however, they are
not fully separable (for some examples see
Refs.~\cite{bennett-1999-82, PhysRevLett.87.040401,
optimalspsq}).\footnote{This counterintuitive behavior can
lead to surprising applications, like the possibility of
distributing entanglement using separable states \cite{cubitt-2003-91}.}

The question to which class a given mixed state belongs, is,
as the separability problem for the bipartite case, not easy
to decide. Some conditions will be discussed in Section
\ref{section:multipartitesepcrit}.
Methods to distinguish  between the mixed W class and GHZ class
are, however, very rare. As we will see later, witnesses give the
possibility to prove that a state belongs to the GHZ class.
However, it is not clear how one can show that a state is
tripartite entangled and belongs to the W class.
This can not be done with witnesses, since they are designed to
show that a state lies {\it outside} a convex set, they fail to
prove that a state is {\it inside} a convex set.

\subsection{Multipartite entanglement distillation}

\label{section:distillability}

\subsubsection{Multipartite distillability}
It is also possible to define
distillability for three and more qubits. For instance,
one may aim at the distillation (also called purification)
of the GHZ state $\ket{GHZ_3},$ that is, one asks whether
the transformation
\be
\underbrace{\vr \otimes \vr \otimes ... \otimes \vr}_{ k\;\;\rm copies}
\stackrel{LOCC}{\longrightarrow} \ket{GHZ_3}
\ee
is possible. If this is possible, the state $\vr$ is called multipartite
distillable. The fact that the target state is a GHZ state, is not important
for the question, whether or not $\vr$ is distillable, since it is always
possible to transform a GHZ state in a W state and vice versa, if many copies
of the state are available.

Many protocols for the distillation of different multipartite states
have been proposed (for a review see Ref.~\cite{dur-2007-70}). Also,
multipartite bound entangled states exist. For instance, the above
mentioned states that are separable with respect to each
bipartition, but not fully separable, are multipartite bound
entangled.

\subsubsection{Multipartite distillability vs. multipartite entanglement}
\label{distillabilityversusentanglement}

Let us compare the notion of multipartite distillability with the
notion of genuine multipartite entanglement and their importance for
the analysis of experiments. First, one has to stress that both
notions are inequivalent. There are genuine multipartite entangled
states that are not distillable. Examples of such states were given
in Ref.~\cite{piani-2007-75}.

To investigate the relation better, consider the mixed three-qubit state
\be
\vr = \frac{1}{3}\big(
\ketbra{\phi^+}_{AB} \otimes \ketbra{0}_C +
\ketbra{\phi^+}_{AC} \otimes \ketbra{0}_B +
\ketbra{\phi^+}_{BC} \otimes \ketbra{0}_A
\big),
\ee
where $\ket{\phi^+}=(\ket{00}+\ket{11})/\sqrt{2}$ is a two-qubit Bell
state. $\vr$ is a mixture of biseparable states with respect to different
partitions, and therefore biseparable. However, as can be easily checked,
$\vr$ is entangled with respect to each bipartition, and also any reduced
two-qubit state is NPT and hence entangled and two-qubit distillable. So,
if many copies of $\vr$ are available, the three parties can first distill
singlets between
$A-B$ and $A-C.$ Then, Alice may prepare a GHZ state locally in her lab, and
teleport two of its parties via the existing singlets to Bob and Charly.
By this simple procedure one sees that $\vr$ is multipartite distillable,
showing that multipartite distillability and genuine multipartite entanglement
are really inequivalent.

The important point for our discussion in the above example
is that the state $\vr$ is extremely easy to prepare experimentally.
First, one prepares a product state $\ket{000}$ and then one
entangles {\it randomly} two of the three qubits.  Equivalently,
one can also prepare the state $\ketbra{\phi^+}_{AB}\otimes\ketbra{0},$
and then one {\it forgets} which pair of qubits was the pair
$A,B$.

{From} this example, we can conclude three things. First,
multipartite distillability is not a good criterion for claiming
success of an experiment, since some multipartite distillable states
are very easy to prepare. Second, since $\vr$ is also NPT with
respect to each bipartition, it is not sufficient to investigate all
bipartitions, and the PPT criterion is a very limited tool to
investigate multipartite entanglement. Finally, this example
justifies that we did not define biseparability with respect to a
fixed partition, and considered also mixtures of biseparable states
that are biseparable with respect to different partitions as not
highly entangled. Due to these facts, we will mainly investigate how
genuine multipartite entanglement can be verified in experiments.

\subsection{Entanglement classes for the general case}
\label{sectiongeneralentanglementclasses}
Here we shortly discuss the classification of entanglement for
general multipartite systems.  This is a simple generalization of
the three-qubit case: First, we distinguish  different types of
entanglement for pure states. Then we extend this definition to
mixed states by considering convex combinations of pure states.

Let us assume that a pure $N$-partite state $\ket{\psi}$ is given.
We call this state {\it fully separable} if it is a product state of
all parties, that is if 
\be 
\ket{\psi}=\bigotimes_{i=1}^N \ket{\phi_i} 
\ee 
holds. A mixed state is called fully separable if
it can be written as a convex combination of pure 
fully separable states, that is, it can be written as 
\be 
\vr = \sum_k p_k \vr_k^{(1)} \otimes  \vr_k^{(2)} \otimes 
... \otimes \vr_k^{(N)}.
\ee 
If a pure state is not fully separable, it contains some
entanglement. Again, as in the three-qubit case, this does not have
to be true $N$-partite entanglement. Thus, we call a pure state {\it
$m$-separable}, with $1< m < N$, if there exist a splitting of the
$N$ parties into $m$ parts $P_1,...,P_m$ such that \be
\ket{\psi}=\bigotimes_{i=1}^m \ket{\phi_i}_{P_i} \ee holds. There
are $(m^N)/m!$ possible partitions of the $N$ parties into $m$
parts. Note that an $m$-separable state still may contain some
entanglement. Again, we call mixed states $m$-separable, if they can
be written as convex combinations of pure $m$-separable states,
which might belong to different partitions. One can refine these 
definitions, by considering separability with respect to fixed 
bipartitions 
\cite{dur-1999-83,dur-2000-61-distill,seevinck-2007}.
Finally, we call a state
truly $N$-partite entangled when it is neither fully separable, nor
$m$-separable, for any $m > 1.$ 

The above classification distinguished different classes, according
to the question: ``How many partitions are separable?''.
Alternatively, one may ask the question ``How many particles are
entangled?'' A classification in this sense was proposed in
Refs.~\cite{SeevinckUffinkThreeParticle, 1367-2630-7-1-229,
PhysRevA.73.052319}. According to that, a pure state $\ket{\psi}$ is
said to contain only $m$-party entanglement, if $\ket{\psi}$ can be
written as 
\be 
\ket{\psi}=\bigotimes_{i=1}^K \ket{\phi_i}
\label{producibilitydefinition} 
\ee where the $\ket{\phi_i}$ are states of maximally $m$ qubits. 
Consequently, $K \geq N/m$ has to
hold. If $\ket{\psi}$ is not of this form, it contains (at least)
$m+1$-party entanglement. Again, this definition can be extended to
mixed states via convex combinations. A mixed state that is
$k$-producible requires only the generation of $k$-party pure
entangled states and mixing for its production. Consequently, 
a mixed state $\vr$ contains $k$-party entanglement, iff the 
density matrix cannot be obtained by mixing pure states that
are $(k-1)$-producible. This classification of
separability classes does not play a significant role in the
analysis of experiments, since there typically only few particles
are considered, and the aim is to prove genuine multipartite
entanglement, which coincides with the statement that $\vr$ contains
$N$-partite entanglement. However, for spin models with a large
number of particles, this notion can be a reasonable property to
study \cite{1367-2630-7-1-229, PhysRevA.73.052319} (see also Section
\ref{sectionenergywit}).

Another characterization is possible based on the number of
unentangled particles. An $N$-qubit pure product state contains $n$
unentangled particles if it can be written as
$\ket{\Psi}=\bigotimes_{k=1}^{n}\ket{\phi_k} \otimes
\ket{\Psi_{n+1..N}}$ where $\ket{\phi_k}$ are single-qubit states,
while $\ket{\Psi_{n+1..N}}$ is a $(N-n)$-qubit state vector.
Consequently, a mixed state contains at most $n$ unentangled spins,
if it can be constructed mixing pure states with $n$ or less
unentangled spins. There are entanglement conditions that can
provide upper bounds on the number of unentangled spins (see
Sec.~\ref{sec_singletcrit}). If there are no unentangled spins, the
state is not necessarily genuine multi-partite entangled: It can be
a tensor product of two-qubit entangled states. However, creating
very many (e.g., $10^5$) two-qubit singlets can be a goal of quantum
control in itself.

Finally, it is worth mentioning that a classification of the pure truly
$N$-partite entangled states like in the three-qubit case is not straightforward.
At least a classification via equivalence classes under SLOCC is not simple,
since it has been shown that already for four qubits there are infinitely
many equivalence classes under SLOCC \cite{verstraete-2002-65,lamata-2006-74}.
This continuous set can then be separated into nine or eight different classes.
Other approaches try to identify different classes via a generalization of the
Schmidt rank \cite{eisert-2001-64} or tensor ranks \cite{chitambar-2008}, but
there it is difficult to determine in which class a pure state lies.

\subsection{Families of multipartite entangled states}

\label{Sec_FamMultipartiteEntStates}

In order to further investigate multipartite entangled states, it is
useful to identify interesting families of multipartite pure states.
Here, one might be interested in maximally entangled states, or
states that are useful for certain applications. For the three-qubit
case, we already know the GHZ state and the W state as such
examples. There are, however, several families of interesting
multi-qubit states with an arbitrary number of qubits:

\subsubsection{GHZ states}
The GHZ state for $N$ qubits is defined as
\be
\ket{GHZ_N} = \frac{1}{\sqrt{2}} \big(\ket{0}^{\otimes N} + \ket{1}^{\otimes N}\big)
\label{ghzndef}
\ee
and is a simple generalization of the three-qubit GHZ state.
It is a superposition of two maximally distinct states, and
therefore also sometimes called a ``Schr{\"o}dinger cat state''.
GHZ states have been studied intensively for a long time and it
turned out that they are useful for many applications. These
include entanglement enhanced spectroscopy and quantum metrology
\cite{ghzspectroscopy, giovannetti-2004-306},
quantum secret sharing \cite{PhysRevA.59.1829}, open destination
teleportation \cite{zhaofiveghz}, quantum computation \cite{gottesmanchuang}
and cryptographic protocols \cite{chen-2004, christandl-2005-3788}.
Furthermore, it has been shown
that GHZ states lead to an extremal violation of local realism
\cite{greenberger-1989,ghz-argument,PhysRevLett.65.1838}. In fact,
it can be shown that for many Bell inequalities,
the unique state leading to a maximal violation is the GHZ state
\cite{scarani-2001-34}.

These properties make the GHZ states interesting from many
perspectives. In some sense they may be considered as the maximally
entangled multi-qubit  states. Therefore, many experiments aim at
the generation of GHZ states, and GHZ states with up to ten qubits
have already been prepared using ions
\cite{sackettfourghz,roosthreequbits, leibfriedsixghz} or photons
\cite{ThreeQubitGHZ1, ThreeQubitGHZ, FourPhotonGHZ, zhaofiveghz,
lu-2007-3, gao-2008}.

\subsubsection{Dicke states and W states}

These states have been investigated already in 1954 by R.H. Dicke,
while studying light emission from a cloud of atoms \cite{PhysRev.93.99}.
Dicke found that, if the atoms are in certain highly entangled states, then the
intensity of radiation is much larger than if the atoms were
emitting light independently, starting all from an excited state. 
In particular, the intensity scales
quadratically with the number of atoms, while for independent atoms
it scales linearly.

Dicke states, in general, are simultaneous eigenstates of the
collective angular momentum operators $J_z = 1/2 \sum_k
\sigma_z^{(k)}$ and $J^2.$\footnote{We will discuss these operators
and their properties later in Section \ref{Sec_collspinent} in more
detail.} Typically, for a certain eigenvalue for $J_z$ and $J^2$
there are several eigenstates and additional parameters are needed
to distinguish them from each other. Since very often many-particle
systems start from a symmetric initial state and they evolve under
symmetric dynamics, perhaps the most important examples for quantum
information are the symmetric Dicke states. Such states are all
eigenstates of $J^2$ with the maximal eigenvalue
$\tfrac{N}{2}(\tfrac{N}{2}+1)$ and they can be characterized
uniquely by their eigenvalues $j_z$ for $J_z.$ Beside $j_z,$ it is
also common to use for their characterization $k=N/2-j_z.$ With this,
an $(N,k)$ symmetric Dicke state  on $N$ qubits is defined as an
equal superposition of $k$ excitations, i.e., \be \ket{D_{k,N}} = {N
\choose k}^{-\frac{1}{2}} \sum_j P_j \Big\{\ket{1}^{\otimes k}
\otimes \ket{0}^{\otimes N-k} \Big\}, \ee where $\sum_j P_j \{...\}$
denotes the sum over all possible permutations of the qubits.

The W state corresponds to the Dicke state $\ket{D_{1,3}}$ and is therefore
an example of a Dicke state. Another important Dicke state is the
four-qubit state $\ket{D_{2,4}} =
(\ket{0011}+\ket{0101}+\ket{0110}+\ket{1001}+\ket{1010}+\ket{1100})/\sqrt{6}.$
In general,
the state $\ket{D_{\tfrac{N}{2},N}}$ is the state for which the superradiance
is the strongest. Also, since its overlap with biseparable states is
the minimum $\tfrac{1}{2}$ for large $N$ \cite{DickeEntanglementJOSAB2007},
for an experimental investigation of genuine multipartite entanglement only
a fidelity of $\tfrac{1}{2}$ is required. Finally, it should be noted that the
entanglement in the Dicke state $\ket{D_{2,4}}$ is  relatively robust against
decoherence \cite{guehneblaauboer}. Dicke states and W states have been prepared
in many experiments \cite{roosthreequbits,DickeExperimentPRL2007,haeffner-2005-438,
PhysRevLett.92.077901,wieczorek:010503}. We will discuss two of them 
in Section \ref{sectionionexp}
and \ref{sectiondickeexp}.

\subsubsection{Graph states and cluster states}
\label{Sec_GraphStatesClusterStates}

\begin{figure}[t]
\centerline{\epsfxsize=0.95\columnwidth
\epsffile{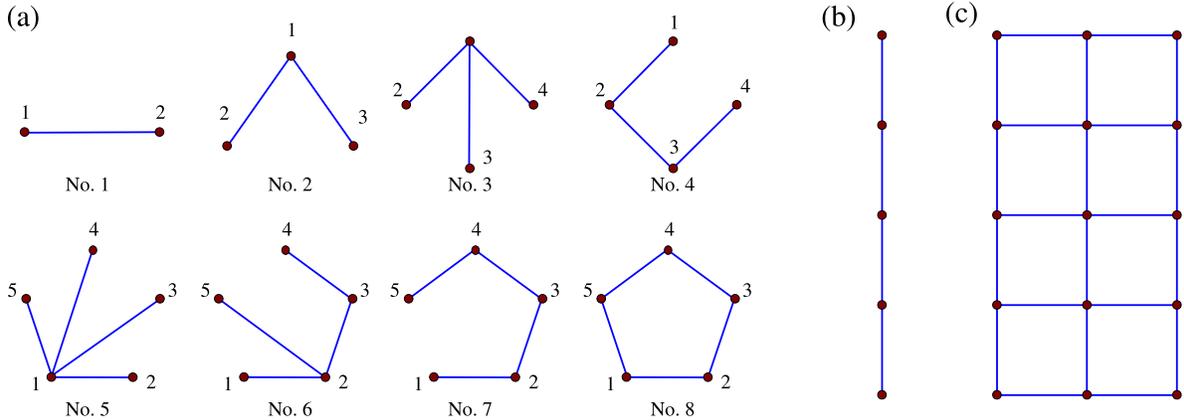} }
\caption{
(a) Several types of graphs for up to five qubits. The shown graphs
represent all inequivalent graph states up to five qubits \cite{hein-2006}.
(b) The graph for a five qubit linear cluster state. (c) The graph for a
2D cluster state (for the case of 15 qubits). This type of graph states
is a universal resource for measurement based quantum computation.
}
\label{grafig}
\end{figure}

Let us come now to the family of graph states \cite{dur-2004-92,
hein-2004-69, hein-2006}. These states are in a mathematical way
defined as follows:
First one considers a graph, i.e., a set of $N$ vertices and some
edges connecting them. Examples of graphs are  shown in
Fig.~\ref{grafig}. For each vertex $i$ the neighborhood $\NN(i)$
denotes the vertices that are connected with $i.$ Then one can
associate to each vertex $i$ a {correlation operator} or
{stabilizing operator} $g_i$ by
\begin{equation}
g_i= X_i\bigotimes_{j\in \NN(i)} Z_j.
\label{stabilizeredefiningequation}
\end{equation}
Here $X_i, Y_i, Z_i$ denote the Pauli matrices
$\sigma_x,\sigma_y,\sigma_z,$ acting on the $i$-th qubit;
we will use this notation often when working in the
graph state formalism. For instance, for the three
vertex graph No.~2 in Fig.~\ref{grafig}(a), the stabilizing
operators are $g_1=Z_1 X_2 \eins_3, g_2=X_1 Z_2 Z_3$
and $g_3=Z_1 \eins_2 X_3.$ The {graph state} $\ket{G}$ associated
with the graph $G$ is the then the $N$-qubit state fulfilling \be
g_i \ket{G}= \ket{G}, \mbox{ for } i=1,...,N.
\label{graphstatedefiningequation} \ee Clearly, $\ket{G}$ is then
also an eigenstate of all possible products of the $g_k,$ the
commutative group of this operators is called the stabilizer
\cite{PhysRevA.54.1862}. Since the $g_i$ form a maximal set of
commuting observables, the graph state $\ket{G}$ is uniquely
determined by these eigenvalue equations.\footnote{For a given
graph, one can consider a whole family of $2^N$ states that are
eigenstates of the $g_i$ with eigenvalues $\pm 1.$ This orthogonal
basis is the so-called graph state basis, we will use it in Section
\ref{sectionstabilizerwitnesses} in more detail.}

The physical meaning of the graph is that it describes the perfect
correlations in the state $\ket{G},$ since $
\mean{g_i}=\mean{X_{i}\bigotimes_{j\in N(i)} Z_{j}}=1.$ At the same
time, the graph denotes a possible interaction history leading to
$\ket{G}:$ the state $\ket{G}$ can be produced from the product
state $[(\ket{0} + \ket{1})/\sqrt{2}]^{\otimes N}$ by an Ising
type interaction $U_{kl}=\exp\{-i t (\eins_k-Z_l)(\eins_k - Z_l)\}$
acting for the time $t=\pi/4$ between the connected qubits.\footnote{If
the interaction time depends on $k$ and $l$, the resulting state
is a so-called weighted graph state, see Ref.~\cite{lorenzpap}
for discussions and applications.}
Finally, the graph state can also directly
be written in the stabilizer elements as
\be
\ketbra{G}=\prod_{k=1}^{N}\frac{\eins + g_k}{2}.
\ee

There are several interesting properties of graph states.
First, it should be noted that the GHZ states belong to this class.
If we consider the state in Eq.~(\ref{ghzndef}),
it can be easily checked that this is an eigenstate of
\be
g_1 = X_1 X_2 ...X_N; \;\;\;
g_k = Z_1 \eins_2 ...  Z_k ...  \eins_{N};\;\;\;  \mbox{  for  } k=2,...,N.
\ee
Up to some local rotations, these are nothing but the stabilizing
operators of the star graphs [Nos.~2, 3, and 5 in Fig.~\ref{grafig}(a)].
Therefore, graph states can be seen as an extension of the GHZ states.

Second, it is important to note that different graphs may lead to
graph states, which can be transformed into each other by local
unitary transformations, implying that the entanglement properties
of the states are the same. In general, it is not known yet, under
which conditions two graphs yield equivalent graph states
\cite{gross-2008-8,ji-2007}. However, it is known that some graph
transformations (called local complementation) do not change the
graph state, and for a small number of qubits ($N \leq 8$) a full
classification has been achieved \cite{hein-2006,2008arXiv0812.4625C}.

This leads directly to the question, what other graph states exists
beside the GHZ state. For 2 qubits, there is only one graph [no.~1
in Fig.~\ref{grafig}(a)], corresponding to a Bell state. For three
qubits, there is again only one independent graph state,
corresponding to the three-qubit GHZ state (No.~2). For four qubits,
there are already two independent graph states: the four-qubit GHZ
state (No.~3), and the four-qubit cluster state, given by \be
\ket{CL_4}= \frac{1}{2} \big( \ket{0000}+
\ket{0011}+\ket{1100}-\ket{1111} \big), \label{fourqubitclustereq}
\ee which corresponds (up to a local change of the basis) to the
graph no.~4. For five qubits there are already four different graph
states (nos. 5-8), and with increasing number of qubits, the number
of inequivalent graph states increases rapidly.

Third, it should be stressed that graph states are of eminent
importance for applications and protocols of quantum information
processing. First, a special graph state, the 2D cluster state
[see Fig.~\ref{grafig}(c)]
can be used as a resource for measurement based
quantum computation \cite{PhysRevLett.86.910,PhysRevLett.86.5188}. In addition,
graph states occur as codewords of quantum error correcting codes
\cite{gottesman-1996-54}, violate certain Bell inequalities maximally
\cite{PhysRevLett.95.120405,PhysRevA.71.042325,cabello-2007}, and are
also robust against
decoherence \cite{dur-2004-92}. Finally, graph states can be used for
the simulation of anyonic statistics \cite{han-2007-98} and can be
related to the computation of the partition function
of classical spin models \cite{nest-2007-98}.

Due to this importance, it is not surprising that graph states have been
prepared in many experiments \cite{PhysRevLett.95.210502, walther-2005-434,
TwoPhotonClusterState, chen-2007-99, lu-2007-3,lu-2007-anyon}.
We will discuss in Section \ref{clusterexp} an experiment
for the generation of the four-qubit cluster state.

\subsubsection{Multi-qubit singlet states}
\label{sectionsinglets}

Another interesting family of multi-qubit states are multi-qubit
singlet states. These are pure states $\ket{\psi},$ that are
invariant under a simultaneous unitary rotation on all qubits, i.e.,
we have $(U)^{\otimes N}\ket{\psi} = e^{i\phi}\ket{\psi}$ (or,
equivalently $ [\ketbra{\psi}, (U)^{\otimes N}] =0$). Such states
exist only for the case that $N$ is even. For two qubits, the only
state of this form is the two-qubit singlet state \ket{\psi^-}. For
four qubits, these states form a two-dimensional subspace of the
total Hilbert space, spanned by the vectors $\ket{\Psi_1}=
\ket{\psi^-}\otimes\ket{\psi^-}$ and \be \ket{\Psi_2} =
\frac{1}{\sqrt{3}} \big[ \ket{0011} +\ket{1100} -
\frac{1}{2}(\ket{01}+\ket{10})\otimes(\ket{01}+\ket{10}) \big].
\label{fourqubitsingletstate} \ee This is the so-called four-qubit
singlet state \cite{PhysRevA.64.010102}. In general, there are $d_N
= N!/[(N/2)!(N/2+1)!]$ linearly independent states in this family
\cite{cabello-2007-75, cabello-2003-50}.

The importance of these states comes from the fact that they are by
definition robust against certain types of noise, where all qubits
undergo the same (random or unknown) unitary transformation. They
lie in a so-called decoherence free subspace \cite{lidar-1998-81,
zanardi-1997-79}. This can be used to encode quantum information,
e.g., one qubit can be encoded in superpositions of the four-qubit
states  $\ket{\Psi_1}$ and $\ket{\Psi_2}.$

Experimentally, it is remarkable that the state $\ket{\Psi_2}$ can
simply be prepared by a second order process in parametric
down-conversion \cite{PhysRevA.64.010102} (see also Section
\ref{mohamedexp}). This has been done in
Ref.~\cite{PhysRevLett.90.200403}, also protocols like secret
sharing or the encoding of quantum information as described above
have been performed \cite{gaertner-2007-98,bourennane-2004-92}.

The expression {\it singlet} is also used in another context,
denoting states with zero total angular momentum. That is, these are
the states that are the eigenstates of the total angular momentum
components $J_x,$ $J_y$ and $J_z$ with eigenvalue zero. Such states
appear very often in condensed matter physics, for example, the
ground state of the anti-ferromagnetic Heisenberg chain is such a
multi-qubit singlet state. We will discuss entanglement detection in
the vicinity of such states in Section~\ref{Sec_collspinent}.

The connection between the two definitions can be seen as
follows.\footnote{We thank Sofyan Iblisdir for discussions about
this point.} One can rewrite a simultaneous multilateral unitary
transformation as \be U\otimes U\otimes...\otimes U
=\exp(i\vec{\alpha}\cdot \vec{J}), \ee where $\vec{J}=(J_x,J_y,J_z)$
and $\alpha$ is a three-element real vector. Hence, we can see that
if a pure state is an eigenstate of $\exp(i\vec{\alpha}\vec{J})$ for
any $\alpha,$ then it is an eigenstate of $J_k$ for $k=x,y,z,$ and
vice versa. Thus, the two definitions coincide for pure states.

\subsubsection{Werner states}
\label{sectionwernerstates}

In the general case, $N$-qudit mixed states invariant under
$U\otimes U\otimes...\otimes U$  are called Werner states \cite{werner89}.
They do not necessarily have zero total angular momentum variance. Such
multi-qudit states naturally arise as the result of the twirling
operation
\be
{\rm Twirl} [\vr] :=\int U^{\otimes
N}\vr(U^{\dagger})^{\otimes N}\mu(dU),
\ee
where the integral is
over the unitary group of $d\times d$ matrices $U(d),$ and $\mu$ is
the Haar measure. Multi-qudit twirling can efficiently be realized
experimentally \cite{tothgarciaripoll}. The resulting Werner states
are the linear combination of relatively few basis matrices
\cite{werner89,eggeling2001spt,gross2007edu}. For example,
for $N=2$ there are two such matrices for any dimension $d,$
while for $N=3$ there are $5$ for $d=2$ and $6$ for any $d>2.$
Thus, such states can be characterized with few and simpler
measurements than the measurements needed for full tomography.
For example, for the two-qubit case the single free parameter of a two-qubit Werner state
can be obtained by measuring $\exs{\sigma_x^{(1)}\sigma_x^{(2)}}.$
Twirling has also been suggested for simplifying process tomography
\cite{emerson2007scn}.

\subsubsection{Maximally entangled states for comb monotones}
As already mentioned, it is a natural question to consider the
maximally entangled states for $N$ qubits. This, however, depends on
the entanglement measure chosen: As we will discuss later, there are
several inequivalent measures of multipartite entanglement, and the
maximally entangled states differ for different measures.

In Ref. \cite{osterloh-2005-72, osterloh-2006-4} a class of
entanglement measures has been introduced, which are based on
anti-linear operators and combs. There are three different measures
of this type for four qubits, resulting in  three different
maximally entangled states. The first one is the GHZ state, the
second one the cluster state, and the third one is the state \be
\ket{\chi}=\frac{1}{\sqrt{6}} \big(\sqrt{2}\ket{1111} + \ket{0001}+
\ket{0010}+\ket{0100}+\ket{1000} \big). \label{osterlohsiewertstate}
\ee This state has also been mentioned in Ref.~\cite{krausdiss},
where it has been noticed that it is the symmetric four-qubit state
 that maximizes certain bipartite entanglement properties. Also,
$\ket{\chi}$ is probably the most entangled symmetric state for the
geometric measure of entanglement \cite{muraoposter} (see also
Section \ref{geometricmeasuresection}).

As it was shown in Ref.~\cite{osterloh-2006-4} one can generalize
this state to five and six qubits, where it is also a maximally
entangled state for some comb measure. It is not yet clear, which
physical properties follow from this fact. Also, the generation of
this state is still an experimental challenge.

Finally, it should be added that there are further types of states,
which deserve attention. For example, there are states for optimal
cloning via teleportation \cite{bruss-1998-57}, or the family of
multipartite states which can be maximally entangled with local
auxiliary systems \cite{kruszynska-2008}.

\subsection{Separability criteria for the multipartite case}
\label{section:multipartitesepcrit}
As for the bipartite case, one may ask how to decide whether a given
mixed multipartite state is entangled or not. This problem is not so
well understood as for the bipartite case, however, several criteria
are known:

\begin{enumerate}

\item {\it Permutation criteria.} As discussed already for the bipartite case,
one can view the PPT as well as the CCNR criterion as norm
conditions on a density matrix where some indices are permuted [see
Eq.~(\ref{permcrit})]. This approach can be generalized to more
parties \cite{horodecki-2006-13,wocjan-2005-12,clarisse-2006-6}. One
first writes the density matrix in some product basis,
$\vr=\sum_{i_1, j_1, ..., i_N, j_N}\vr_{i_1, j_1, ..., i_N, j_N}
\ket{i_1}\bra{j_1} \otimes ... \otimes \ket{i_N}\bra{j_N},$ and then
for separable states we have \be \Vert\vr_{\pi(i_1, j_1, ..., i_N,
j_N)}\Vert_1 \leq 1 \ee where $\pi(...)$ is an arbitrary permutation
of the indices. This approach generalizes the CCNR and PPT criterion
to the multipartite case.\footnote{For a different generalization of
the CCNR criterion see Ref.~\cite{chen-2002-306}.} Clearly, not all
permutations $\pi$ result in different criteria. As we have already
seen, for two parties there are only two inequivalent permutations,
resulting in the PPT and CCNR criterion. Further, it has been shown
in Refs.~\cite{wocjan-2005-12,clarisse-2006-6} that for the case of
three parties there are six independent permutations, and for four
parties there are 22 independent permutation criteria. All these
criteria can only rule out full separability.

\item {\it Quadratic Bell-type inequalities.} It was already demonstrated in
Ref.~\cite{PhysRevLett.88.230406} that quadratic Bell-type
inequalities can be a useful tool for the investigation of
multipartite entanglement. Recently, a general method to derive
quadratic inequalities for the investigation of the different
classes of multipartite entanglement has been found
\cite{seevinck-2007, seevinckthesis}. This method can be sketched as
follows: for two qubits and $i=0,1$ one considers the families of
observables $A^{(2)}_i = (A_1 \otimes A_2 \mp B_1 \otimes B_2)/2$,
$B^{(2)}_i = (B_1 \otimes A_2 \pm A_1 \otimes B_2)/2$, $C^{(2)}_i =
(C_1 \otimes \eins_2 \pm \eins_1 \otimes C_2)/2$, and $I^{(2)}_i =
(\eins_1 \otimes \eins_2 \pm C_1 \otimes C_2)/2$, where $A_k, B_k,
C_k$ denote orthogonal spin observables, i.e., Pauli matrices in
arbitrary directions. The two-qubit observables $A^{(2)}_i, ...,
I^{(2)}_i$ fulfill similar relations as the Pauli matrices, and for
separable states it can be shown that \be \max_i\{\mean{A_i^{(2)}}^2
+ \mean{B_i^{(2)}}^2\} \leq \min_i\{\mean{I_i^{(2)}}^2 -
\mean{C_i^{(2)}}^2 \} \ee holds. This is in fact a necessary and
sufficient condition for two-qubit separability
\cite{uffink-2008-372}. This construction can then be iterated: One
defines four sets of three-qubit observables $A^{(3)}_i, ...,
I^{(3)}_i$ as $A^{(3)}_i = (A_1 \otimes A^{(2)}_k \mp B_1 \otimes
B^{(2)}_k)/2$ etc., and it is possible to prove similar relations as
above for the different classes of separability. These conditions
can be translated into conditions on the density matrix elements by
fixing $A_i, B_i$ and $C_i.$ For instance, for three qubits it can
be shown that the matrix elements of biseparable states fulfill
\be |\vr_{18}| \leq
\sqrt{\vr_{22} \vr_{77}} + \sqrt{\vr_{33}
\vr_{66}}+\sqrt{\vr_{44} \vr_{55}}.
\ee
These and the other
conditions improve some already known conditions
\cite{laskowski-2005-72}. In addition, they are relatively easy to
check in experiments.

\item {\it Algorithmic approaches.} Also some of the computational criteria
can be extended to the multipartite scenario. The method of the
symmetric extensions can be formulated for the multipartite case
\cite{doherty-2005-71}, and it can be shown again that the hierarchy
is complete. However, this method is practically not very feasible.
Also, the method of Brand{\~a}o and Vianna \cite{brandao-2004-93} can
be applied to show the entanglement of some multipartite bound entangled
states.

\item {\it Further criteria.} There is a series of other approaches
that try to extend bipartite criteria to the multipartite case.
These include positive maps \cite{horodecki-2000-pla} and
conditions on the generalization of the Bloch vector
\cite{yu-2005-72,hassan-2007}. There is also a variety of
entanglement witness like criteria for full separability, these will
be discussed in Section \ref{fullsepwitnesses}. Finally, it should
be noted that any bipartite separability criterion can also be used
to rule out full separability, since a fully separable state is also
separable with respect to each partition.

\end{enumerate}

Finally, it should be added that also criteria for the
distillability of a multipartite quantum state are known. First, in
a similar argument as in Section (\ref{section:distillability}) one
can conclude that a mixed state is distillable, if enough reduced
two-qubit states are NPT, allowing the distillation of singlet pairs
between them. For some special families of states (i.e., states that
are diagonal in a graph state basis or the GHZ basis) more elaborate
criteria have been worked out
\cite{dur-2000-61-distill,dur-2001-34}.

\subsection{Entanglement witnesses for multipartite entanglement}
\label{multiewsection}

Let us now discuss entanglement witnesses for multipartite entanglement.
Clearly, witnesses can be used to distinguish the different classes
of multipartite entanglement. This is due to the fact that all the classes
defined above are defined via convex combinations, thus they are convex
sets. Contrary to the bipartite case, one has several types of witnesses,
for the different types of multipartite entanglement.

\subsubsection{Three qubits}

First, there are GHZ class witnesses $\WW_{GHZ}$ that allow to
detect mixed states belonging to the GHZ class. Thus, their
expectation value is positive on all fully separable, biseparable,
and W-type states: \bea Tr(\WW_{GHZ} \vr) & < & 0  \; \Rightarrow
\;\vr \mbox{ is in the GHZ class. } \nonumber
\\
Tr(\WW_{GHZ} \vr) & \geq & 0 \; \Rightarrow \; \vr \mbox{ is not detected. }
\label{mewit1}
\eea
An example for such a witness is
\be
\WW_{GHZ} = \frac{3}{4} \eins - \ketbra{GHZ_3}.
\ee
This witness is constructed as in
Eqs.~(\ref{projwitdef}, \ref{projwitdef2}), and the constant $3/4$
is the maximal overlap between the state $\ket{GHZ_3}$ and the pure
W class states \cite{PhysRevLett.87.040401}. In other words, this witness
expresses the fact
that if the  fidelity of the GHZ state is larger than $3/4$, then the
state belongs to the GHZ class.

In a similar fashion, other witnesses can be defined: Witnesses for
genuine tripartite entanglement, denoted by $\WW_3,$ have a positive
expectation  value on all biseparable states thus a negative expectation
value indicates the presence of true tripartite entanglement. Witnesses for
biseparable entanglement $\WW_e$ have a positive expectation value on all fully
separable states, thus a negative expectation value is still a signature
of entanglement, which might be only biseparable entanglement.

\begin{figure}[]
\centerline{\includegraphics[width=0.55\columnwidth]{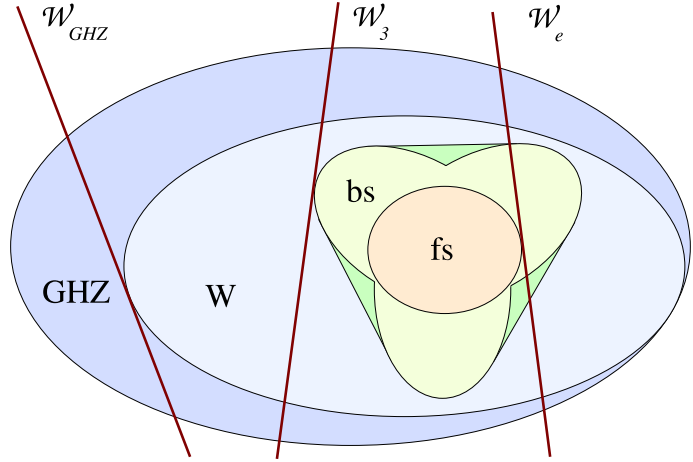}}
\caption{
Schematic picture of the set of mixed three-qubit states with three
witnesses: The witness $\WW_{GHZ}$ detects GHZ-type states, the
witness $\WW_3$ detects three-qubit entanglement,
and the witness $\WW_e$ rules out full separability.
\label{3qubitwitnessfigure}
}
\end{figure}

Such witnesses can be constructed in a similar manner as above: one makes
the ansatz
\be
\WW = \alpha \eins -\ketbra{\psi},
\label{standardwitness}
\ee
where $\alpha$ is the maximal overlap between $\ket{\psi}$ and the
biseparable (or fully separable, respectively) states. Note that
it suffices to consider pure biseparable states, as they are the
extremal points of the convex set of mixed biseparable states.
The calculation of $\alpha$ depends on the type of states,
one wishes to exclude. All the types of witnesses
can be cast into a schematic picture as shown in
Figure \ref{3qubitwitnessfigure}.

The computation of the maximal overlap between a given state
$\ket{\psi}$ and the pure biseparable states is straightforward
\cite{bourennane:087902}: Any pure biseparable state is separable
with respect to some fixed bipartition, and for this case we know
already that this overlap is given by the maximal squared Schmidt
coefficient [see Eq.~(\ref{projwitdef})]. So one can take for the
construction of witnesses for genuine tripartite entanglement \be
\alpha = \max_{{\rm bipartitions}\;\; bp} \big\{ \max_{{\rm Schmidt
\;\;coefficients} \;\; s_k(bp)} \{[s_k(bp)]^2 \} \big\} \ee as the
maximal squared Schmidt coefficient over all bipartitions. The
computation of other overlaps, e.g., the overlap with fully
separable states is, in general, not so simple. However, for many
cases, results on this problem are known, especially in the context
of the so-called geometric measure of entanglement
\cite{wei-2003-68}, which will be explained in Section
\ref{geometricmeasuresection}\footnote{However, such witnesses are
not so relevant for this review, as the detection of genuine
multipartite entanglement has more experimental relevance.}. To give
an example, a witness for genuine multipartite entanglement around
the state $\ket{W_3}$ would be $\WW_3=2/3 \cdot  \eins -
\ketbra{W_3}$ and a witness excluding full separability would be
$\WW_3=4/9 \cdot \eins - \ketbra{W_3}.$

\subsubsection{The general case}

For the case of more than three qubits similar methods can directly be
applied. Especially for the detection of genuine multipartite entanglement
the method from above provides a standard method for the construction of
witnesses.
In addition, the required overlaps are for many classes of states known.

For instance, it has been shown in Ref.~\cite{PhysRevLett.94.060501}
that the maximal overlap between graph states and biseparable states
is always $1/2$, hence
\be
\WW_{G_N} = \frac{\eins}{2}-\ketbra{G_N}
\label{WGN}
\ee
is for any graph state $\ket{G_N}$ a suitable
witness. Especially, \be
\WW_{GHZ_N}= \frac{\eins}{2}-\ketbra{GHZ_N} 
\mbox{  and  } \WW_{CL_N}= \frac{\eins}{2}-\ketbra{CL_N} \label{WCN}
\ee are suitable witnesses for GHZ states and cluster states,
respectively. For some symmetric Dicke states, the maximal overlap can also be
computed straightforwardly \cite{DickeEntanglementJOSAB2007}, e.g.,
a witness for the $\ket{W_N}$ state reads
 \be
 \WW_{W_N} = \frac{N-1}{N}\eins -\ketbra{W_N}.
 \label{WN}
 \ee
and for the $\ket{D_{\tfrac{N}{2},N}}$ state it is
\begin{equation}
\WW_{D_N}=\frac{1}{2}\frac{N}{N-1}\eins-\ketbra{D_{\tfrac{N}{2},N}},
\label{W_Dicke}
\end{equation}
Other witnesses for the states in Eqs.~(\ref{fourqubitsingletstate},
\ref{osterlohsiewertstate}) read \bea \WW_{\Psi_2} =
\frac{3}{4}\eins - \ketbra{\Psi_2} &\;\;\;\; \mbox{and} \;\;\;\; &
\WW_{\chi}= \frac{1}{2}\eins - \ketbra{\chi}.
\label{psivierwitnesseq}
\eea
Note that these are only the basic constructions for witnesses for multipartite states,
more sophisticated constructions and  their experimental implementation will be discussed
in great detail in Section 6.

\subsubsection{Witnesses for full separability}
\label{fullsepwitnesses}

Let us finally mention some other ways of constructing
witnesses (or witness like inequalities) for multipartite
systems which result in conditions excluding full separability.

First, there are witnesses for multipartite entanglement that are
derived in a similar way as Bell inequalities
\cite{PhysRevLett.94.010402,PhysRevLett.88.230406}. Then, one can
use geometrical arguments to construct entanglement witnesses
\cite{durkin-2005-95,badziag-2007} or use properties of correlation
functions \cite{kaszlikowski-2007}. Also, similarly as demonstrated
above, one can construct witnesses, if the expectation value for a
given observable can be bounded for product states. To compute such
bounds, linear programming can be used
\cite{BellStateDiagEntWit,MultiQubitStabClusterEntWit,jafarizadeh-2006}.

Finally, other ways of constructing witnesses will be discussed in
the other parts of this review. Bell inequalities will be discussed
in Section 5. In Section \ref{sectionstabilizerwitnesses} we explain
the construction of witnesses using stabilizer theory. In Section
\ref{Sec_collspinent} we will discuss separability criteria based on
collective angular momentum observables (spin squeezing inequalities).
In Section \ref{sectionenergywit} we will review the possibility
to use the Hamiltonian of a spin model as an entanglement witness.

\section{Entanglement measures}
\label{sectionentanglementmeasures}
In this Section, we explain the basic notions of entanglement
measures. We restrict our attention to the basic definitions and
requirements of entanglement measures that are needed for the
methods presented later to analyze entanglement in experiments. For
a deeper discussion, we can recommend several excellent general
reviews on entanglement measures \cite{horodecki-2007,plenio-2005-1,horoqicreview}.

\subsection{General properties of  entanglement measures}

\subsubsection{Requirements for entanglement measures}

By definition, entanglement measures (or entanglement monotones)
should quantify the amount of entanglement in a given state. As 
such, a general entanglement measure $E(\vr)$ should have several 
desired properties \cite{vedral-1997-78}, which are listed below. 
It should be noted, however, that not all of the following properties 
are fulfilled by all discussed entanglement quantifiers.

\begin{enumerate}

\item First, a  natural requirement is that the entanglement
measure $E(\vr)$ vanishes, if $\vr$ is separable.

\item Second, an entanglement measure should be invariant
under a local change of the basis. This means that it
should be invariant under local unitary transformations,
\be
E(\vr) = E(U_A \otimes U_B \vr U_A^\dagger \otimes U_B^\dagger).
\ee

\item As entanglement can not be created by LOCC, it
is reasonable to require that $E(\vr)$ does not increase under such
transformations. That is, if $\Lambda^{\rm LOCC}$ is a positive map
that can be implemented by LOCC, then \be E[\Lambda^{\rm LOCC}(\vr)]
\leq E(\vr). \label{locc1} \ee Often, this condition is replaced by
a different requirement, namely that $E(\vr)$ should not increase on
average under LOCC. That is, if a LOCC transformation maps $\vr$ to
states $\vr_k$ with probabilities $p_k,$ then \be \sum_k p_k
E(\vr_k) \leq E(\vr). \label{locc2} \ee This condition is stronger
than Eq.~(\ref{locc1}), but many entanglement measures also fulfill
this stronger condition. The monotonicity under LOCC in
Eq.~(\ref{locc1}), implies invariance under local unitary
transformations.

\item A further property that is often demanded and that is
fulfilled by most entanglement measures is convexity. That is, one
requires that entanglement decreases under mixing of two or more
states, \be E(\sum_k p_k \vr_k) \leq \sum_k p_k E(\vr_k). \ee This
inequality expresses the fact that if one starts from an ensemble
$\vr_k$ of states, and looses information about the single instance
$\vr_k$, then the entanglement decreases.

Not all measures fulfill this property, and it has  been argued that
the convexity condition may be relaxed by requiring only that
$E(\vr)$ should not increase if locally distinguishable states are
mixed (see Ref.~\cite{plenio-2005-95} for a discussion).

\item
Further questions arise, if more than two copies of a state
are available. For instance, if Alice and Bob share $n$ copies
of the same state $\vr,$ it may be reasonable to require
additivity, that is
\be
E(\vr^{\otimes n}) = n E(\vr).
\ee
Even stronger, one may require full additivity. This means, that if
Alice and Bob share two different states, $\vr_1$ and $\vr_2,$ then
\be
E(\vr_1 \otimes \vr_2) = E(\vr_1)+ E(\vr_2).
\ee
Such additivity requirements are not fulfilled for some measures,
or they are difficult to prove \cite{plenio-2005-1}.

\end{enumerate}

\subsubsection{Convex roof constructions}

There a several strategies to define an entanglement monotone. A
first choice is to take the usefulness of a state for a certain task
as a measure of entanglement. For instance, one may define for a
state $\vr$ the optimal distillation rate  $E_D(\vr)$ as a measure
of entanglement, as we will discuss in Section
\ref{entanglementofdistillationsection}. Clearly, such quantities
are difficult to compute, especially for mixed states, as their
computation involves a complicated optimization over all
distillation protocols. Only for special cases, exact results are
known (see Refs.~\cite{horodecki-2007, plenio-2005-1} for details).

Another strategy for defining entanglement measures uses the
so-called convex roof construction. For that, one first defines
the measure  $E(\ket{\psi})$ for pure states, and then defines
for mixed states
\be
E(\vr) = \inf_{\small p_k, \ket{\phi_k}} \sum_k p_k E(\ket{\phi_k}),
\label{convexroofdefinition}
\ee
where the infimum is taken over all possible decompositions of $\vr,$
i.e., over all $p_k$ and $\ket{\phi_k}$ with $\vr=\sum_k p_k \ketbra{\phi_k}.$
In other words, $E(\vr)$ is defined as the largest convex function smaller
than $E(\ket{\psi}).$

The advantage of the convex roof construction lies in the fact that
the resulting entanglement measure has by construction some
desirable properties, as $E(\vr)$ is convex.
Moreover, from the properties of $E(\ket{\psi})$ for pure states one
can often directly see, whether the measure for mixed states
fulfills the conditions for entanglement monotones, e.g., whether it
is non-increasing under LOCC \cite{vidal-2000-47,horodecki-2005-12,demkowiczdobrzanski-2006-74}.

Clearly, the optimization in Eq.~(\ref{convexroofdefinition}) is also
not straightforward to compute. Again, only for special cases  results
are known. There are however, some general recipes to give lower bounds
on the convex roof \cite{chen-2005-95a,chen-2005-95b}. The main idea is
to take an easily computable convex function $F(\vr)$ and to derive a
lower bound in terms of $F$ for pure states. Then, as the convex roof
is the largest convex function smaller than $E$ for pure states, the same
lower bound also holds for mixed states. For instance, the trace norm of
the partially transposed state $F(\vr) = \Vert \vr^{T_A} \Vert_1$ is such a
convex function, and from this the lower bound on the concurrence
[see Eq.~(\ref{concdef})] in a $d \times d $-system
\be
C(\vr)\geq \sqrt{\frac{2}{d(d-1)}} \Vert \vr^{T_A} \Vert_1
\label{cafformel}
\ee
can be established. The resulting lower bounds are often surprisingly
good \cite{chen-2005-95a,chen-2005-95b}. In the Sections
\ref{estimationofentanglementmeasures} and \ref{walbornexp}
we will investigate how entanglement measures can be evaluated
or estimated  in experiments.

\subsection{Examples of entanglement measures}
Now we discuss some examples of entanglement measures. We start
with measures for the bipartite case, and then explain some
measures for the multipartite case.

\subsubsection{Entanglement cost and entanglement of distillation}
\label{entanglementofdistillationsection}
The entanglement cost
$E_C(\vr)$ is defined as the minimal rate of singlets that have to
be used to create many copies of the state $\vr$ via LOCC
\cite{PhysRevA.54.3824}. More formally, it is given by \be E_C(\vr)=
\inf_{\rm{LOCC}} \lim_{n_{\rm out} \rightarrow \infty}
\frac{n_{\rm{in}}}{n_{\rm out}}. \ee Here, $\inf_{\rm{LOCC}}$
denotes the minimization over all LOCC protocols that map
$n_{\rm{in}}$ input singlets onto $n_{\rm{out}}$ output copies of
the state $\vr.$\footnote{To be more precise, one should
also include in this definition that the LOCC protocols do not
generate prefect copies of the state $\vr,$ but states arbitrary
close to it \cite{plenio-2005-1}.}

Conversely, the entanglement of distillation $E_D(\vr)$ is defined
as the optimal singlet distillation rate from many copies of $\vr,$
\be
E_D(\vr)= \sup_{\rm{LOCC}} \lim_{n_{\rm in} \rightarrow \infty}
\frac{n_{\rm{out}}}{n_{\rm in}},
\ee
where the LOCC protocols map now $n_{\rm{in}}$ input copies of $\vr$
onto  $n_{\rm{out}}$ output singlets.

With these definitions, $E_D(\vr) \leq E_C(\vr)$ has to hold. Since
there are bound entangled states that require entanglement for their
generation ($E_C > 0$), but from which no entanglement can be
distilled ($E_D = 0$), it can happen that $E_D(\vr) \neq E_C(\vr).$ For
pure states, however, it was shown in Ref.~\cite{PhysRevA.54.3824}
that the entanglement cost equals the entanglement of distillation
and they are given by the von Neumann entropy of the reduced state
$\vr_A,$ \be E_C(\ket{\psi}) = E_D(\ket{\psi}) = S(\vr_A) =
-Tr[\vr_A \log_2(\vr_A)]. \ee Therefore, pure states can be transformed
reversibly into singlet states.

\subsubsection{Entanglement of formation}

The entanglement of formation is defined as the convex roof of the
von Neumann entropy, \be E_F(\vr) = \inf_{\small p_k, \ket{\phi_k}}
\sum_k p_k S[(\vr_A)_k], \label{eofdefinition} \ee where $(\vr_A)_k$
is Alice's reduced state of the state $\ket{\psi_k},$ and the
optimization is defined as in Eq.~(\ref{convexroofdefinition})
\cite{PhysRevA.54.3824}. Physically, the entanglement of formation
may be interpreted as a minimal number of singlets that are required
to build a single copy of the state.

A central problem in the study of entanglement is whether or not the
entanglement of formation is fully additive. This question is not
solved yet, but its solution would have wide ranging consequences
also on other topics, like the additivity of classical capacity of
quantum channels (see Ref.~\cite{plenio-2005-1} for a discussion).

\subsubsection{Concurrence}
\label{sectionconcurrence}
A very popular measure for the quantification of bipartite quantum
correlations is the concurrence \cite{PhysRevLett.78.5022,
PhysRevA.64.042315}. This quantity can be defined for pure states as
\be C(\ket{\psi}) = \sqrt{2[1-Tr(\vr_A^2)]}, \label{concdef} \ee
where $\vr_A$ the reduced state of $\ket{\psi}$ for Alice. For mixed
states this definition is extended via the convex roof construction.

The popularity of the concurrence stems from the fact that for two
qubits the convex roof can analytically be computed
\cite{wootters98}. Namely, one has \be C(\vr)=\max\{0, \lambda_1 -
\lambda_2 -\lambda_3 -\lambda_4\}, \ee where the $\lambda_i$ are the
decreasingly ordered eigenvalues of the matrix
$X=\sqrt{\sqrt{\vr}(\sigma_y \otimes \sigma_y) \vr^* (\sigma_y
\otimes \sigma_y) \sqrt{\vr}}$ and the complex conjugation $\vr^*$
is taken in the standard basis. Moreover, for two qubits the
entanglement of formation can be expressed in terms of the
concurrence as \be E_F(\vr) = h
\left(\frac{1+\sqrt{1-C^2(\vr)}}{2}\right), \ee where
$h(p)=-p\log(p)- (1-p)\log(1-p)$ is the binary entropy function. For
other dimensions, however, such a relation does not hold and the
physical interpretation of the concurrence is not so clear. Moreover,
the concurrence is not additive.

\subsubsection{Negativity}
\label{negativitymeasure}
The negativity is just given as the violation of the PPT criterion
\cite{vidal-2002-65, zyczkowski-1998-58} \be N(\vr) = \frac{\Vert
\vr^{T_B} \Vert_1 - 1}{2}, \ee where $\Vert ... \Vert_1$ denotes the
trace norm (i.e., the sum of all singular values) of the partially
transposed state. Two of the main advantages are that negativity is
very easy to compute and it is convex. In order to make this
quantity additive, one can consider the logarithmic negativity
$E_N(\vr) = \log_2 \Vert \vr^{T_B} \Vert_1.$ This gives an upper
bound on the entanglement of distillation, $E_N(\vr)\ge E_D(\vr)$
\cite{vidal-2002-65}. The logarithmic negativity is, however, not
convex anymore \cite{plenio-2005-95}. By construction, the
negativity fails to recognize entanglement in PPT states.

\subsubsection{Distance measures}
\label{sectionrobustness}
Another class of entanglement quantifiers uses the distance to the
separable states as a measure for entanglement. As there are several
possible notions of a distance, several entanglement parameters
arise.

One of the first distance  measures is the relative entropy of entanglement
\cite{vedral-1998-57},
\be
E_R(\vr) = \inf_{\sigma} S(\vr \Vert \sigma),
\ee
where the infimum is taken over all separable states, and the distance
$S(\vr \Vert \sigma)$ is the relative entropy,
$S(\vr \Vert \sigma) = Tr[\vr \log(\vr) - \vr \log(\sigma)].$

Another class of distance-like measures that are relevant in our
future discussion are the so-called robustness measures. Roughly
speaking, these measures quantify, how much noise must be added, in
order to make the state separable. For instance, the robustness of
entanglement $R(\vr)$ is defined as the minimal $s,$ such that the
state \be \chi = \frac{1}{1+s}(\vr + s \sigma) \ee is separable
\cite{vidal-1999-59}. Here, $\sigma$ denotes an arbitrary separable
state. One can study variations of this measure, by allowing
$\sigma$ to be an arbitrary state (the generalized robustness
$R_g(\vr)$ \cite{PhysRevA.67.054305,harrow-2003-68}) or by fixing it
to be the maximally mixed state (the random robustness $R_g(\vr)$
\cite{vidal-1999-59}). The interesting point for our purpose is,
that the negative mean value of an entanglement witness can be
directly turned into a lower bound on the robustness (see Section
\ref{estimationofentanglementmeasures}).

\subsubsection{The geometric measure of entanglement}
\label{geometricmeasuresection}

A further interesting entanglement measure for multipartite systems
is the geometric measure of entanglement $E_G$ \cite{wei-2003-68,
shimonygeo, barnumgeo}. This also quantifies the distance to the 
separable states. The
geometric measure is defined for pure states as \be E_G(\ket{\psi})
= 1- \sup_{\small \ket{\phi}=\ket{a}\ket{b}\ket{c}...}
|\braket{\phi}{\psi}|^2, \label{gmdefinition} 
\ee 
i.e., as one minus
the maximal squared overlap with pure fully separable states, and
for mixed states via the convex roof construction. In many cases, the 
geometric measure is similarly defined as 
$E_G^{\rm log}(\ket{\psi}) = 
-2 \log_2 
(\sup_{\small \ket{\phi}=\ket{a}\ket{b}\ket{c}...}|\braket{\phi}{\psi}|).$
This is then a lower bound on the relative entropy of 
$\ket{\psi}$, however, it is  still not additive \cite{werner:4353}
and its convex roof is not an entanglement monotone 
anymore \cite{wei-2004-4}.

The geometric measure is a {\it multipartite} entanglement
measure, as it is not only a summation over bipartite
entanglement properties. Despite of its abstract definition, 
it has turned
out that $E_G$ can be used to quantify the distinguishability
of multipartite states by local means \cite{hayashi-2006-96}.
Furthermore, the optimization involved in Eq.~(\ref{gmdefinition})
can be performed in many cases, e.g. for W-type states 
\cite{wei-2003-68,tamaryan-2008-77} or certain graph states
\cite{markham-2007-9,hayashi-2008-77}. Also the convex roof can 
be computed in important cases \cite{wei-2003-68,guehneblaauboer}. 
We will see in Section 
\ref{estimationofentanglementmeasures} how the geometric measure 
can be estimated from the mean values of entanglement witnesses.

\subsubsection{The three-tangle and comb monotones}

The three-tangle $\tau$, introduced in Ref.~\cite{PhysRevA.61.052306}, is
an entanglement measure for three-qubit states. 
For states in the
form of Eq.~(\ref{acinform}) it is given by 
\be 
\tau (\ket{\psi})= 4 \lambda_0^2 \lambda_4^2. 
\ee 
Interestingly, an arbitrary pure three-qubit state
fulfills the monogamy relation $C^2_{A|BC}(\ket{\psi}) =
C^2_{AB}(\vr_{AB}) + C^2_{AC}(\vr_{AC}) +  \tau(\ket{\psi})$ where
$C^2_{A|BC}(\ket{\psi}) = 2 \sqrt{\det(\vr_A)}$ is the concurrence
between A and the other two qubits, and $C^2_{AB}(\vr_{AB})$ is the
concurrence between A and B
\cite{PhysRevA.61.052306,osborne:220503}. For mixed states, the
tangle is defined via the convex roof construction, which can, in
important cases, be performed analytically \cite{lohmayer-2006-97,fan-2007}.
For more than three qubits, the tangle can be generalized via combs
and filters \cite{osterloh-2005-72,osterloh-2006-4}.

\section{Bell inequalities}
\label{sectionbellinequalities}

In this Section, we review Bell inequalities as  the oldest tool to
detect entanglement. Originally, Bell inequalities were designed to
rule out local hidden variable (LHV) models. We will therefore first
explain LHV theories and then present different Bell inequalities
for various situations. While Bell inequalities were violated in
many experiments, several loopholes make it still possible that 
\emph{in principle} LHV models could describe the measurement results.
We will discuss the detection efficiency loophole and the locality 
loophole. Finally, we will discuss a two-photon experiment, where 
the locality loophole has been closed.

\subsection{Bipartite systems}

The first Bell inequality was published by John Bell in 1964 in
Ref.~\cite{Bell64}. The goal of this paper is to capture the
Einstein-Podolsky-Rosen paradox in a quantitative way. The basic
idea behind Bell inequalities is the following. Let us consider a
bipartite system, and perform measurements on both parties. If we
assume that measurement results existed locally at the parties
before the measurement then it is possible to obtain bounds on
certain quantities composed from two-body correlation terms. In
quantum mechanics, it is possible to design experiments in which a
higher value is measured than these bounds. This, on the one hand,
shows that quantum physics violates \emph{local realism}: The
measurement is not simply reading out a pre-existing local quantity
and  its results cannot be described by a LHV model.  On the other
hand, as we will see later, if measurements on a quantum state
violate a Bell inequality, it implies that the state is entangled.

\subsubsection{Local hidden variable models and CHSH-inequality}

\begin{figure}
\centerline{\epsfxsize=5.0in\epsffile{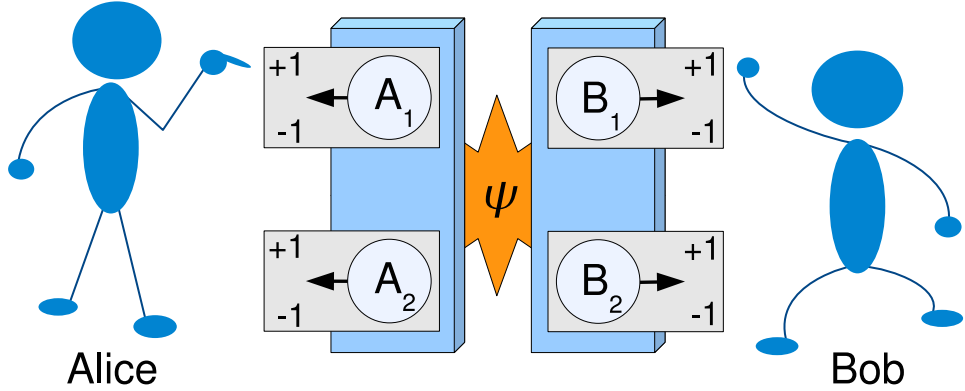}}
\caption{The basic setup for the CHSH inequality: in a bipartite
system Alice can choose to measure two quantities, denoted by $A_1$ and $A_2,$
each giving the results $a_i \in \{-1,+1\}.$ Similarly, Bob can measure
$B_1$ and $B_2$ on the right party.} \label{lhv_chsh}
\end{figure}

Let us consider a simple example of a bipartite system. We assume
that Alice can measure two quantities at her party, called
$A_1$ and $A_2$, while Bob can also measure two quantities at
his party, called $B_1$ and $B_2,$ as shown in Fig.~\ref{lhv_chsh}.
The results of these experiments are $a_1, a_2$ and $b_1, b_2$
and we assume that these results can take the values $+1$ or $-1$.
Now let us perform simultaneous measurements of these four quantities
$M$ times.
Expectation values can be obtained
simply by averaging the measurement results,
$\exs{A_i B_j}=\tfrac{1}{M}\sum_{k=1}^M a_i(k)b_j(k).$
If the probabilities of the outcomes are given by expressions
like $p(a_1^+,b_2^-) := p(a_1=+1, b_2 = -1)$ this can be rewritten
as
\begin{equation}
\exs{A_i B_j}= p(a_i^+, b_j^+) - p(a_i^-, b_j^+)-
p(a_i^+, b_j^-)+p(a_i^-, b_j^-).
\label{exval}
\end{equation}

Now one can ask whether these probabilities and expectation values can be
described by a \emph{local hidden variable (LHV) model.} Latter assumes that the
probabilities of the measurement results are independent of whether they
are indeed measured or not (the assumption of reality) and that Alice's
probabilities do not depend on the choice of the observables by Bob and
vice versa (locality assumption). So one makes the ansatz
\begin{equation}
p(a_i^\alpha, b_j^\beta)
= \int\!d\lambda\;  p(\lambda) \mathcal{A}_\lambda(a_i^\alpha)
\mathcal{B}_\lambda(b_j^\beta),
\label{hvar}
\end{equation}
where $\alpha, \beta$ denote the possible outcomes $\pm 1$ and
$\lambda$ is the hidden variable, occurring with a probability
$p(\lambda).$ The response function $\mathcal{A}_\lambda(a_i^\alpha)
\mathcal{B}_\lambda(b_j^\beta)$ factorizes due to the locality
assumption, as for a fixed $\lambda$ Alice's probabilities should
not depend on Bob's choice of observables. Furthermore, one can
always assume the LHV model to be deterministic (i.e., for a
fixed $\lambda$ the $\mathcal{A}_\lambda(a_i^\alpha)$ and
$\mathcal{B}_\lambda(b_j^\beta)$ take only the values $0$ or $1$)
as a nondeterministic model corresponds to a deterministic one, where the
hidden variable $\lambda$ is not known \cite{PeresConj,wernerwolfbell}.
Finally, it should be noted that if the probabilities are of the form
as in Eq.~(\ref{hvar}), then the expectation values can be written as
$\mean{A_i B_j} = \int\!d\lambda\;p(\lambda) \mathfrak{A}_\lambda (A_i)
\mathfrak{B}_\lambda(B_j)$ with
$\mathfrak{A}_\lambda (A_i) =
\mathcal{A}_\lambda(a_i^+)-\mathcal{A}_\lambda(a_i^-)$ etc.

{From} the fact that the probabilities are described by a LHV model,
bounds on the correlations follow. These are the Bell inequalities.
On the level of probabilities, the Clauser-Horne (CH) inequality
\cite{PhysRevD.10.526} reads\footnote{This is the simplest
possible form from Ref.~\cite{PeresConj}, the original CH 
inequality is equivalent, but contains six terms.}
\be p(a_1^-, b_1^-) + p(a_1^+, b_2^-) +
p(a_2^-, b_1^+) - p(a_2^-, b_2^-) \geq 0. 
\label{pp}
\ee
Eq.~(\ref{pp}) reflects the fact that the last event ($a_2^-,
b_2^-$) can only occur if at least one of the three other possibilities also
occurs, and follows directly from the structure of the response
function.

Using the normalization of the probabilities, one can derive a
similar inequality for the mean values, which is the
Clauser-Horne-Shimony-Holt (CHSH) inequality
\cite{PhysRevLett.23.880,PhysRevLett.24.549}. It reads \be \mean{A_1
B_1} + \mean{A_2 B_1} + \mean{A_1 B_2} - \mean{A_2 B_2} \leq 2.
\label{CHSH} \ee Eq.~(\ref{CHSH}) can also be directly understood by
considering all possible measurement results $a_1, a_2, b_1, b_2$, as
for all possible of them $a_1 b_1 +a_1 b_2 +a_2 b_1 -a_2 b_2  \le
2.$

Naturally, inequality Eq.~(\ref{CHSH}) is satisfied if there is a
hidden variable model for the probabilities of the
measurement results. Such an
inequality is called \emph{Bell inequality}. Note at this point that
for defining Bell inequalities quantum physics is not used at all,
only simple basic assumptions lying behind classical physics were used.

Let us connect these ideas to quantum physics. Assuming that Alice
and Bob both have a quantum mechanical spin-$1/2$ particle one can
identify $A_i$ and $B_j$ with measurement results of the Pauli spin
operators $X = \sigma_x$ and $Y=\sigma_y$, respectively. Then, the
left hand side of Eq.~(\ref{CHSH}) defines a quantum mechanical
observable $\mathcal{B} = A_1 \otimes B_1 + {A_2 \otimes B_1} + {A_1
\otimes B_2} - {A_2 \otimes B_2}$, the so-called Bell operator
\cite{PhysRevLett.68.3259} that
can be measured. One finds that there are quantum states that
violate the inequality and give $2\sqrt{2}$ for the left-hand side.
Taking the observables $A_1 = -\sigma_x, A_2 = -\sigma_y,  B_1 =
(\sigma_x+\sigma_y)/\sqrt{2} $ and
$B_2=(\sigma_x-\sigma_y)/\sqrt{2}$ the quantum state with the
highest violation is the  eigenstate corresponding to the largest
eigenvalue of this Bell operator. It is given by
\begin{equation}
\ket{\Psi_{\rm CHSH}}=\frac{1}{\sqrt{2}}(\ket{01}-\ket{10}),
\label{CHSH_Phi}
\end{equation}
which is just the two-qubit singlet state. It can be shown that
quantum mechanics does not allow for violations larger than
$2\sqrt{2},$ a fact which is known as the Tsirelson bound
\cite{tsirelsonbound}. Furthermore, the question whether a given
two-qubit state violates the CHSH inequality for some optimized
settings can directly be answered \cite{horodeckibellpla}: If
one considers the $3\times 3$ matrix
$\tau=\mean{\sigma_i \otimes \sigma_j}$ with $i,j \in \{x,y,z\}$
then the state violates the CHSH inequality iff the sum of the two
largest eigenvalues of the matrix $\tau^T \tau$ is larger than one.
For general states one can show that any entangled state violates
the CHSH inequality, after local filterings and adding an auxiliary
system that does not violate the CHSH inequality itself
\cite{masanes-2008-090403}. In general, the problem whether a given
state violates violates a Bell inequality for some choice of settings
can be tackled using semidefinite programming \cite{liang-2007-042103}.

How is a violation of the Bell inequality possible? The reason is
that the simple assumptions we made for the Bell inequality are not
valid for quantum mechanics. Indeed, the violation of Bell
inequalities in experiment shows that either locality or realism
must be abandoned, however it is not clear which one of them.
Interestingly, one can derive Bell-like inequalities that rely only
on realism and on some assumptions about the correlation functions
$\mean{A_i B_j}$ \cite{leggettnonlocal} and are also found to be 
violated in experiments \cite{groeblacher-2007-446} (for 
generalizations see Ref.~\cite{branciard-2008-4}).
This might be interpreted as if the assumption of realism must be
dropped, but it should be noted that there are realistic (but
non-local) theories such as Bohmian mechanics that can describe
these experiments. Another interesting non-local realistic model
is described in Ref.~\cite{cerf2005smq}. It is based on simulating 
the correlations arising from quantum mechanics without communication 
between the parties, using Popescu-Rohrlich non-local boxes 
\cite{popescu1994qna}.

\subsubsection{Violation of the CHSH inequality and entanglement}

The violation of a Bell inequality implies non-locality,
i.e., that the measurement probabilities cannot be described by a LHV model.
In quantum mechanics, the violation implies that the state is entangled.
This can be seen as follows. By construction, a separable state in
a bipartite system can be written in the form
$
\varrho = \sum_k p_k \varrho_k^{A} \otimes \varrho_k^{B}.
$
Hence, one can write
\begin{equation}
\langle A_i B_j \rangle = \sum_k p_k Tr(A_i\varrho_k^{A})
Tr(B_j\varrho_k^{B}),
\end{equation}
which is, if we define $\mathfrak{A}_k (A_i) =
Tr(A_i\varrho_k^{A})$ and
$\mathfrak{B}_k(B_j)=Tr(B_j\varrho_k^{B}),$ directly a LHV model.
This means that if some measurements on a state violate a Bell
inequality then it cannot be separable, i.e., it must be entangled.
Note, however, that the converse is not true: there are entangled
states for which a local realistic model for all von Neumann
measurements can be written down explicitly (see also below)
\cite{werner89}.\footnote{The question whether a state allows for a
LHV model or not can also be tackled by looking for symmetric
extensions of the state using semidefinite programming
\cite{terhal-2003-90}.} Therefore, they do not violate any Bell
inequality. The same statement holds even if one allows
positive-operator-valued measurements (POVMs)
\cite{PhysRevA.65.042302}. For pure states, however any bipartite
entangled state violates a Bell inequality
\cite{carpasso,1991PhLA..154..201G,1992PhLA..162...15G}. Due to that, Bell
inequalities may be viewed as non-optimal entanglement witnesses
\cite{hyllus:012321}. (Concerning the comparison of Bell inequalities and witnesses
see Sec.~\ref{subsec_comp_wit_bell}.)

Finally, before considering further Bell inequalities, let us mention
ways to characterize them from the point of view of their
usefulness in experiments. We have seen that the CHSH inequality is
maximally violated by the state Eq.~(\ref{CHSH_Phi}). However, in
a real experiment one can never prepare the state Eq.~(\ref{CHSH_Phi})
perfectly, it is always mixed with noise
\begin{equation}
\vr_{\rm noisy singlet}(p):=
p\ketbra{\Psi_{\rm CHSH}}+(1-p)\frac{\ensuremath{\mathbbm 1}}{4},
\end{equation}
which is a two-qubit Werner state \cite{werner89}.
The CHSH inequality is violated if $p>1/\sqrt{2}\approx 0.71$ and,
using the PPT criterion, the state is entangled for $p > 1/3.$ Moreover,
it is known that this state allows for a LHV model (for all von Neumann
measurements) for $p \leq 0.6595$ \cite{werner89,AcinGrothendiecksConstant}
and it violates a Bell inequality (different from the CHSH one) for
$p \geq 0.7056$ \cite{vertesi-2008}.\footnote{See also Problem 19
on {\tt http://www.imaph.tu-bs.de/qi/problems/}.}

The bound for the CHSH inequality characterizes the minimum
\emph{visibility} that the CHSH inequality requires. In general,
the visibility is the ratio between the maximum of the Bell operator
$\BB$ for quantum states and the maximum for LHV models
\begin{equation}
\VV(\BB):=\frac {\max_{\Psi} |\exs{\BB}_\Psi|} {\max_{LHV}
|\exs{\BB}_{LHV}|}.\label{VV}
\end{equation}
Clearly, a large visibility is an advantage from the experimental
point of view.


\subsection{Multipartite systems}

\subsubsection{Multipartite LHV models}
\label{subsection_multiLHV}

In the previous subsection we discussed Bell inequalities for
bipartite systems. The ideas can straightforwardly be generalized
to multipartite systems. Let us consider an $N$-partite system,
in which each party can measure some observables.
Then the measurement results fit a local hidden
variable model if for any choice of measured variables we have
\begin{equation}
\exs{A_i B_j C_k D_l ... } =
\int \!\! d\lambda p(\lambda)
\mf{A}_\lambda(A_i) \mf{B}_\lambda(B_j)
\mf{C}_\lambda(C_k) \mf{D}_\lambda(D_l) ...,
\label{hvarmulti}
\end{equation}
where $-1\le \mf{A}_\lambda(A_i) \le +1$ etc. If a multipartite
Bell inequality is violated then there is no local hidden variable
model that could explain the measurement results.

Similarly to the definition of biseparability one can also ask,
whether it is possible to explain our measurement results by
bipartite models in which several of the parties join
\cite{PhysRevD.35.3066,PhysRevLett.89.060401,collins-2002-88}.
For example, for the three-partite case this would mean that any
expectation values could be explained by
\begin{equation}
\exs{A_i B_j C_k}=
p_1 \int \!\!  d\lambda p(\lambda)
\mf{A}_\lambda(A_i)\mf{X}_\lambda(B_j C_k)+
p_2 \int \!\! d\lambda p'(\lambda)
\mf{B}_\lambda(B_j)\mf{Y}_\lambda(A_i C_k)+
p_3 \int  \!\!  d\lambda p''(\lambda)
\mf{C}_\lambda(C_k)\mf{Z}_\lambda(A_i B_j),
\label{hvargenuine}
\end{equation}
where $p_1+p_2+p_3=1.$ If there is not such a description for the
measurement results, we say that there is genuine three-partite
non-locality in the system.

\subsubsection{Mermin and Ardehali inequalities}
In the previous subsection we discussed the CHSH
inequality for bipartite systems. It needed
the measurement of two quantities for each party and these
measurements had two outcomes. Similar {\it two-setting}
inequalities exist also for multipartite systems. The most famous
one is the Mermin inequality \cite{PhysRevLett.65.1838}.
For $N$ qubits it is given by
\begin{equation}
\sum_\pi \mean{X_1 X_2 X_3 X_4 X_5\cdot\cdot\cdot X_N}
-\sum_\pi \mean{Y_1 Y_2 X_3 X_4 X_5\cdot\cdot\cdot X_N}
+\sum_\pi \mean{Y_1Y_2Y_3Y_4X_5\cdot\cdot\cdot X_N}
- ... + ...\leq L_{\rm Mermin},
\label{mermin}
\end{equation}
where $X_i$ and $Y_i$ represent the Pauli matrices $\sigma_x, \sigma_y$ on
the $i$-th qubit\footnote{Of course, in the sense of LHV models, they can
be replaced by arbitrary observables with two outcomes $\pm 1$.},
$\sum_\pi$ represents the sum of all possible permutations of the qubits
that give distinct terms and $L_{\rm Mermin}$ is the maximum for local
states. It is $L_{\rm Mermin}= 2^{N/2}$ for even $N$ and
$L_{\rm Mermin}=2^{(N-1)/2}$ for odd $N.$

The Mermin inequality is maximally violated by the GHZ state defined
in Eq.~(\ref{ghzndef})\footnote{Note that the original Mermin
inequality is slightly different and is maximally violated by the
state $(\ket{000...}+i\ket{111...})/\sqrt{2}.$}. The Bell operator
is the sum of $2^{(N-1)}$ terms that all give $+1$ for the GHZ
state. Hence the maximum of the Bell operator is $2^{(N-1)}$ for
quantum states.  Thus, the visibility  is $\VV_{\rm
Mermin}=2^{N/2-1}$ for even $N$ and $2^{(N-1)/2}$ for odd $N.$ Note
that the visibility increases exponentially with the number of
qubits.  It can be proved that for odd number of parties the Mermin
inequality has the largest visibility possible among two-setting
inequalities with full correlation terms. Latter means that the Bell
operator is the sum of correlation terms that contain a variable for
each spin. For example, for $N=4$, $X_1Y_2X_3Y_4$ is a full
correlation term, however, $X_1 Y_2 \eins_3 X_4$ is not. The
$\eins_3$ should indicate that on the third particle no measurement
has been done.

The Mermin inequality consists of stabilizing operators of the GHZ
state (see Sec.~\ref{Sec_GraphStatesClusterStates}). This implies
that there is a quantum state that gives $+1$ for all terms, while
LHV models have to result in $-1$ for at least one term. This is a
form of the GHZ argument \cite{greenberger-1989}, which leads to an
obvious contradiction to local realism. Similarly, one can derive
Mermin-like inequalities for other graph states
\cite{PhysRevLett.95.120405,PhysRevA.71.042325,
PhysRevA.73.022303,cabello-2007,hsu:042308,cabello-2007-99}.

Let us now discuss the Ardehali inequality \cite{PhysRevA.46.5375}.
For an even number of parties this inequality is superior to the
Mermin inequality. It is defined as
\begin{eqnarray}
&& \mean{({A_1^{(+)}}- {{A}_1^{(-)}})
\big( - \sum_\pi  {X_2 X_{3} X_{4}X_{5} \cdot\cdot X_{N}} +
\sum_\pi  {Y_{2} Y_{3} X_{4} X_{5} \cdot\cdot X_{N}} - \sum_\pi
{Y_{2}Y_{3}Y_{4}Y_{5}X_6 \cdot\cdot X_{N}} +...-... \big)}
\nonumber
\\
&&+\mean{({A_1^{(+)}}+{{A}_1^{(-)}})\big(\sum_\pi {Y_{2}X_{3}X_{4}X_{5} \cdot\cdot X_{N}} -
\sum_\pi {Y_{2}Y_{3}Y_{4}X_{5} \cdot\cdot X_{N}} + \sum_\pi
{X_{2}Y_{3}Y_{4}Y_{5} Y_6 X_7 \cdot\cdot X_{N}} -...+... \big)}
\nonumber
\\
&&
\leq L_{\rm
Ardehali},\label{ardehali}
\end{eqnarray}
where $A_1^{(\pm)}$ are operators corresponding to measuring the
first spin along directions corresponding to the quantum operators
$A_1^{(\pm)} = ({\mp X_1-Y_1})/{\sqrt{2}}.$ Note that on the other
spins $X$ or $Y$ is measured as in the case of the Mermin
inequality. The constant $L_{\rm Ardehali}$ appearing in
Eq.~(\ref{ardehali}) is the maximum for local states. It is
$2^{N/2}$ for even $N$ and $2^{(N+1)/2}$ for odd $N.$ It is again
maximally violated by the GHZ state.\footnote{Note that the original
Ardehali inequality has a different definition for $A_1^{(\pm)}$ and
is maximally violated by the state
$(\ket{000...}-\ket{111...})/\sqrt{2}.$} The Bell operator is the
sum of $2^N$ terms that all give $1/\sqrt{2}$ for this state. Thus,
the visibility $\VV_{\rm Mermin}=2^{N/2}$ for even $N$ and
$2^{(N-1)/2}$ for odd $N.$

An interesting feature of the Ardehali inequality is that the  Bell
operator is essentially the same as for the Mermin inequality, but
the rewriting  with $A_1^{(\pm)}$ leads to a higher violation of
local realism. This shows that the violation of local realism of GHZ
states with an even number of qubits can increase, if non-stabilizer
observables are considered. This also holds for some other graph
states \cite{PhysRevA.73.022303,guhne-2008-77bell}.

Note that so far all the inequalities needed the measurement of two
operators for each qubit. Also, they were the sum of full correlations
terms. Among such inequalities, it can be shown that for any $N$ the
Mermin-Ardehali construction, also discovered independently by
Klyshko \cite{BelinskiiKlyshko93,Klyshko93} provides Bell inequalities
with an optimal violation. It turns out that the full set of such Bell
inequalities can be generated and written down concisely in the form
of a single nonlinear inequality
\cite{PhysRevA.64.032112,PhysRevLett.88.210401}.
Surprisingly, there are multi-qubit pure entangled states that do not violate
any of these Bell inequalities \cite{PhysRevLett.88.210402}.

The situation is more complicated with inequalities that are not sums
of full correlation terms. Then, it has been shown that such inequalities
can detect any pure entangled multi-qubit state \cite{PopescuRohrlich}. Also,
inequalities of this type can be constructed such that they are maximally
violated by cluster states and graph states \cite{PhysRevLett.95.120405,
PhysRevA.71.042325,PhysRevA.73.022303} (see
Sec.~\ref{Sec_GraphStatesClusterStates} for the definition of these
states). In particular, for the four-qubit cluster state
this inequality looks like
\cite{PhysRevA.71.042325, PhysRevLett.95.020403}\footnote{This
is the inequality for the cluster state in the basis
as defined via Eqs.~(\ref{stabilizeredefiningequation},\ref{graphstatedefiningequation}),
for the cluster state as in Eq.~(\ref{fourqubitclustereq}) it reads
$\mean{Z_1 \eins_2 X_3 X_4} + \mean{X_1 Y_2 Y_3 X_4}
- \mean{Z_1 \eins_2 Y_3 Y_4} + \mean{X_1 Y_2 X_3 Y_4} \leq 2$.}
\begin{equation}
\mean{X_1 \eins_2 X_3 Z_4} + \mean{Z_1Y_2Y_3Z_4} + \mean{X_1\eins_2 Y_3 Y_4}
- \mean{Z_1 Y_2 X_3 Y_4} \le 2.
\label{c4bell}
\end{equation}
Note that on all of the qubits two operators are measured except for
the second qubit for which only $Y_2$ is measured. One can show that
for a large class of graph states, e.g., for linear cluster states, it
is possible to construct two-setting Bell inequalities that have a
visibility increasing exponentially with $N$
\cite{PhysRevA.73.022303}.

\subsubsection{Bell inequalities detecting genuine multipartite non-locality}
\label{bellgenuine}
So far all the Bell inequalities presented could be used to confirm
that a state is nonlocal, that is they ruled out local hidden
variable models as in Eq.~(\ref{hvarmulti}), but did not give any
information on whether or not the state possesses genuine
multipartite non-locality [see Eq.~(\ref{hvargenuine})]. However,
some Bell inequalities exist that can rule out such hybrid LHV
models. Let us consider the three-particle Svetlichny inequality as
an example \cite{PhysRevD.35.3066}. It reads 
\be \mean{A_1 B_1 C_2}
+ \mean{A_1 B_2 C_1} + \mean{A_2 B_1 C_1} - \mean{A_2 B_2 C_2} +
\mean{A_2 B_2 C_1} + \mean{A_2 B_1 C_2} + \mean{A_1 B_2 C_2} -
\mean{A_1 B_1 C_1} \leq 4, 
\ee 
where the $A_i, B_j$ and $C_k$ are
two outcome measurements (e.g., Pauli matrices) on each particle.
This Bell inequality is a sum of two three-particle Mermin
inequalities. However, one can directly see that for hybrid LHV
models (e.g., for an $1|23$ case, where the particles 2 and 3 are
combined, the Svetlichny inequality can be written as a sum of two
CHSH inequalities) the same bound holds. For GHZ states, the Bell
operator can take values of $4 \sqrt{2}$ 
\cite{collins-2002-88,PhysRevA.70.060101}, however, not 
all genuine multipartite entangled states lead to a violation 
of the  Svetlichny inequality \cite{seevinckthesis}, even if 
they are pure \cite{ghose-2008}.

A generalization of this inequality for an  arbitrary number of 
parties is presented in Ref.~\cite{PhysRevLett.89.060401}. 
Ref.~\cite{collins-2002-88} gives  bounds for Bell inequalities 
both for genuine multi-qubit non-locality and genuine multi-qubit 
entanglement.

\subsubsection{Further Bell inequalities}

There are Bell inequalities that use more than two operators for each
particle (i.e., more than two settings) or need the measurement of
observables with more than two outcomes. Ref.~\cite{kaszlikowski2000vlr}
considers bipartite systems with qudits having dimension $d>2,$ and studies
two-setting Bell inequalities with observables with $d$ outcomes. It finds
that with the increase of $d,$ it is possible to find Bell inequalities with
increasing maximal violation. Ref.~\cite{collins2002bia} presents a family
of Bell inequalities, called the Collins-Gisin-Linden-Massar-Popescu (CGLMP)
inequalities for bipartite quantum systems of arbitrarily high dimensionality,
which are strongly resistant to noise. They are for systems with two
measurement settings per site, with more than two measurement
outcomes. Ref.~\cite{ito2006bis} presents Bell inequalities stronger
than the CHSH inequality for isotropic states of two three-state
particles.

Systematic study of such inequalities were also presented for small
number of outcomes. Such an inequality can be characterized by four
numbers $ijmn$, where Alice measures one of $i$ operators that have
$m$ outcomes. Similarly, Bob measures one of $j$ operators that have
$n$ outcomes. In the bipartite case, for two two-outcome observables
the CHSH inequality is the only Bell inequality apart from simple
transformations \cite{PhysRevLett.48.291}. That is, it is the only tight Bell
inequality, where tight means that the inequality is a facet of the polytope
of correlations allowed by LHV models \cite{pitowsky1989qpq}. 
Non-tight inequalities also exist, but they
do not detect more states than the tight inequalities.

For the $3322$ case, that is, when three two-outcome observables are
measured at each party, then there is a single new inequality
\cite{froissart,sliwa,collins2004rtq}. It reads
\be
\mean{A_1 \eins} - \mean{A_2 \eins}
+
\mean{\eins B_1} - \mean{\eins B_2}
-
\mean{A_1 B_1} + \mean{A_1 B_2}
+ \mean{A_2 B_1} -\mean{A_2 B_2}
+
\mean{A_1 B_3} + \mean{A_2 B_3}
+
\mean{A_3 B_1} + \mean{A_3 B_2}
\leq 4.
\ee
This is a relevant inequality in
the sense that it detects two-qubit states that cannot be detected
by the CHSH inequality \cite{collins2004rtq}.
For the $4322$ case there are
three new inequalities \cite{collins2004rtq}. For systems with
measurements with more outcomes the numerical search is exceedingly
difficult. A set of $26$ inequalities for the $4422$ case are
presented in Ref.~\cite{brunner2008plb}. Ref.~\cite{pal-2008}
determines the quantum violation of $241$ tight bipartite
Bell inequalities with up to five two-outcome measurement
settings per party, for up to eight dimensional complex and six
dimensional real Hilbert spaces, using semidefinite programming.
The violation of bipartite
inequalities with arbitrary number of two-outcome observables
is bounded by the Grothendieck constant \cite{AcinGrothendiecksConstant}.

For  three parties and for two-outcome observables, inequalities
with arbitrarily large violation can be designed
\cite{perezgarcia-2007}. Ref.~\cite{PhysRevA.56.R1682} presents
three-setting two-outcome inequalities for GHZ states that have a
higher violation than the Mermin inequality. A set of multipartite
three-setting Bell inequalities were studied, with two-outcome
observables in Refs.~\cite{zukowski2006tmb,wiesniak2007efc}.
Finally, Ref.~\cite{sliwa} presents the full list of Bell inequalities for three parties,
two measurements and two outcomes.

\subsection{Consequences of a Bell inequality violation}
\label{sectionbellconsequences}

We have seen already that for the bipartite case a violation
of a Bell inequality implies entanglement. For the multipartite
case similarly non-locality implies entanglement, however, only
ruling out LHV models as in Eq.~(\ref{hvargenuine}) can prove
genuine multipartite entanglement. As in the bipartite case,
entanglement does not imply non-locality. For example,
there are three-qubit states with genuine multipartite
entanglement that are local if von Neumann measurements
are considered \cite{toth2006gte}.

Interestingly, while Bell inequalities are primarily designed to rule
out LHV models (as described in Sec.~\ref{subsection_multiLHV}) they can sometimes be
used to detect genuine multipartite entanglement. For example, the
three-qubit Mermin inequality used in the photonic experiments of
Ref.~\cite{ThreeQubitGHZ,ThreeQubitGHZBook} is
\begin{equation}
\mean{X_1X_2X_3}-\mean{Y_1Y_2X_3}-\mean{X_1Y_2Y_3}-\mean{Y_1X_2Y_3} \le 2.
\end{equation}
As a Bell inequality, violation of this inequality does not imply
three-qubit entanglement. In quantum mechanics, however, it can be
proved that, if $X_k$ and $Y_k$ refer to the measurement of Pauli
spin matrices at qubit $k$, then for biseparable states the maximum
is also two \cite{SeevinckUffinkThreeParticle,Addendum05}. Thus, the
violation of this inequality implies genuine multi-qubit
entanglement and in the experiment of Ref.~\cite{ThreeQubitGHZ}
genuine multi-qubit entanglement was detected.\footnote{This has
also been shown in Ref.~\cite{Nagata2} based on a different
argument.} In another experiment creating a four-qubit GHZ state
\cite{FourPhotonGHZ}, a bound for multipartite entanglement for the
Ardehali inequality was used to detect four-qubit entanglement. In
general, the detection of multi-qubit entanglement with many-qubit
Bell inequalities was studied in
Refs.~\cite{Gisin98,collins-2002-88}. These papers presented bounds
for various forms of multipartite entanglement.
Refs.~\cite{Nagata,PhysRevLett.88.230406} present nonlinear Bell
inequalities for the same aim.

Besides, there are some other interesting connections between the
violation of a Bell inequality and the properties of a state:

\begin{enumerate}

\item{ \it Quantum communication.} 
It has been observed that in the BB84
protocol the error thresholds where a Bell inequality can be
violated and where the mutual information between Alice and Bob and
Alice and Eve are the same, coincide
\cite{PhysRevA.56.1163,gisinhuttnera,gisinhuttnerb}. Similar results
have been found in the multipartite scenario
\cite{scarani-2001-87,PhysRevA.65.012311}. This suggests a
connection between Bell inequality violation and security of quantum
cryptography. Moreover, the Ekert protocol is based on distributing entangled particles \cite{PhysRevLett.67.661}. 
Its security is directly related to the violation of a Bell inequality.
Note that Bell inequality violation can imply security even in
the device independent case when no assumption is made on the quantum system we
are using for communication (e.g., dimension of the system) \cite{acin2007dis}.

\item {\it Communication complexity.} The violation of a Bell
inequality means that measurement results on a quantum state
cannot be replaced by shared classical randomness. Because of that,
using the quantum state as a resource has an advantage in quantum
communication. Indeed, for any state violating a Bell inequality one
can write down a communication complexity problem, for which the quantum
state improves the solution compared with classical resources
\cite{PhysRevLett.92.127901,buhrman-1999-60,galvao-2000}.

\item {\it Distillability.} Quantum states violating two-setting
Bell inequalities are bipartite distillable \cite{PhysRevLett.88.027901,
PhysRevA.66.042323}. This fact, however, relies on the spectral
decomposition of the Bell operator, and not on the nonexistence
of a LHV model. In the multipartite case, this means distillability
at least with respect to some bipartitioning of the qubits, which
arises if two groups of parties join. On the other hand, if the
distillation protocol has to be local with respect to each party,
there exist multipartite non-distillable states which violate a
Bell inequality \cite{dur-2001-87}.

This directly leads to the conjecture of Peres, stating
that bipartite states with a positive partial transposition
allow a LHV description. While no proof or counterexample
has been found, in several special cases it is shown that such states
do not violate large classes of Bell inequalities, e.g., for the
Mermin-Klyshko inequalities \cite{PhysRevA.61.062102} and for a wide
range of multi-setting inequalities \cite{PhysRevA.74.062109}.

\end{enumerate}

\subsection{Loopholes}
\label{Sec_loopholes}
While Bell inequalities have been violated many times in experiments,
several loopholes make it still possible {\it in principle} that the
experimental results can be explained with a LHV model. There are mainly
two loopholes:

\begin{enumerate}

\item
{\it The fair sampling assumption or  detector efficiency loophole.}
In many setups based on photons, photo-detectors detect photons only in
part of the experiments, while those experiments in which there are no
photons detected (no "clicks") go unnoticed. We assume, naturally,
that the unnoticed experiments have the same statistical properties as the
noticed ones. However, a so far unknown mechanism could cheat us in
this respect. It could happen, that all the experiments together
would not violate a Bell inequality, however, the subset for which
our detectors click, violate it \cite{PhysRevD.2.1418,PhysRevA.46.3646}.

The question, for which detector efficiency the loophole becomes
closed depends on the Bell inequality, and is, in general, not
straightforward to answer. For several important experimental
situations, results have been obtained in
Refs.~\cite{PhysRevA.47.R747,PhysRevA.57.3304,brunner-2007-98,cabello-2008-101,
cabello-2007-98}. The detection loophole was eliminated in
experiments with trapped ions with very large detection efficiency
in Ref.~\cite{rowe2001evb}. In a recent experiment, also eliminating the detection loophole, the ions were $1$m apart \cite{matsukevich2008biv}.

\item
{\it The locality loophole.} Ideally, before each correlation
measurement the circuits connected to the detectors should decide
locally which operator to measure. This has to be done such that the
detectors cannot communicate their choice if we assume that the
speed of this communication cannot be larger than the speed of
light. In practice this means that the event of deciding the
measurement direction for one of the detectors must be space-like
separated from event of reading out the measurement on the other
detector. This requirement is very challenging. In
Ref.~\cite{aspect1982etb}, an experiment is described in which the
directions for detection are determined after the photons left the
source. In Ref.~\cite{PhysRevLett.81.5039}, a photonic experiment is
described that eliminates the locality loophole, however, in this
realization the detection loophole is still present. We will discuss
it in the following section.

\end{enumerate}

Another assumption in the derivation of Bell inequalities is that 
Alice's and Bob's choices of the measurements do not depend on the 
hidden variables. This corresponding loophole is sometimes called 
the freedom-of-choice-loophole \cite{PhysRevA.49.3209,scheidl-2008} 
and has been addressed experimentally in Ref.~\cite{scheidl-2008}.


\subsection{Experimental violation of a Bell inequality}

\label{expbell}

In this Section we describe as a first experiment the violation
of Bell inequalities in a two-photon experiment by G. Weihs {\it et al.}
\cite{PhysRevLett.81.5039}.

\subsubsection{Generation of entangled photons}

First, we review briefly the fundamentals of the experimental techniques,
since there are numerous similar many-photon experiments. An excellent and
exhaustive review of this topic can be found in Ref.~\cite{pan-2008}.
As is well known, a photon can be used to store a single qubit
information since it can be in a superposition of the horizontally
and vertically polarized states, denoted by $\ket{H}$ and $\ket{V},$
respectively, as
\begin{equation}
\ket{\Psi}=\alpha\ket{H} + \beta\ket{V},\label{psistate}
\end{equation}
where $\vert\alpha\vert^2+\vert\beta\vert^2=1.$ Other basis states
can be also chosen. The left and right circularly polarized states
can be expressed as $\ket{L}:=\tfrac{1}{\sqrt{2}}(\ket{H}+i\ket{V})$
and $\ket{R}:=\tfrac{1}{\sqrt{2}}(\ket{H}-i\ket{V}).$

The basic linear optical elements for the manipulation of a single
photon are as follows:

\begin{enumerate}

\item {\it Wave plates} are devices that delay one of the polarization
components with respect to the other, and from the state
Eq.~(\ref{psistate}) they lead to
\begin{equation}
\ket{\Psi}=\alpha\ket{H} + e^{i\phi}\beta\ket{V},\label{psistate2}
\end{equation}
where $\phi$ describes the delay. For half-wave plates (HWP) and
quarter-wave plates (QWP) the delays are $\phi=\pi = 180^\circ$
and $\phi=\pi/2 = 90^\circ,$ respectively.

\item A {\it beam splitter} (BS) is a device that has two {\it spatially}
separated input modes. The state of the two spatially separated
output modes can be obtained through linear transformations. That
is, let us denote the two spatially separated modes $\{\ket{T}$ and
$\ket{B}\}$ referring to Top and Bottom. Then a $50-50\%$ beam
splitter that does not introduce any additional phase shift makes
the transformation
\begin{eqnarray}
\ket{T} \rightarrow  \tfrac{1}{\sqrt{2}}\big(\ket{T}+\ket{B}\big), \nonumber\\
\ket{B} \rightarrow  \tfrac{1}{\sqrt{2}}\big(\ket{T}-\ket{B}\big).
\end{eqnarray}

\item {\it Polarizing beam splitters.}
If photons carry also a polarization degree of freedom, then a
polarizing beam splitter (PBS) can be used to direct photons with $H$
polarization into one direction, while photons with $V$ polarization
into the other direction. Using of polarizing beam splitters is
typical before detecting the state of the photons. That is, after
the polarizing beam splitter two detectors detect the photons of the
two different polarizations.

\end {enumerate}

Next we discuss, how to realize multi-photon experiments. The
difficulty of using photons, rather than atoms or ions, for creating
entangled states is that they do not interact with each other.
Thus, special techniques are needed to create entangled states. The
most often used method is using spontaneous parametric
down-conversion and post-selection.

\begin{enumerate}

\item {\it Spontaneous parametric down-conversion (SPDC)} is 
a process in which a photon with frequency $\nu$ is converted 
to two photons with lower frequencies $\nu_1$ and $\nu_2$ 
\cite{PhysRevLett.75.4337}. Due 
to conservation of energy, $\nu=\nu_1+\nu_2.$ In a typical 
experiment, by filtering out the unwanted frequencies one 
sets the frequencies of the outgoing photons equal, that 
is, $\nu_1=\nu_2=\tfrac{\nu}{2}.$  Due to conservation of 
momentum, the sum of the momenta of the outgoing photons 
equal the momentum of the incoming photon. The down-conversion 
is collinear, if the outgoing photons proceed in the same 
direction as the incoming one, while it is non-collinear if 
this is not the case. Non-collinear down-conversion allows 
the spatial separation of the two photons right after the 
down-conversion. In case of type-I down-conversion the two 
photons have the same polarization, while for type-II 
down-conversion they have perpendicular polarizations.

\item {\it Conditional detection} or {\it post-selection.}
Since the SPDC happens with {\it some small probability},
e.g., when a laser beam is directed to a $\beta$-Barium-Borate
(BBO) crystal, and the photon detectors do not have perfect 
efficiency, only the events where two photons are detected 
are relevant. In experiments with more than two photons, 
the photons are usually generated by higher order emissions
of SPDC (see Sec.~\ref{mohamedexp}) or by interaction of 
entangled photon pairs via beam splitters. Then, it  
happens that the $N$ photons are not equally distributed over
the $N$ spatial modes, i.e., with some probability one mode
may contain more than one photon and some modes may contain 
none. In this case, the observation of a multipartite state 
is usually based on {conditional detection} (or post-selection): 
only if each spatial mode contains one photon, i.e., all detectors 
register one photon, the desired state is observed 
(for an example see Sec.~\ref{mohamedexp}). 

\end{enumerate}

\subsubsection{The experiment}

Now we are in the position to review the experiment of
 Ref.~\cite{PhysRevLett.81.5039}. In the experiment they created a
two-qubit singlet state [Eq.~(\ref{CHSH_Phi})] and used a CHSH
Bell's inequality [Eq.~(\ref{CHSH})] to prove the non-locality of
the state. The key point of the experiment is that the two detection
events are space-like separated. Thus, what is measured at one of
the qubits cannot influence the results of the measurement of the
other qubits, even if we allow for a physical mechanism, so far
unknown, that could make communication possible between the two
detectors with the speed of light. Thus, this experiment closed the
locality loophole discussed in Sec.~\ref{Sec_loopholes}. Note,
however, that it did not close the detection loophole since the
detection efficiency for photonic experiments is relatively low (in
the present case it was $5 \%$).

\begin{figure}
\begin{minipage}{0.45\textwidth}
\centerline{ \epsfig{file=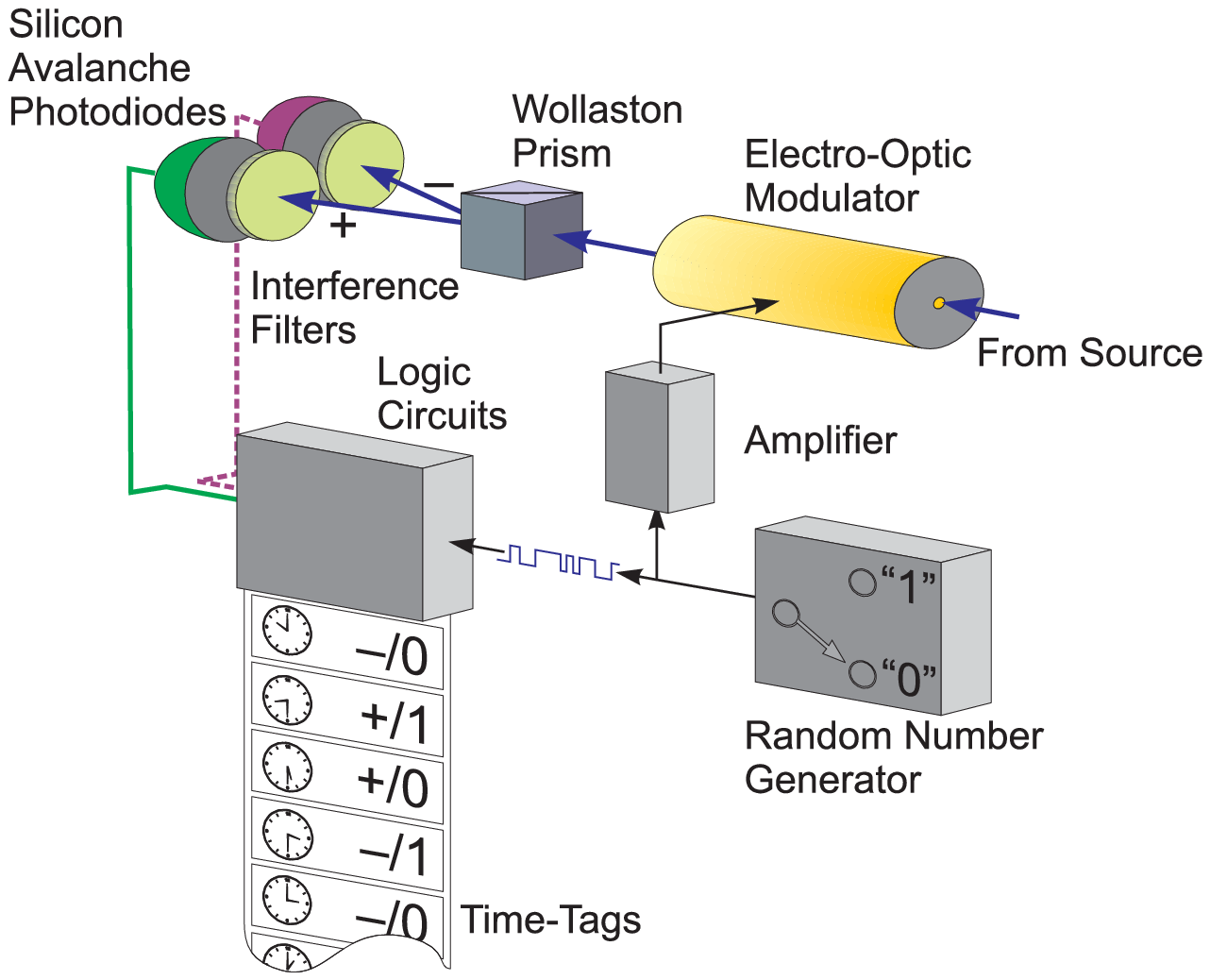,bb=-60 155 307
450,clip=,width=2.5in}} \caption{One of the two observer stations.
All alignments and adjustments were pure local operations that did
not rely on a common source or on communication between the
observers. Figure from Ref.~\cite{PhysRevLett.81.5039}.}
\label{expbell_fig2}
\end{minipage}
\begin{minipage}{0.1\textwidth}
\mbox{ }
\end{minipage}
\begin{minipage}{0.45\textwidth}
\centerline{ \epsfig{file=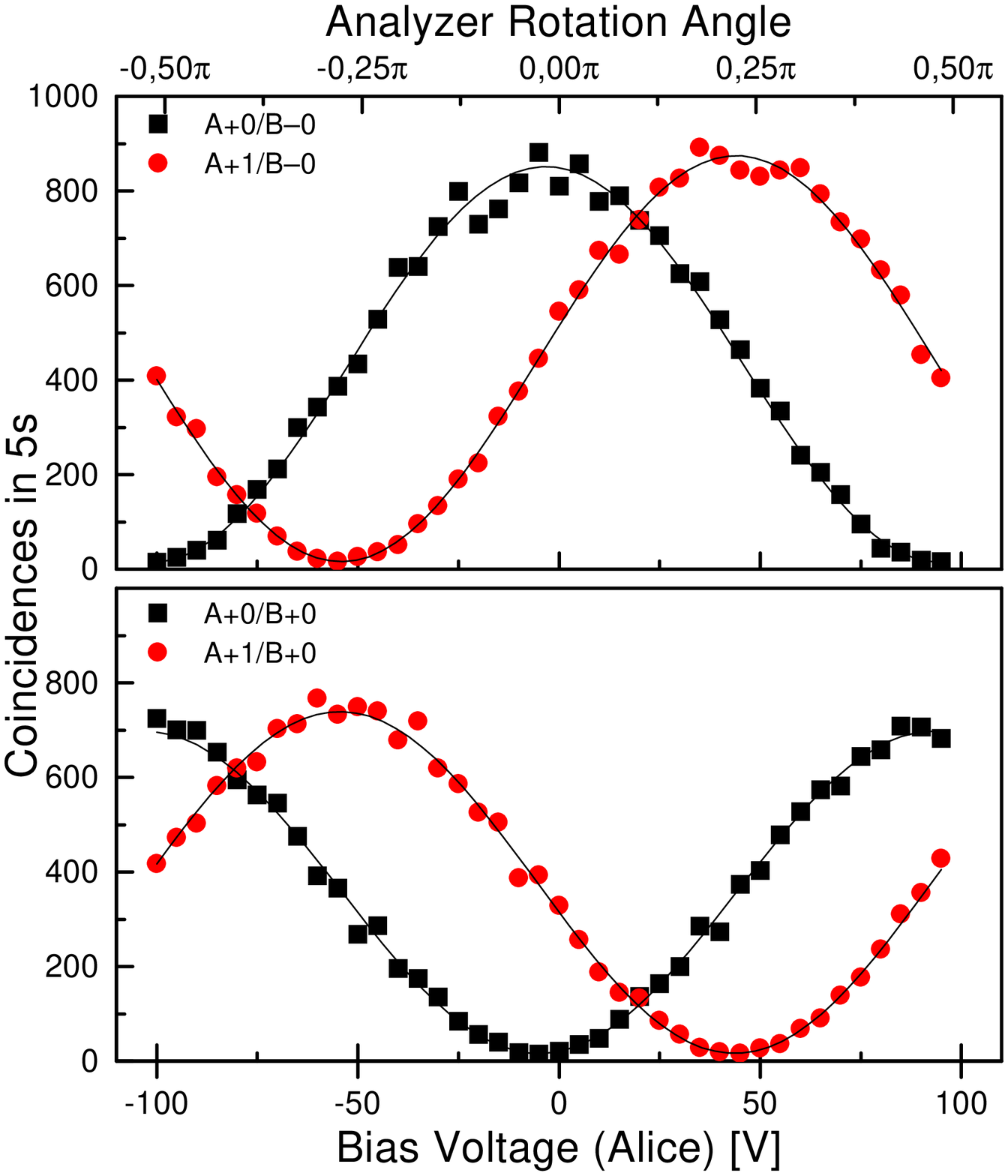,bb=44 196 551
786,clip=,width=2.1in}} \caption{Four out of sixteen coincidence
rates between various detection channels as functions of bias
voltage (analyzer rotation angle) on Alice's modulator. A$+$1/B$-$0
for example are the coincidences between Alice's ``$+$'' detector
with switch having been in position ``1'' and Bob's ``$-$'' detector
with switch position ``0''. The difference in height can be explained by
different efficiencies of the detectors. Figure from
Ref.~\cite{PhysRevLett.81.5039}.} \label{expbell_fig3}
\end{minipage}
\begin{minipage}{0.1\textwidth}
\mbox{ }
\end{minipage}
\end{figure}

In the experiment, type-II parametric down-conversion was used to
generate an entangled photon pair in state Eq.~(\ref{CHSH_Phi})
\cite{PhysRevLett.75.4337}. Then the two photons were led to the
two observers stations (see Fig.~\ref{expbell_fig2}) with optical
fibers. In order to achieve space-like separated detector events ("clicks"), the
observer stations were $400$m away from each other.

It was decided locally, which one of the two operators to measure at
each qubit. This was done using a quantum random generator, rather
than a pseudo-random number generator. Latter would be based on a
deterministic process and in principle one party could predict what
the other measures knowing the algorithm of the generator. Moreover,
the timing of the generation of the random decision for the
detectors was such that, again, communication that is not faster
than light, could not bring information on the choice of what to
measure at one detector to the photon at the other detector. This
means that the generation of a random bit, the choice between the
two possible measurements and even detection of the result was
realized locally at the two parties, as can be seen in
Fig.~\ref{expbell_fig2}. The operator to be measured was set by an
electro-optical modulator. The result was recorded in a logic
circuit together with a time tag and the information on which one of
the two operators were measured.

If the two polarization analyzers are set into directions $\alpha$
and $\beta,$ respectively, then the coincidence rates show a
sinusoidal dependence on $\alpha-\beta,$ which can be seen in
Fig.~\ref{expbell_fig3}. The solid lines are the theoretical curves,
while the dots correspond to the measurement values. The
experimental results fit very well the theoretical calculations.
The mean value of the Bell operator was $2.73\pm0.02$ and the visibility is
$97\%.$ For that $14700$ coincidence events were collected in 10s.
Note that the coincidence rates in multi-qubit experiments based on
down-conversion and post-selection are much lower, and the
experimental time is much longer. Such experiments are presented in
Sec.~\ref{mohamedexp} and Sec.~\ref{sectiondickeexp}.

\section{Entanglement witnesses}
\label{sectionentanglementwitnesses}

In this Section, we will explain how entanglement witnesses can be
used for the {\it experimental} detection of  entanglement. The
theoretical concept of entanglement witnesses has already been
introduced in the Sections \ref{ewsection} and \ref{multiewsection},
and some major construction methods were shown. Now we will discuss
how witnesses can be implemented in experiments. Furthermore, we will describe advanced
methods to construct witnesses for special experimental situations
and will also discuss several recent experiments from the point of
view of entanglement detection. Finally, we will discuss how
entanglement witnesses can be used to derive {\it quantitative}
statements about the entanglement in a quantum state and will explain
applications of witnesses in quantum key distribution.

\subsection{Witnesses for two qubits}
Let us start our discussion with a simple two-qubit
example \cite{PhysRevA.66.062305, guehne-2003-50}.
From an experimental point of view, the
investigation of two qubits seems trivial, as most
experiments nowadays aim at the generation of multiparticle
entangled states. However, the investigation of the two-qubit
case  will show already the main features of entanglement
detection via witnesses in a simple setting.

\subsubsection{Construction of the witness}

Let us consider a setup that is intended to produce a particular
pure state $\KetBra{\psi},$ but due to the imperfections some noise
is added. As a simple approximation, one may consider a
mixed state $\varrho$ of the form
\begin{equation}
\varrho(p) := p \KetBra{\psi} + (1-p)\frac{\eins}{4},
\label{rhodefinition}
\end{equation}
where $\eins/4$ is the maximally mixed state of two qubits. The aim
is to give a scheme to decide whether $\varrho(p)$ is entangled or
not. Clearly, as the form of the state in Eq.~(\ref{rhodefinition})
is only a rough approximation, the conclusion that the state is
entangled should not depend on any assumption on the form of the
state. However, it is reasonable to choose the detection method such
that it detects states of the form Eq.~(\ref{rhodefinition}) even
for small $p.$ The optimal entanglement witness for an entangled
$\varrho(p)$ is easy to construct from the results of Section
\ref{ewsection}. First, one has to compute the eigenvector
corresponding to the negative eigenvalue of $\varrho(p)^{T_B},$ then
the witness is given by the partially transposed projector onto this
eigenvector.

If the Schmidt decomposition of $\Ket{\psi}$ is
$\Ket{\psi} = a \Ket{01} + b \Ket{10}$  the eigenvalues
of $\varrho(p)^{T_B}$ are given by
$\{ (1-p)/4 + pa^2; (1-p)/4 + pb^2; (1-p)/4+p ab;
(1-p)/4-pab \}.$  Therefore $\varrho(p)$ is entangled
iff $p > 1/(1+4ab).$ The eigenvector corresponding
to the minimal eigenvalue $\lambda_-$ is given by
\begin{equation}
\Ket{\phi_-} = \frac{1}{\sqrt{2}}(\Ket{00}-\Ket{11}),
\end{equation}
and thus the witness $\WW$ can be written in several ways as
\begin{equation}
\WW = \KetBra{\phi_{-}}^{T_B}
= \frac{1}{2}\eins - \ketbra{\psi^+}
    = \frac{1}{2}\left( \begin{array}{cccc}
1 & 0 & 0 & 0\\
0 & 0 &-1 & 0\\
0 &-1 & 0 & 0\\
0 & 0 & 0 & 1
\end{array}\right),
\label{twoqubitwitness}
\end{equation}
with $\Ket{\psi_+} = (\Ket{01}+\Ket{10})/\sqrt{2}.$
Note that this witness does neither depend on $p,$ nor on the
Schmidt coefficients $a,b.$ Due to the special assumption about
the state, it detects  $\varrho(p)$ iff it is entangled,
since  we have $Tr[\KetBra{\phi_{-}}^{T_B} \varrho(p)] =
Tr[\KetBra{\phi_-} \varrho(p)^{T_B}] = \lambda_-.$

\subsubsection{Local decomposition of the witness}

In the form of  Eq.~(\ref{twoqubitwitness}) the witness $\WW$ 
cannot easily be implemented in an experiment. The reason is 
that $\WW$ is up to a constant a projector onto one of the Bell 
states and such a nonlocal observable is not straightforward 
to measure in practice (for a theoretical proposal to do that 
see Ref.~\cite{filip-2002-65}). The observables that can easily
be measured in any experiment are local observables like 
$\mean{\sigma_z \otimes \sigma_z}$ or projectors like
$\mean{\ketbra{0}\otimes \ketbra{0}}.$
Therefore, for the experimental implementation it is necessary to decompose
the witness into operators that can be measured locally
\cite{PhysRevA.66.062305, terhal-2002-287}. Thus, we need a
decomposition into projectors onto product vectors of the form
\begin{equation}
\WW=\sum_{i=1}^{k} c_{i}\KetBra{e_{i}} \otimes
\KetBra{f_{i}}.
\label{pvdecomposition}
\end{equation}
Such a decomposition can be measured locally: Alice and Bob measure
the expectation values of the $\KetBra{e_{i}} \otimes \KetBra{f_{i}}$
and add their results with the weights $c_i.$ One can construct such
a decomposition in many ways, but it is reasonable to do it in a way
that corresponds to the smallest possible experimental effort for
Alice and Bob.

The experimental ``cost'' that Alice and Bob have to pay when
measuring the witness $\WW$ via such a decomposition is the number
of measurements they have to perform. As already mentioned, local
measurements are easily possible, and by measurements we always mean
von Neumann (or projective) measurements. One measurement on Alice's
side in this sense consists of a choice of one orthonormal basis for
Alice's Hilbert space. For qubits, this corresponds to a
polarization measurement in a certain direction. For a particle with
spin $s$ one may interpret this as the choice of a direction for a
Stern-Gerlach-like apparatus. With this setup, Alice can measure the
probabilities for all the outcomes.

A similar argument holds for Bob, thus, if one considers a $d \times
d$ system a term of the form
\begin{equation}
\MM = \sum_{k,l=1}^{d}c_{kl}\KetBra{a_k}\otimes\KetBra{b_l}
\label{lvnmdefinition}
\end{equation}
with $\BraKet{a_s}{a_t}=\BraKet{b_s}{b_t}=\delta_{st}$ can be
measured with one collective setting of measurement devices
of Alice and Bob. Alice and Bob can discriminate between the
states $\Ket{a_k b_l},$ measure the probabilities of these states
and add their results with the weights $c_{kl}$ using one
collective setting and some classical communication. See also
Fig.~\ref{psi4results} for a four-qubit example of such 
measurement results.

It is therefore reasonable to find a decomposition of the form
\begin{equation}
\WW=\sum_{i=1}^{m} \MM_i = \sum_{i=1}^{m}
\sum_{k,l=1}^{d} c^{i}_{kl}
\KetBra{a^{i}_{k}} \otimes\KetBra{b^{i}_{l}}
\label{problemdefinition}
\end{equation}
with $\BraKet{a^i_s}{a^i_t}=\BraKet{b^i_s}{b^i_t}=\delta_{st}$ and
minimal number of measurements, {i.e.,} a minimal $m.$ Note that a
decomposition of the form (\ref{problemdefinition}) is more general
than a decomposition into a sum of tensor products of operators $
\WW=\sum_{i=1}^{m}\gamma_{i} A_i \otimes B_i.
$ The difference is that a measurement as in
Eq.~(\ref{lvnmdefinition}) consist of a choice of observables (say,
$\sigma_z \otimes \sigma_z$ for definiteness), but then all
expectation values like $\mean{\sigma_z \otimes \sigma_z}$,
$\mean{\sigma_z \otimes \eins}$ and $\mean{\eins \otimes \sigma_z}$
can be determined from the same data. These terms would occur as
different $A_i \otimes B_i$ in a decomposition like
above.\footnote{One can also ask for a decomposition of $\WW$ that
minimizes the number of product vectors in
Eq.~(\ref{pvdecomposition}) \cite{sanpera-1998-58}, although this is
not optimal from an experimental point of view. This has been
discussed in Refs.~\cite{PhysRevA.66.062305, guehne-2003-50}.}

The optimal decomposition of the witness can be obtained as follows:
$\WW$ is a partially transposed projector $\KetBra{\psi}^{T_B},$
and we consider the Schmidt decomposition of $\Ket{\psi}=
\alpha\Ket{00}+\beta\Ket{11}.$ The actual  $\WW$ is
the special case  $\alpha=1/\sqrt{2}=-\beta,$ but one can directly
deal with the most general $\Ket{\psi}.$ We define  as usual the
spin directions by $\Ket{z^+}=\Ket{0},\Ket{z^-}=\Ket{1},
\Ket{x^\pm}=\frac{1}{\sqrt{2}}(\Ket{0}\pm \Ket{1},
\Ket{y^\pm}=\frac{1}{\sqrt{2}}(\Ket{0}\pm i \Ket{1}$
and obtain the decomposition
\begin{eqnarray}
\label{witdec1}
{\KetBra{\psi}^{T_B}}
&=& \alpha^2\KetBra{z^+ z^+}+\beta^2\KetBra{z^- z^-}+
\alpha\beta\left(\KetBra{x^+ x^+}+\KetBra{x^- x^-}-\right.
\nonumber\\
& & \left.-\KetBra{y^+ y^-}-\KetBra{y^- y^+}\right)
\\
&=&\frac{1}{4}\left[\Eins \otimes \Eins +\sigma_z\otimes \sigma_z
+(\alpha^2-\beta^2)(\sigma_z\otimes\Eins+\Eins\otimes\sigma_z)
+2\alpha\beta(\sigma_x\otimes \sigma_x+\sigma_y\otimes
\sigma_y)\right].
\label{witdec}
\end{eqnarray}
This decomposition into six product vectors requires only a measurement
of three settings: Alice and Bob only have to measure the coincidence
probabilities of $\sigma_x \otimes \sigma_x$, $\sigma_y \otimes \sigma_y$,
and $\sigma_z \otimes \sigma_z$ to measure $\KetBra{\psi}^{T_B}.$ We will
discuss an implementation of this witness in Section \ref{sectionmartiniexp}.

The question arises, whether this decomposition is really optimal. Indeed,
this is the case. It can be shown
(see Refs.~\cite{PhysRevA.66.062305, guehne-2003-50}) that
the witness $\WW$ can not be decomposed in two measurements, hence the
decomposition is optimal.

Finally, it should be noted that for the case that the target state
is given by $\ket{\psi}=(\ket{01}+\ket{10})/\sqrt{2}= \ket{\psi^+},$
the witness $\WW$ also allows for a determination of the fidelity of
the state. This quantity is defined as $F_\psi=\bra{\psi}\vr
\ket{\psi}$ and quantifies to which extent the desired state was
produced. Due to Eq.~(\ref{twoqubitwitness}) we have \be F_{\psi^+}
= \frac{1}{2} - \mean{\WW} \label{fidwit} \ee and the decomposition
of $\WW$ in three local measurements just means that the fidelity of
$\ket{\psi^+}$ can be determined with three measurements. This
connection between mean values of witnesses and the fidelity of the
target state will also occur in the multipartite setting.

\subsubsection{Discussion}

We have seen that with the help of a local decomposition
entanglement witnesses can be used for the experimental detection of
entanglement. For two qubits, this required only three local
measurements, which is less than quantum state tomography that would
require $3^2=9$ measurements, namely the measurement of all
correlations $\mean{\sigma_i \otimes \sigma_j}.$ In general, two
criteria determine the efficiency of an entanglement detection
scheme:

\begin{enumerate}

\item {\it Robustness to noise:} The robustness to noise is given
by the minimal $p$ in Eq.~(\ref{rhodefinition}) that has to be
reached for the state to be detected. Equivalently, one can define a
$p_{\rm noise}=1-p$ and ask for the maximal $p_{\rm noise}$ that can
be tolerated. This robustness quantifies the noise that can be
allowed in an experiment, and if some amount of noise is expected
one can estimate, whether there is a chance to confirm the success
of the experiment at all.

Clearly, in an experimental setting
the noise does not have to be white and the state may not be
of the form as in Eq.~(\ref{rhodefinition}). However, this
is often a reasonable approximation, hence the robustness
to noise is a good indicator to compare different entanglement
witnesses or different entanglement detection methods.

\item {\it Number of measurement settings:} Another crucial
point is the number of local measurement settings that are required
for an implementation. To use as few measurements as possible is
important from several points of view: First, in multi-photon
experiments the count rates (i.e., the number of successfully
generated states per time unit) is often low. Therefore, the
measurement of a single setting with a given accuracy needs a
certain time, and only a restricted set of measurements can be done
while keeping the setup stable. But even if measurements can be
repeated fast, and a high rate of states is available (as it is in
ion trap experiments), state tomography would require an
exponentially increasing effort, making it practically impossible
for ten or more qubits \cite{haeffner-2005-438}. Finally, for
experiments using hyper-entangled photon states, the switching of
one measurement setting to the other requires the adjustment of an
interferometer, which needs a considerable amount of time and work
\cite{TwoPhotonClusterState,chen-2007-99} (see also Section
\ref{clusterexp}).

In the example above, three measurement settings suffice, which is a
modest improvement compared with full state tomography. However, the
difference becomes large, if the number of particles increases. For
tomography of an $N$-qubit state, $3^N$ measurements are necessary,
which amounts to an exponentially growing effort. For the
measurement of the witness, typically only a few measurements are
required. As we will see in Section \ref{sectionstabilizerwitnesses}, for important states, even
for arbitrarily many qubits, only two measurements are required for
the detection of entanglement and an estimation of the fidelity.
\end{enumerate}

\subsubsection{Further results for bipartite systems}

Let us finally mention further results on the decomposition of
witnesses for bipartite systems. As a first question, one may ask
what happens if the witness is of the form
$\WW=\ketbra{\psi}^{T_B},$ where
\be
\ket{\psi}= \sum_i \lambda_i\ket{ii}
\ee
is a state on a $d \times d$-system. It has been shown
in Ref.~\cite{guehne-2003-50} that if $\ket{\psi}$ has a Schmidt
rank of $k,$ then at least $k+1$ measurements are necessary. Also,
if $k$ is even (odd), an explicit decomposition into $2k$ ($2k-1$)
measurements has been found. For the case that $\ket{\psi}$ is a
projector onto a maximally entangled state in a $d \times d$-system
and $d$ is prime, it has been shown that $\WW$ can indeed be
decomposed into $d+1$ measurements \cite{pittenger-2003-67}, hence
for this special case the problem of the optimal decomposition
is solved.

A different interesting problem, which is somehow reverse to the
original problem is the following: If we assume that some
correlation measurements like $\sigma_x \otimes \sigma_x, \sigma_x
\otimes \sigma_z, \sigma_z \otimes \sigma_x$ and $\sigma_x \otimes
\sigma_x$ have been done, which states can then be detected with
that data? This question is important in the context of quantum key
distribution, and we will discuss it in Section \ref{sectionqkd}.


\subsection{Implementation of a two-qubit witness}
\label{sectionmartiniexp}

In this Section we describe the experiment by M. Barbieri {\it et
al.} \cite{PhysRevLett.91.227901}, which was one of the first 
implementations of entanglement witnesses.\footnote{For a similar experiment see Ref.~\cite{schuckthesis}.} In the experiment they created
two-qubit Werner states (see Sec.~\ref{sectionwernerstates}) of
photons using parametric down-conversion and linear optics 
(see Sec.~\ref{expbell}).

\begin{figure}
\begin{minipage}{0.45\textwidth}
\centerline{ \epsfig{file=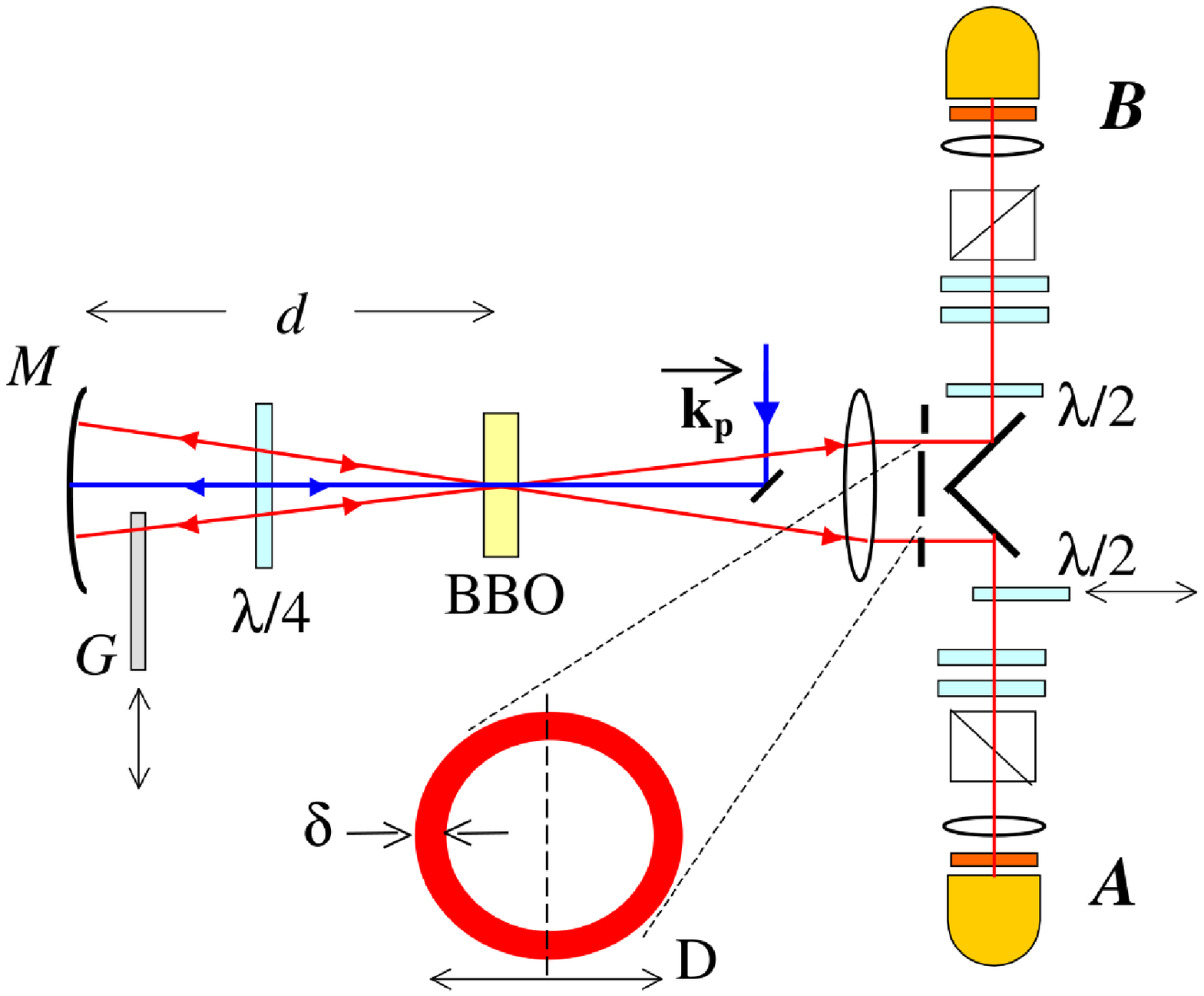,width=2.9in}}
\caption{Layout of polarization entangled photon source. For details
see the text. Figure taken from Ref.~\cite{PhysRevLett.91.227901}.}
\label{martini_fig1}
\end{minipage}
\begin{minipage}{0.1\textwidth}
\mbox{ }
\end{minipage}
\begin{minipage}{0.45\textwidth}
\centerline{ \epsfig{file=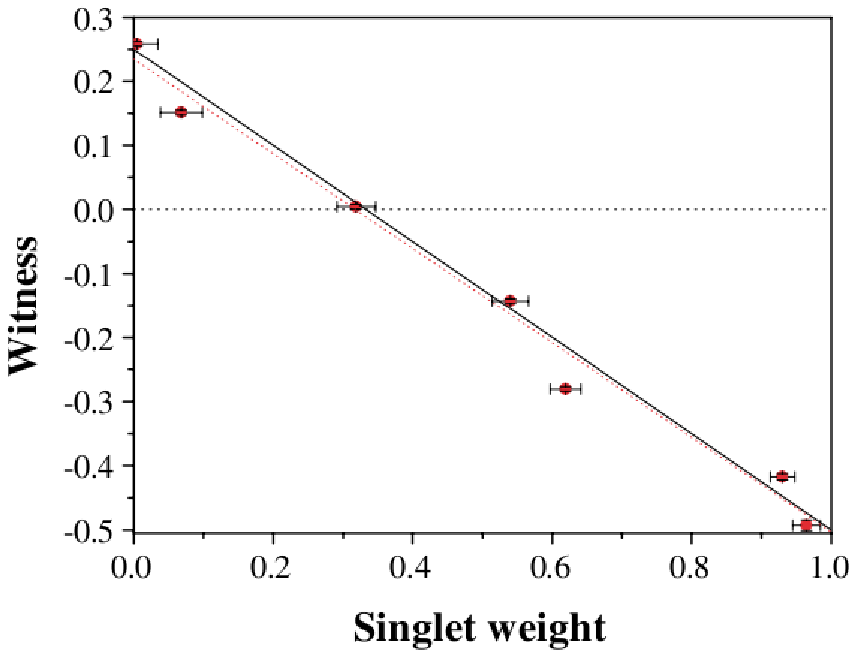,width=2.9in}}
\caption{Experimental results of the entanglement witness for Werner
states: Theoretical curve (solid), experimental results (dots),
and  experimental best fit (dotted). The horizontal dashed line indicates
the transition between separable and entangled states. Figure taken
from Ref.~\cite{PhysRevLett.91.227901}.} \label{martini_fig2}
\end{minipage}
\begin{minipage}{0.1\textwidth}
\mbox{ }
\end{minipage}
\end{figure}

The experimental setup can be seen in Fig.~\ref{martini_fig1}. The
incoming photon with wavelength $\lambda_P$ and $V$-polarization
arrives from the top from the laser. Then, after a mirror, it
propagates to the $-k_p$ direction towards the BBO crystal. Here
with some probability it is converted down (type I) into two photons
with $\lambda:=\lambda_P/2$ and $H$ polarization. The two photons
must have total momentum $-k_p$ due to momentum conservation. Thus,
if one of them is propagating in a direction slightly up, the other
has to propagate into a direction slightly down. Both of these
photons are reflected on the mirror M, go back through the crystal
and a lens. Another beam of photon pairs arises from the photons
that went through the crystal without down-conversion, got reflected
on the mirror and then converted down on the BBO crystal when
propagating from the left to right. These photons also arrive to the
same lens. Due to the wave plate between the mirror M and the BBO
crystal, the photon pair from the first down-conversion arrived in
$\ket{HH}$ state to the lens, while the other arrives in the
$\ket{VV}$ state. As the contributions are indistinguishable,
finally the setup generates photons in the state
\begin{equation}
\ket{\Psi}=\frac{1}{\sqrt{2}}(\ket{HH}+e^{i\phi}\ket{VV}).
\label{demartinistate}
\end{equation}
There is a $\lambda/2$ wave plate inserted before
detector A, which leads to a $\ket{\Psi_{\rm
singlet}}=\tfrac{1}{\sqrt{2}}(\ket{HV}-\ket{VH})$ state.

It is also possible to "spoil" the entangled state
Eq.~(\ref{demartinistate}) and create a mixture
\begin{equation}
    \varrho(p)=p\ketbra{\Psi_{\rm
singlet}}+(1-p)\frac{\openone}{4}.
    \label{demartinistate2}
\end{equation}
This can be done using the coated glass plate G and the wave plate
before the detector A. This partially spoils the indistinguishability
and leads to the Werner state. For large values of $p$ the state created
is entangled, while for small values it is separable. The entanglement of
the state is detected by the following witness operator [see Eq.~(\ref{witdec})]
\begin{equation}
    \WW =\frac{\openone}{2}-\ketbra{\Psi_{\rm singlet}}=
    \frac{1}{4}\big(\openone+\sigma_x\otimes\sigma_x
    +\sigma_y\otimes\sigma_y+\sigma_z\otimes\sigma_z\big),
\end{equation}
that can efficiently be measured locally. The experimental results
are shown in Fig.~\ref{martini_fig2}. Note that the witness detects the states also
for $p \leq 0.65$, where no Bell inequality can detect the entanglement of
that state.

\subsection{Witnesses for multi-qubit systems}

Let us now discuss witnesses for multipartite systems. Here, the
general strategy is quite similar to the two-qubit case: The
experimenter aims to prepare a state that is affected by some noise.
The problem is to find a witness for this scenario that is robust
against noise, and can be measured with a moderate effort. In this
section we will mainly discuss the ``standard'' witnesses for
genuine multipartite entanglement as described in
Eq.~(\ref{standardwitness}). Different witness constructions will be
discussed in the following sections.

\subsubsection{Three-qubit GHZ state}

Let us discuss first the three-qubit GHZ state
$\ket{GHZ_3}=(\ket{000}+\ket{111})/\sqrt{2}$ as a simple example. The
experimentally created state may be of the form
\be
\vr = (1-p_{\rm noise}) \ketbra{GHZ_3} + p_{\rm noise} \frac{\eins}{8}
\ee
As we already know from Eq.~(\ref{WCN}) the standard witness would
read $\WW = \eins/2-\ketbra{GHZ_3}$ and this detects the state for
$p_{\rm noise} < 4/7 \approx 0.57.$

For this witness, a decomposition into local measurements can be
found as follows: first one decomposes $\WW$ into tensor products of
Pauli matrices, then, one optimizes this decomposition in order to
minimize the measurement settings. So one can write
\cite{guehne-2003-42}: \bea \WW &=& \frac{1}{8}\big( 3
\cdot\Eins\otimes\Eins\otimes\Eins-
\Eins\otimes\sigma_z\otimes\sigma_z- \sigma_z \otimes \Eins \otimes
\sigma_z- \sigma_z \otimes \sigma_z \otimes \Eins - \sigma_x \otimes
\sigma_x \otimes \sigma_x + \sigma_x \otimes \sigma_y \otimes
\sigma_y +
\nonumber \\
& &
+
\sigma_y \otimes \sigma_y \otimes \sigma_x +
\sigma_y \otimes \sigma_x \otimes \sigma_y \big).
\nonumber \\
&=&
\frac{1}{8}\Big(
3 \cdot\Eins\otimes\Eins\otimes\Eins-
\Eins\otimes\sigma_z\otimes\sigma_z-
\sigma_z \otimes \Eins \otimes \sigma_z-
\sigma_z \otimes \sigma_z \otimes \Eins
- 2 \cdot \sigma_x^{\otimes 3} +
\nonumber \\
& & +\sqrt{2}\cdot \Big(\frac{\sigma_x +
\sigma_y}{\sqrt{2}}\Big)^{\otimes 3} + \sqrt{2}\cdot
\Big(\frac{\sigma_x -\sigma_y}{\sqrt{2}}\Big)^{\otimes 3} \Bigr).
\eea This decomposition requires only the measurement of the four
correlations $ \sigma_Z^{\otimes 3}$, $ \sigma_x^{\otimes 3}$ and
$({\sigma_x \pm \sigma_y}/{\sqrt{2}})^{\otimes 3}.$ These are simple
measurements on the Bloch sphere that can be implemented
experimentally. From the data, also the fidelity of the GHZ state
can be determined, as in Eq.~(\ref{fidwit}). It can be further shown
that this decomposition is optimal, i.e., it is not possible, to
determine $\mean{\WW}$ with three local measurement settings
\cite{guehne-2003-42,guehnediss}.

\subsubsection{General states}

Let us discuss now general states. Using the presented methods one
can calculate the witness and determine its robustness to noise
straightforwardly for an arbitrary state $\ket{\psi}$. To obtain a
good local decomposition requires often some effort, especially
proving that a given decomposition is optimal, is often very
difficult \cite{guehnediss}. However, for many interesting states
results on this problem are known, and we summarize them in Table
\ref{witnesstable}. Interestingly, one can also make some general
statement about the required measurement settings: For any pure
state there exists a witness that requires $2N-1$ measurements
\cite{chen-2007,ToolBox}, but the robustness to noise may be small.
Furthermore, there exist observables, for which the local
decomposition requires $2 \cdot 3^{N-1}/(N+1)$ local measurements, 
which means that a local measurement of these observables requires
nearly the same effort as state tomography, but specific examples 
of such observables are not known \cite{guehnediss}.

\begin{table}
\caption{Results on local decompositions of different entanglement
witnesses for different states.
\label{witnesstable}}

\begin{minipage}{\textwidth}
\centering
\begin{tabular}{|c|c|c|c|c|c|c|c c c}
\hline
\# of qubits & state & witness & maximal $p_{\rm noise}$ & local measurements & references & remarks\\
\hline
3 & $\ket{GHZ_3}$ & $\tfrac{1}{2}\eins - \ketbra{GHZ_3}$ & 4/7 & 4 (optimal) & \cite{guehne-2003-42} &
\footnote{Witnesses that tolerate less noise but require less settings exist. See Section \ref{sectionghzstabwit}.}
\\
\hline
3 & $\ket{W_3}$ & $\tfrac{2}{3}\eins - \ketbra{W_3}$ & 8/21 & 5 (optimal) & \cite{guehne-2003-42}  &
\footnote{Witnesses that tolerate more noise with the same measurements exist. See
Sections \ref{sectionionexpwit} and \ref{sectiondickeexp}.}
\\
\hline
4 & $\ket{CL_4}$ & $\tfrac{1}{2} \eins - \ketbra{CL_4}$ & 8/15 & 9 (optimal)& \cite{guehnediss,TokunagaFidelityEstimation}  & $^{\rm a}$
\\
\hline
4 & $\ket{\Psi_2}$ & $\tfrac{3}{4} \eins - \ketbra{\Psi_2}$ & 4/15 & 15 & \cite{bourennane:087902,ToolBox}  &
\footnote{Witnesses that tolerate more noise and require less settings exist \cite{ToolBox,seevinck-2007}.}
\\
\hline
4 & $\ket{D_{2,4}}$ & $\tfrac{2}{3} \eins - \ketbra{D_{2,4}}$ & 16/45 & 21 & \cite{DickeEntanglementJOSAB2007}  &
\footnote{For witnesses that tolerate less noise with less settings see Section \ref{sectiondickeexp},
for witnesses which tolerate more noise see Ref.~\cite{seevinck-2007}.}
\\
\hline
$N$ & $\ket{GHZ_N}$ & $\tfrac{1}{2} \eins - \ketbra{GHZ_N}$ & $1/2 \cdot [1/(1-1/2^N)]$ & $N+1$ & \cite{ToolBox}  & $^{\rm a}$
\\
\hline
$N$ & $\ket{W_N}$ & $\tfrac{N-1}{N}  \eins - \ketbra{W_N}$ & $1/N \cdot [1/(1-1/2^N)]$ & $2N-1$ & \cite{haeffner-2005-438,ToolBox}  & $^{\rm b}$
\\
\hline
$N$ & $\ket{G_N}$ & $\tfrac{1}{2}\eins - \ketbra{G_N}$ &  $1/2\cdot [1/(1-1/2^N)]$& depends on the graph &
\cite{PhysRevLett.94.060501}  & $^{\rm a}$
\\
\hline $N$ & $\ket{D_{\frac{N}{2},N}}$ & $\tfrac{N}{2N-2} \eins -
\ketbra{D_{\frac{N}{2},N}}$ & $1/2 \cdot (N-2)/[(N-1)(1-1/2^N)]$& not known
& \cite{DickeEntanglementJOSAB2007}  &
\\
\hline
\end{tabular}
\end{minipage}
\end{table}

\subsection{Comparing witnesses with Bell inequalities}
\label{subsec_comp_wit_bell}

Let us now compare the entanglement detection via Bell
inequalities and entanglement witnesses in some more
detail. This will also shed light on the underlying
assumptions of different entanglement verification methods
(for a detailed treatment on this see Ref.~\cite{vanenk-2007-75}).

When comparing different entanglement verification schemes, two
main questions are of interest. First, one may ask for the power
of the method in an experiment, i.e., to which extent it gives
useful and relevant information about the state. Here, also the
question for the experimental effort becomes relevant. A second
main question concerns the underlying assumption for the scheme,
and to which extent these assumptions are realistic in a given
experimental situation.

\subsubsection{Usefulness}

Concerning the first question, we have seen in Section \ref{bellgenuine} 
that a violation of most of the Bell inequalities for multipartite systems 
does only exclude full separability of the state. But, as seen in the same
Section, there are some Bell inequalities that can be used for proving 
genuine multipartite
entanglement. However, up to now such constructions work only for
few states (like the GHZ states) and there is no construction known
for other important states, such as cluster states. Furthermore, the
existing Bell inequalities for the detection of genuine multipartite
non-locality require a high fidelity of the target state and an
exponentially growing number of measurement settings. In contrast to
that, we will see in Section \ref{sectionghzstabwit} that one can
write down witnesses for the GHZ state that tolerate more noise and
need only two measurement settings, independently from the number of
qubits.

\subsubsection{Underlying assumptions}

A further interesting question concerns the underlying assumptions
of the different entanglement detection methods. As the loopholes
are in most experiments not closed, Bell inequalities require the
assumption that the parties act like as if they were space-like
separated and that the detector efficiencies do not play a role.
Apart from that, however, no more assumptions are needed. Especially,
the measurement directions do not have to be fixed, a misalignment of
the measurement directions does not invalidate the conclusion that the state
was entangled.

For entanglement witnesses, more assumptions are needed. First,
loopholes can also arise for entanglement witnesses
\cite{skwara-2007-76}. Then, the validity of quantum
mechanics is assumed. Most importantly, entanglement
witnesses require that the experimentalist knows what
he is doing: The measurements have to be made in the
proper direction and the measurement apparatuses must
work as they are intended to work.\footnote{For some 
schemes to deal with non-ideal detectors see 
Refs.~\cite{seevinck-2007-76,beaudry-2008-101}.}

A further question is whether the use of entanglement witnesses
requires an assumption on the dimension of the underlying Hilbert space
\cite{brunner-2008,acin-2006-97}. In many experiments one works with
multi-level systems like ions in a trap, however, for the
theoretical description one considers it as a two level system and
disregards the levels that are not occupied. The question arises,
whether this might lead to false detection of entanglement.
To discuss this in a simple setting, consider two qubits and
the witness from Eq.~(\ref{witdec1}). A measurement of this
witness consists of the determination of the outcome probabilities
for the eigenvectors $\ketbra{x^\pm}, \ketbra{y^\pm},$ and
$\ketbra{z^\pm}$ for Alice and Bob. These probabilities are
defined on an arbitrary dimensional Hilbert space, and if
these probabilities violate the witness inequality, the corresponding
reduced two-qubit state must be entangled. This implies entanglement
of the total state.\footnote{To be precise, if $\vr_{\rm tot}$
is the total state on a $d \times d$ system, violation of the
inequality implies entanglement of the (not normalized) state
$\vr_{\rm red}=
\PP_A^{(2)}\otimes \PP_B^{(2)} \vr_{\rm tot} \PP_A^{(2)}\otimes \PP_B^{(2)}$
with $\PP^{(2)} = \ketbra{0} + \ketbra{1}.$ This implies entanglement
of $\vr_{\rm tot},$ as the projection is local.}

Finally, it should be noted that due to the finite number of copies
available in any experiment, Bell inequalities and entanglement
witnesses can only detect entanglement with a finite statistical
significance. This is quantified by the standard deviation of the
experimentally measured mean value.\footnote{For an interesting
proposal to determine the probability of entanglement via
Bayesian updating see Sec.~\ref{furtherestimationmethods}}

\subsection{Implementation of a multi-qubit witness}
\label{mohamedexp}

Now we discuss an experiment by M. Bourennane {\it et al.}
\cite{bourennane:087902}, which was the first experiment, where a
multi-qubit witness has been implemented. In this experiment, the
four-qubit singlet state $\ket{\Psi_2}$
[Eq.~(\ref{fourqubitsingletstate})] has been generated using the
second order emission of parametric down-conversion and postselection as
proposed in Ref.~\cite{PhysRevA.64.010102}.

\begin{figure}[t]
\begin{minipage}{0.47\textwidth}
\centerline{\psfig{figure=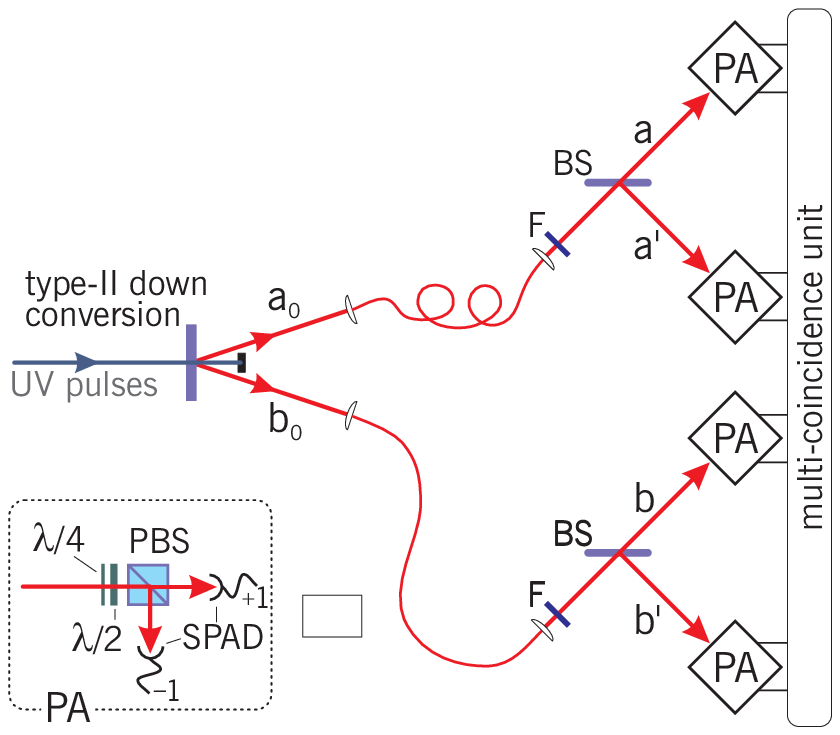,width=0.98\columnwidth}}
\caption{Setup for the generation of the four-photon singlet state
$\ket{\Psi_{2}}.$ See the text for details. The figure is taken from
Ref.~\cite{PhysRevLett.90.200403}.
 \label{psi4setupfigure}
}
\end{minipage}
\begin{minipage}{0.05\textwidth}
\mbox{ }
\end{minipage}
\begin{minipage}{0.47\textwidth}
\centerline{\psfig{figure=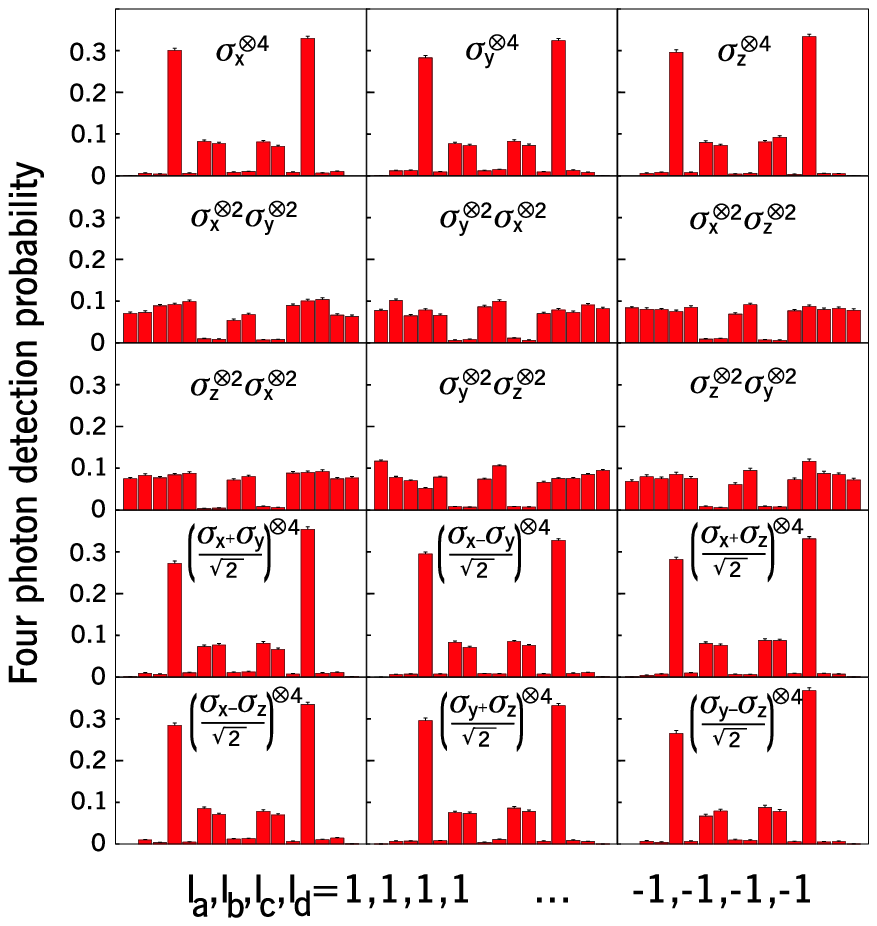,width=0.8\columnwidth}}
\caption{
Fourfold coincidence probabilities for the fifteen measurement
settings, needed for the fidelity measurement of the $\ket{\Psi_{2}}$-state.
The figure is taken from Ref.~\cite{bourennane:087902}. \label{psi4results}
}
\end{minipage}
\end{figure}

One possibility to generate entangled states of polarized photons in
a laboratory uses  type-II spontaneous parametric down-conversion
(SPDC) \cite{PhysRevLett.75.4337}.  In this process, an ultraviolet
(UV) laser pumps a nonlinear $\beta$-Barium-Borate (BBO) crystal
(see Fig.~\ref{psi4setupfigure}). Multiple emission events during
one pump pulse lead to the emission of the state \be \ket{\psi} \sim
\exp[-i \alpha (a^\dagger_{0,H} b^\dagger_{0,V} + a^\dagger_{0,V} b^\dagger_{0,H})]\ket{0}
\ee distributed onto two modes, $a_0$ and $b_0.$ Here, $a^\dagger_{0,H}$
denotes a creation operator, creating a horizontally polarized
photon in mode $a_0;$ $b^\dagger_{0,V}$ creates a vertically polarized
photon in mode $b_0$ and $\ket{0}$ denotes the vacuum. In the second
order of this process only the quartic terms are relevant, leading
to $ \ket{\psi} \sim (a^\dagger_{0,H} b^\dagger_{0,V} + a^\dagger_{0,V} b^\dagger_{0,H})^2
\ket{0}. $ This results in the state \be \ket{\psi} \sim
\ket{2H}_{a_0} \ket{2V}_{b_0} +
                \ket{2V}_{a_0} \ket{2H}_{b_0} +
                2 \ket{1H,1V}_{a_0} \ket{1H,1V}_{b_0},
\label{ex2}
\ee
where $\ket{2H}_{a_0}$ denotes a state of two horizontally polarized
photons in mode $a_0,$ and $\ket{1H,1V}_{b_0}$ denotes one horizontally
and one vertically polarized photon in mode $b_0,$ etc.

Two beam-splitters are used to distribute  the photons in
the four modes $a,b,c$ and $d.$  Then postselection is used
and one can directly calculate that
fourfold coincidences in all the four modes $a,b,c,$ and $d$ can
only occur when the state
\be
\ket{\psi} \sim  \ket{H}_{a}\ket{H}_{b}\ket{V}_{c}\ket{V}_{d}
+\ket{V}_{a}\ket{V}_{b}\ket{H}_{c}\ket{H}_{d}
+\frac{1}{2}
\big(\ket{V}_{a} \ket{H}_{b}+ \ket{H}_{a}\ket{V}_{b}\big)
\big(\ket{V}_{c}\ket{H}_{d}+\ket{H}_{c}\ket{V}_{d}\big)
\ee
is produced. This state is, up to a sign which can be removed by a
local change of the basis the four-qubit singlet state
$\ket{\Psi_{2}}$. In the experiment, this state was created
with a count rate of around five four-photon events per minute.

The witness for this state, is given by Eq.~(\ref{psivierwitnesseq})
as $ \WW^{\Psi_{2}}  =  \tfrac{3}{4} \eins  -\ketbra{\Psi_{2}}. $
This witness is equivalent to a measurement of the fidelity and
requires fifteen measurement settings. It should be noted, that in
the meantime witnesses have been found that require significantly
less measurements and tolerate even more noise \cite{ToolBox,
seevinck-2007}. The measurement results are given in
Fig.~\ref{psi4results}. The fact that the coincidence probability
results are similar for different settings is a signature of the
$U\otimes U\otimes U\otimes U$-symmetry of the singlet state. From
these data, one obtains \be \mean{\WW^{\Psi_{2}}}=-0.151 \pm 0.01
\mbox{ and } F_{\Psi_2} = 0.901 \pm 0.01 \ee which clearly proves
the four-partite entanglement. The state $\ket{\Psi_{2}}$ can then
be further used to encode a qubit in a decoherence free subspace
\cite{bourennane-2004-92} or for secret sharing \cite{gaertner-2007-98}.
With a similar setup, three-qubit W states can  be generated
\cite{PhysRevLett.92.077901}.

\subsection{Stabilizer witnesses}
\label{sectionstabilizerwitnesses}

In this Section we will explain a method for constructing witnesses
that are very convenient for experiments, since they can be
implemented with very few, often only two measurements.

\subsubsection{The main idea and GHZ states}
\label{sectionghzstabwit}

\begin{figure}
\begin{minipage}{0.47\textwidth}
\centerline{\epsfxsize=\textwidth
\epsffile{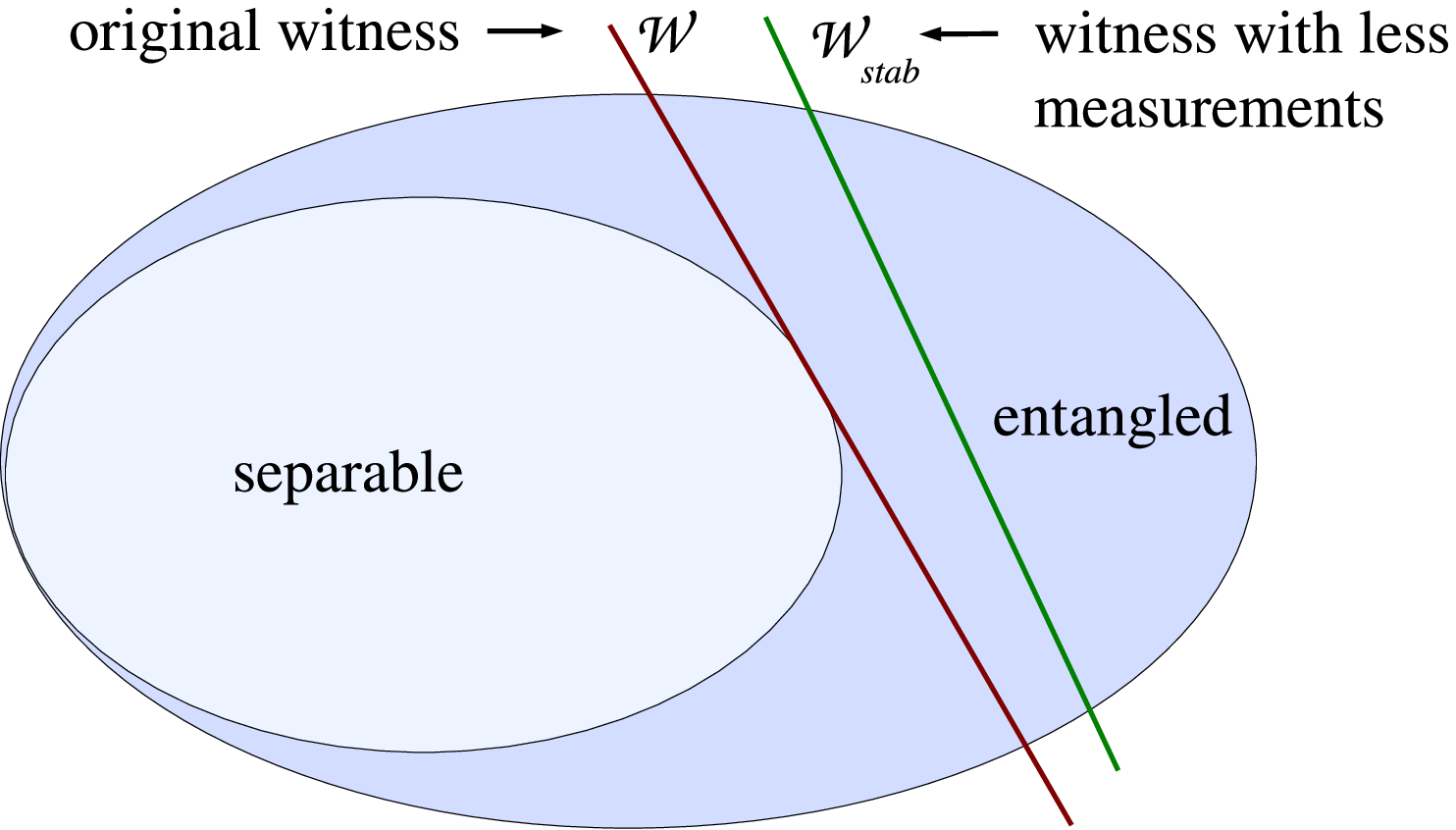} } \caption{Schematic view of
the construction of stabilizer witnesses: Instead of measuring the
original witness $\WW,$ (which might be difficult) one considers the
witness $\WW_{\rm stab}.$ This witness detects fewer states,
however, it is often much easier to measure than the original
witness. The witness $\WW_{\rm stab}$ can be constructed from the
stabilizing observables for the target state.}
\label{stabwit_fig_witnesses}
\end{minipage}
\begin{minipage}{0.06\textwidth}
\mbox{ }
\end{minipage}
\begin{minipage}{0.47\textwidth}
\centerline{\epsfxsize=\textwidth \epsffile{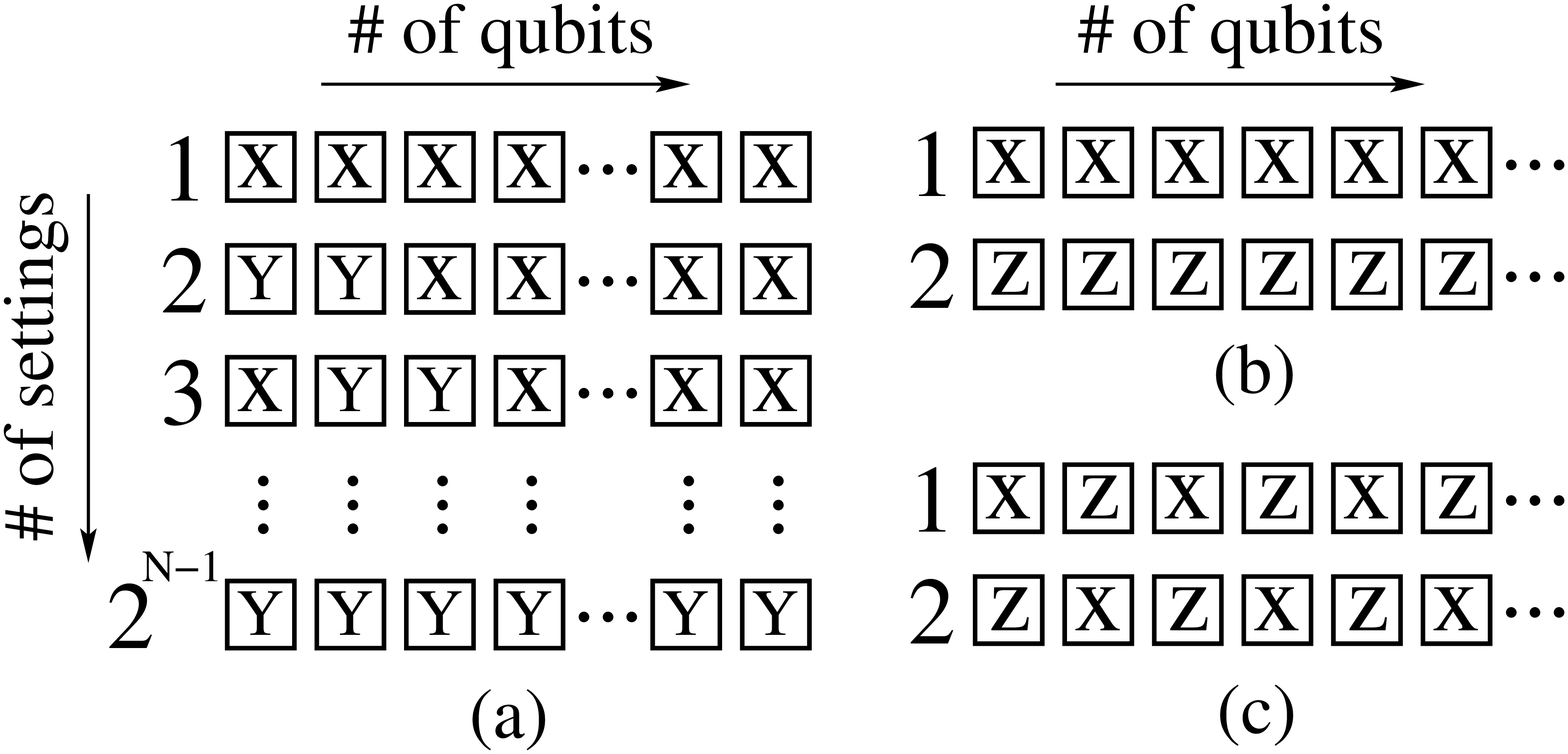} }
\caption{
  (a) Measurement settings needed for detecting genuine multi-qubit
  entanglement close to GHZ states with the Mermin inequality 
  [see Eqs.~(\ref{mermin}, \ref{Merminwit})]. For each qubit the 
  measured spin component is
  indicated.
  (b) Settings needed for the entanglement witnesses detecting entangled
  states close to GHZ states [Eq. (\ref{wghz})] and (c) cluster
states [Eq. (\ref{CN})].}
\label{stabwit_fig_settings}
\end{minipage}
\end{figure}

The main idea of the stabilizer witnesses is the following: If an
existing witness $\WW$ requires many measurement settings, one 
can look for witnesses that might be less powerful, but easier 
to measure. The relationship between the old witness that is 
difficult to measure and the new one is demonstrated in 
Fig.~\ref{stabwit_fig_witnesses}.

How can such a new witness be constructed? Let us first consider a
general observable $\widetilde{\WW}.$ If one can find an $\alpha >
0$ such that
\begin{equation}
\widetilde{\WW}-\alpha \WW \ge 0,
\label{finerequation}
\end{equation}
then  $Tr(\vr \widetilde{\WW}) < 0$ implies $Tr(\vr \WW) <
0.$\footnote{The condition in Eq.~(\ref{finerequation}) is indeed
equivalent to the ``finer'' relation in the sense of Section
\ref{sectionoptimization}, i.e., $\WW$ is finer as $\widetilde{\WW}$
iff a relation as in Eq.~(\ref{finerequation}) holds.} Therefore any
state  $\widetilde{\WW}$ detects is also detected by $\WW.$
Moreover, if we know that $\WW$ detects states with genuine
multipartite entanglement, so does $\widetilde{\WW}.$ The 
question remains, how $\widetilde{\WW}$ can be found. It 
turns out that stabilizer theory is a well suited tool for 
this task, but it should be noted that it works also for 
non-stabilizer states.

Let us shortly recall some facts about it, see also
Section~\ref{Sec_GraphStatesClusterStates}. Stabilizing operators
are locally measurable operators that have the state $\ket{\Psi}$ as
an eigenstate with eigenvalue $+1,$
\begin{equation}
\ket{\Psi}=S_k \ket{\Psi}. \label{stabS}
\end{equation}
It is common  to express Eq.~(\ref{stabS}) saying that operator $S_k$
stabilizes the state $\ket{\Psi}.$ If operators $S_k$ and $S_l$
stabilize $\ket{\Psi}$, then so does their product $S_kS_l.$ It is
easy to see that these stabilizing operators form a commutative {group.}
This group was named stabilizer in Ref.~\cite{PhysRevA.54.1862}.
If $\ket{\Psi}$ is an $N-$qubit quantum state then the group of stabilizing
operators has $2^N$ elements. The group has $N$ generators $g_k$ from
which all elements are possible products of  the $S_k$. Finally, the projector to
$\ket{\Psi}$ can be written as
\begin{equation}
\ketbra{\Psi}=\frac{1}{2^N}\sum_{k=1}^{2^N} S_k=\prod_{k=1}^N
\frac{1+g_k}{2}.
\label{Psi_stab}
\end{equation}
Let us show a simple example and consider the
witness operator Eq.~(\ref{WCN}) detecting genuine multipartite
entanglement around a GHZ state. The elements of the stabilizer
for the GHZ state are products of the generating operators
\begin{eqnarray}
g_1^{(GHZ_N)}:= \prod_{k=1}^N X_k,
&\;\;\;\;\;&
g_k^{(GHZ_N)}:=Z_{k-1} Z_k \mbox{ for } k=2,3,...,N.
\label{eigenGHZ}
\end{eqnarray}
Then, one considers the following observable:
\begin{equation}
\widetilde{\WW}_{GHZ_N} :=
3\openone-2\bigg[\frac{g_1^{(GHZ_N)}+\openone}{2}
+\prod_{k=2}^N\frac{g_k^{(GHZ_N)}+\openone}{2}\bigg]. \label{wghz}
\end{equation}
Knowing that $\WW_{GHZ_N} = 1/2 \cdot \eins - \ketbra{GHZ_N}$ is a
valid witness, one can verify Eq.~(\ref{finerequation}) for
$\alpha=2:$ The left hand side of Eq.~(\ref{finerequation}) is then
diagonal in the GHZ-basis, and it suffices to prove that the
diagonal elements are positive, which can be directly checked
\cite{PhysRevLett.94.060501,PhysRevA.72.022340}. Therefore,
$\widetilde{\WW}_{GHZ_N}$ is a valid witness for the detection of
genuine multipartite entanglement around the GHZ state.  For the
special case of three qubits, it reads $\widetilde{\WW}_{GHZ_3} =
\tfrac{3}{2} \eins - X_1 X_2 X_3 - \tfrac{1}{2}(Z_1 Z_2 \eins_3 + Z_1 \eins_2 Z_3 + \eins_1
Z_2 Z_3).$

The first remarkable thing is that $\widetilde{\WW}_{GHZ_N}$
requires only two measurement settings: The first term in the square
bracket can be measured with a $\{X,X,X,... ,X\}$ setting, the
second with a $\{Z,Z,Z,... ,Z\}$ setting. This is shown in
Fig.~\ref{stabwit_fig_settings}(b). Second, the witness
Eq.~(\ref{wghz}) is quite robust against noise. It detects the GHZ
state as entangled if the fraction of the white noise fulfills
\begin{equation}
p_{\rm noise}<\frac{1}{3-2^{2-N}}.\label{pnoise_ghz}
\end{equation}
Note that for large $N$ the right hand side of
Eq.~(\ref{pnoise_ghz}) converges to $1/3.$\footnote{It can be shown,
that the witness in Eq.~(\ref{wghz}) is optimal within a certain
class of witnesses with two measurements \cite{PhysRevA.72.022340}.}
Therefore, GHZ states with a fixed amount of noise can always be
detected with two measurement settings, {\it independently} from the
number of qubits. This shows that entanglement detection does not
become necessarily more complicated when increasing the number of
qubits.

If more than two settings can be measured then criteria with a
higher noise tolerance can be obtained. At this point it is
interesting to show a connection between stabilizer witnesses and
the Mermin inequality Eq.~(\ref{mermin}). Let us consider the
following witness
\begin{eqnarray}
\WW_{\rm Mermin}&:=&2^{N-2}\openone-\sum_\pi X_{1}X_2X_3X_4\cdots X_N
+\sum_\pi Y_{1}Y_2X_3X_4\cdot\cdot\cdot X_N
-
\sum_\pi Y_{1}Y_2Y_3Y_4X_5\cdot\cdot\cdot X_N+...-....,\nonumber\\
\label{Merminwit}
\end{eqnarray}
where $\sum_\pi$ indicates that every term represents the sum of all
of its permutations. This witness is just constructed from the Bell
operator of the Mermin inequality given in Eq.~(\ref{mermin}). We
used the bound for genuine multipartite entanglement for this
inequality presented in Refs.~\cite{Nagata,PhysRevA.72.022340}.
Thus, $\WW_{\rm Mermin}$ detects genuine multipartite entanglement.
Note that the witness is also a stabilizer witness since all the
terms in Eq.~(\ref{Merminwit}) are stabilizing operators of the GHZ
state. This witness detects the noisy GHZ state as entangled if
$p_{\rm noise} < {1}/{2}.$

\subsubsection{Linear cluster states and other graph states}

Similar methods can be applied to create witnesses with few
measurements for cluster states and in general, graph states,
introduced in Sec.~\ref{Sec_GraphStatesClusterStates}. The
generators for the group of stabilizing operators of the cluster
state are \be g_1^{(C_N)}:=X_{1} Z_{2}, \;\;\; g_k^{(C_N)}:=Z_{k-1}
X_k Z_{k+1} \mbox{ for } k=2,3,...,N-1, \;\;\; g_N^{(C_N)}:=Z_{N-1}
X_N. \label{eigenC} \ee For detecting genuine multipartite
entanglement we have already presented a witness Eq.~(\ref{wghz}).
Based on arguments similar to creating witnesses for the GHZ state,
one can construct a witness that also detects genuine multipartite
entanglement, however, it is easier to measure
\begin{eqnarray} \widetilde{\WW}_{C_N} &:=& 3\eins- 2 \bigg[\prod_{\text{even
k}}\frac{g_k^{(C_N)}+\eins}{2}+ \prod_{\text{odd
k}}\frac{g_k^{(C_N)}+\eins}{2}\bigg]. \label{CN}
\end{eqnarray}
It detects the cluster state as entangled if
\begin{eqnarray} p_{\rm
noise}<\bigg\{
\begin{array}{ll}
(4-4/2^{\frac{N}{2}})^{-1} & \mbox{ for even }N, \\
{[}4-2(1/2^{\frac{N+1}{2}}+1/2^{\frac{N-1}{2}})]^{-1} & \mbox{ for
odd } N.
\end{array}
\end{eqnarray}
For large $N$ the limit is $p_{\rm noise}=\tfrac{1}{4}.$
The measurement settings needed are shown in
Fig.~\ref{stabwit_fig_settings}(c).

The ideas presented for GHZ and cluster states can be
straightforwardly generalized for graph states. This requires the
notion of the colorability of a graph. A graph is called
$k$-colorable, if there exist $k$ disjoint subsets $M_1,...,M_k$ of
the  set of vertices, such that for any $j$ there are no edges
between any pair of vertices in $M_j.$ For instance, the star graph
(see Fig. \ref{grafig} in Section
\ref{Sec_GraphStatesClusterStates}) and the cluster graph are
two-colorable, while a three-vertex ring graph is three-colorable.

Using this notation, for a $k$-colorable graph state $\ket{G_N}$
the following witness detects genuine multi-partite entanglement.
\begin{eqnarray}
\widetilde{\WW}_{G_N} &:=&
3\eins- 2
\bigg[
\sum_{j=1}^{k}
\Big(
\prod_{{i \in M_j}}\frac{g_i^{(G_N)}+\eins}{2}
\Big)
\bigg],
\end{eqnarray}
This witness requires the measurement of $k$ settings, since each of
the terms in the sum can be measured with a single setting, namely
with a setting
measuring $\{X_i\}_{i \in M_j}$ and $\{Z_i\}_{i \not\in M_j}.$
The
robustness to noise depends on $k.$ It should be noted, however,
that the notion of $k$-colorability is a property of the graph (and
not the state) and if a graph state can be represented (up to local
unitaries) by different graphs, the colorability properties of these
graphs may be different \cite{hein-2006}.

Such witnesses have been used in many experiments. That is, the
two-setting witness for cluster states Eq.~(\ref{CN}) was used first
in Ref.~\cite{PhysRevLett.95.210502}, later in Refs.~\cite{
TwoPhotonClusterState,chen-2007-99}. In another experiment, six
photon graph states were created in a photonic experiment and
genuine multipartite entanglement was detected with a witness using
six measurement settings \cite{lu-2007-3}. The advantage of using
six, rather than only two settings is the larger robustness to noise
of the witness. We will discuss in Section \ref{clusterexp} an
experiment for the generation of the four-qubit cluster state in
detail.

\subsubsection{Non-stabilizer states}
Similar ideas can also be used for the detection close to states that
are not stabilizer states. For such states it is not possible to find
$2^N$ operators such that they stabilize the state and also they are
the tensor-products of single-qubit operators. An example for such
a state is the three-qubit W-state $\ket{W_3}.$ If nonlocal operators
are also considered, however, the following operators, which are
the sum of several operator products, stabilize this state:
\begin{eqnarray}
g_1^{(W_3)}&:=&\frac{1}{3}\big(Z_{1}+2Y_{1}Y_2Z_3
+2X_{1}Z_{2}X_3\big),
\;\;\;\;
g_2^{(W_3)}:=\frac{1}{3}\big(Z_{2}+2Z_{1}Y_2Y_3+ 2X_{1}X_2Z_3
\big),\nonumber\\
g_3^{(W_3)}&:=&\frac{1}{3}\big(Z_3+2Y_{1}Z_{2}Y_3
+2Z_{1}X_2X_3\big).
\end{eqnarray}
There is only a single state, the W-state, that gives $+1$ for all
the three operators, as in the case of stabilizer states. These
operators commute with each other. Multiplying these operators gives
further operators that stabilize the W state thus these operators
form an eight-element group. The $g_k^{(W_3)}$ operators are the
generators of the group. However, now the choice of these eight
stabilizing operators is not unique. A simple method for finding
such operators is given in Ref.~\cite{PhysRevA.72.022340}. Using
these operators, and the strategy from above, one can find now
the witness \cite{PhysRevA.72.022340}
\be
\widetilde{\WW}_W = \frac{11}{3}\eins + 2 Z_1 Z_2 Z_3 -
\frac{1}{3}\sum_{k \neq l} (2 X_k X_l + 2 Y_k Y_l - Z_k Z_l),
\ee
which requires only three measurement settings. For another application
of similar ideas see Ref.~\cite{li-2007-76}.

\subsubsection{Fidelity estimation}
Detecting genuine multipartite entanglement is closely related to
measuring how close the experimental state is to a given
multipartite quantum state. This can be seen noting that the
projector-based entanglement witnesses $\WW_{GHZ_N}$ and $\WW_{C_N}$
of Eq.~(\ref{WCN}) were simply based on measuring the projector,
that is, the fidelity $F=\bra{\psi}\vr_{\rm exp} \ket{\psi}$ with
respect to a highly entangled quantum state $\ket{\psi}.$ Thus,
methods similar to the ones for detecting genuine multipartite
entanglement with few measurements also work for estimating the
fidelity with few measurements \cite{PhysRevA.72.022340,
TokunagaFidelityEstimation, SommaLowerBoundsForFidelity}. In
particular, one can consider the relation
\begin{equation}
\ketbra{GHZ_N} \ge  \bigg[\frac{g_1^{(GHZ_N)}+\openone}{2}
+\prod_{k=2}^N\frac{g_k^{(GHZ_N)}+\openone}{2}\bigg]-\openone.
\label{Pghz}
\end{equation}
which is essentially equivalent to Eq.~(\ref{finerequation}) for
the stabilizer witness for the GHZ state. This relation implies
that measurement of the right hand side of Eq.~(\ref{Pghz}) can be
used to give a lower bound on the fidelity of the GHZ state. Equivalently,
the stabilizer witness allows to give a lower bound on the fidelity via
$F_{GHZ_N} \geq (1 - \mean{\widetilde{\WW}_{GHZ_N}})/2.$

There are other  ways to estimate the fidelity, similarly to
Eq.~(\ref{Pghz}) \cite{TokunagaFidelityEstimation, ToolBox}. To see
the main idea, let us consider the example of a four qubit cluster
state $\ket{CL_4} =
(\ket{0000}+\ket{1100}+\ket{0011}-\ket{1111})/2$, for which the
usual fidelity-based witness is given by $\WW = \eins/2 -
\ketbra{CL_4}.$ Then, the observable \bea \tilde{\WW} &=&
\frac{1}{2} \eins - \ketbra{CL_4} + \ketbra{CL'_4} \nonumber
\\
&=& \frac{1}{2}
[\eins -
\ketbra{00} \otimes (\ket{00}\bra{11}+\ket{11}\bra{00})
+
(\ket{00}\bra{11}+\ket{11}\bra{00})\otimes  \ketbra{11})
]
\eea
with $\ket{CL'_4} = (-\ket{0000}+\ket{1100}+\ket{0011}+\ket{1111})/2$
is, due to  Eq.~(\ref{finerequation}), also an entanglement witness
and allows to estimate the fidelity via $F \geq 1/2 - \mean{\tilde{\WW}}.$
As the off-diagonal terms of $\tilde{\WW}$ in the computational basis
are of a simple form, $\tilde{\WW}$  can be measured with four measurements
only \cite{TokunagaFidelityEstimation}, which is significantly less than
the nine measurements for $\WW.$ Similar ideas can be applied to other states
\cite{ToolBox} and have been used in several six-photon experiments
\cite{lu-2007-3,lu-2007-anyon}.

\subsubsection{Witnesses for full separability}

Up to now, we considered stabilizer witnesses that detect only
genuine multi-qubit entanglement. Typically, the noise in an
experiment must be relatively low if we want to detect this form of
entanglement. On the other hand, if we want to detect entanglement
in general, including partial or biseparable entanglement, then
typically in an experiment much higher noise is allowed.

The basic idea behind detecting entanglement with stabilizing
operators is that a product state cannot give $+1$ for two
stabilizing operators that do not commute locally \cite{PhysRevA.72.022340}.
Based on this idea, and after simple algebra, a witness detecting
entanglement around an $N$-qubit GHZ state can be obtained as
\begin{equation}
\WGHZN_m := \eins-g_1^{(GHZ_N)}- g_m^{(GHZ_N)},\label{fsepGHZ}
\end{equation}
where $m=2,3,...,N$. The proof is based on the Cauchy-Schwarz
inequality. Using this and $\mean{X_i}^2+\mean{Z_i}^2 \leq 1$, 
we obtain for pure product states
\begin{align}
& \!\!\!\!\!\!\exs{g_1^{(GHZ_N)}} + \exs{g_m^{(GHZ_N)}}
= \exs{X_{1}}\exs{X_2}...\exs{X_N} + \exs{Z_{m-1}}\exs{Z_m}
\nonumber\\
&\le |\exs{X_{m-1}}|\cdot|\exs{X_m}|+ |\exs{Z_{m-1}}|
\cdot|\exs{Z_m}|
\le
\sqrt{\exs{X_{m-1}}^2+\exs{Z_{m-1}}^2 }
\sqrt{\exs{X_m}^2+\exs{Z_m}^2 }
\le 1. \label{cschwarz}
\end{align}
By convexity, the bound is also valid for mixed separable
states. This proof can straightforwardly be generalized for
arbitrary two locally non-commuting elements of the stabilizer of
any graph state. Witnesses can be constructed with more than two
elements of the stabilizer as
\begin{equation}
\WGHZNPRIME_m:=\eins - g_1^{(GHZ_N)}- g_m^{(GHZ_N)} -
g_1^{(GHZ_N)}g_{m}^{(GHZ_N)} \label{fsepGHZprime}
\end{equation}
for $m =2,3,...,N$ that rule out full separability. The proof is
similar to the proof of Eq.~(\ref{fsepGHZ}). These witnesses
tolerate noise if $p_{\rm noise}<\tfrac{1}{2}$ and $\tfrac{2}{3},$
respectively. As can be seen, more terms make it possible to have
higher noise tolerance. One can even construct a witness from the
Bell operator of the Mermin inequality
\begin{eqnarray}
\WW_{\rm Mermin}'&:=&2^{\frac{N-2}{2}}\openone-\sum_\pi
X_{1}X_2X_3X_4\cdot\cdot\cdot X_N
+\sum_\pi Y_{1}Y_2X_3X_4\cdot\cdot\cdot X_N
-
\sum_\pi Y_{1}Y_2Y_3Y_4\cdot\cdot\cdot X_N +...-....\nonumber\\
\label{Merminwit2}
\end{eqnarray}
This differs from Eq.~(\ref{Merminwit}) in having a smaller constant
term corresponding to the relaxed conditions for biseparable
entanglement compared to genuine multipartite entanglement. The
constant comes from the calculations presented in
Refs.~\cite{PhysRevLett.94.010402,seevinck-2007,badziag-2007}.
While the number of terms
increases with the number of qubits for $\WW_{\rm Mermin}',$ the
noise tolerance is also increasing.

For the cluster state similar witnesses are
\cite{PhysRevA.69.052327,PhysRevA.72.022340} $ \WCN_k:=
\eins-g_k^{(C_N)}-g_{k+1}^{(C_N)} \mbox{ for } k =1,2,...,N-1, $
and, with  a better noise tolerance $ \WCNPRIME_k:=
\eins-g_k^{(C_N)}-g_{k+1}^{(C_N)}- g_k^{(C_N)}g_{k+1}^{(C_N)} \mbox{
for } k=1,2,...,N-1. $ Both witnesses involve only the qubits of a
quadruplet and tolerate noise if $p_{\rm noise}<\tfrac{1}{2}$ and
$\tfrac{2}{3},$ respectively. It is also possible to obtain a series 
of witnesses for which the noise tolerance approaches $1$ for
increasing $N$ \cite{PhysRevA.72.022340}.

\subsubsection{Mixed stabilizer states}

So far, entanglement conditions were presented that detect
entanglement in the vicinity of pure multipartite states. It is also
possible to use witnesses made out of stabilizing operators for
detecting entanglement in the vicinity of mixed stabilizer states
\cite{PhysRevA.72.022340}. The general idea is considering witnesses
of the from
\begin{eqnarray}
\WW_{\rm mixed}:=const.-\sum_{k=1}^{M} g_k, \label{WWmixed}
\end{eqnarray}
where $g_k$ are {\it some} of the generators of the stabilizer for
some graph state. If $M=N$ then there is only a single quantum state
giving the minimum for $\exs{\WW_{\rm mixed}}.$ However, if $M<N$
then the minimum is degenerate and the states giving the minimum
make up a subspace of dimension $(N-M+1).$ Then, if one can
determine the constant in Eq.~(\ref{WWmixed}) such that
$\exs{\WW_{\rm mixed}}$ is positive on product states, yet it is
still negative for some state, then $\WW_{\rm mixed}$ is a valid
witness.

As an example, let us consider the biseparable (but not fully
separable) mixed quantum state \be
\vr_3:=\frac{1}{2}\big(\ketbra{\xi^+}+\ketbra{\xi^-}\big), \;\;\;\;
\ket{\xi^+}:=\frac{1}{\sqrt{2}}(\ket{00}+\ket{11})\otimes\ket{0},
\;\;\;\;
\ket{\xi^-}:=\frac{1}{\sqrt{2}}(\ket{00}-\ket{11})\otimes\ket{1},
\label{rho3} \ee which occurs if the fourth qubit of a cluster state
is traced out. The states $\ket{\xi^\pm}$ are stabilizer states. Any
mixture of states $\ket{\xi^\pm}$ is stabilized by the common
elements of $\mathcal{S}^{(\xi^\pm)}$. The generators of the
stabilizer of these mixed states is given by
$\mathcal{S}^{(\vr_3)}=\{ Z_{1}Z_{2}, X_{1}X_2Z_3\}$. Based on that,
the following witness can be constructed
\begin{eqnarray}
\WW_{\vr_3}:=\eins- Z_{1}Z_{2}- X_{1}X_2X_3. \label{Wrho3}
\end{eqnarray}
The constant term in Eq.~(\ref{Wrho3}) can be obtained via methods
similar to the ones used for obtaining a bound for product states
for the sum of two stabilizing operators used for
Eqs.~(\ref{fsepGHZ}).

\subsection{Experimental generation of cluster states}
\label{clusterexp}

Now we describe an experiment from K.~Chen {\it et
al.}~\cite{chen-2007-99}, where a four-qubit cluster state has been
observed and a stabilizer witness has been used.\footnote{At the
same time, a similar experiment has been performed by Vallone {\it
et al.}~\cite{TwoPhotonClusterState}.} This experiment used
hyper-entanglement of photons for the creation of the cluster state
\cite{Kwiathyper}. This means that several degrees of freedom of
each photon are used to carry the entanglement. In the given
experiment, each photon represented two qubits, one polarization
qubit and one spatial qubit. This technique is becoming more and
more important in multi-photon experiments as it allows to extend
the Hilbert space significantly \cite{barreiro-2005-95, kwiatnatphys,
vallonenewhyper} and also complete Bell state analysis becomes 
possible \cite{schuck:190501}. Quite recently, this technique has 
been used to realize a ten-qubit GHZ state with five photons 
\cite{gao-2008}.

\begin{figure}[t]
\centerline{\includegraphics[width=0.5\columnwidth]{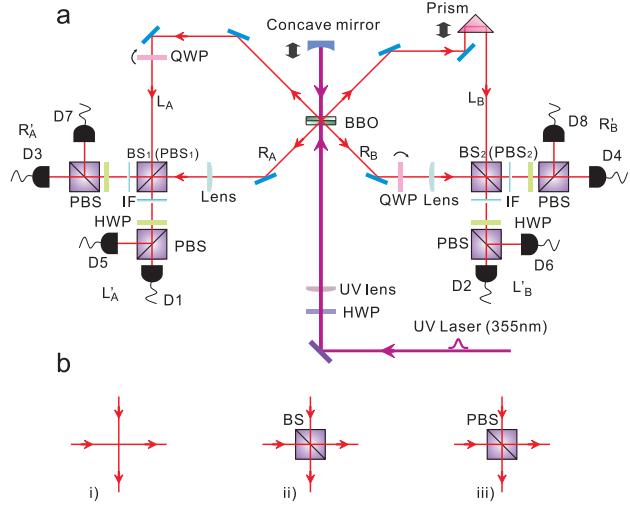}}
\caption{(a) Setup for the generation of four-qubit cluster states using two
photons and hyper-entanglement. (b) Required setups for the measurement of the
polarization qubit and the spatial qubit of one photon. See the text for further
details. Figure is taken from Ref.~\cite{chen-2007-99}.
\label{hypersetup}
}
\end{figure}

The setup of the experiment is shown in Fig.~\ref{hypersetup}. A UV
laser pulse passes two times through two combined BBO crystals. This
leads either to emission of $\ket{H_A H_B} + \ket{V_A V_B}$ in the
upward direction (modes $L_A, L_B$) or to emission of $\ket{H_A H_B}
- \ket{V_A V_B}$ in the downwards direction, i.e., modes $R_A, R_B$.
Here, $H$ and $V$ denote horizontal and vertical polarization. If
the photons overlap perfectly at the beamsplitters $BS_1$ and
$BS_2,$ the state will be a superposition \be \ket{\psi} =
\frac{1}{2} \big[ (\ket{H_A H_B} + \ket{V_A V_B}) \ket{L_A L_B} +
e^{i \phi} (\ket{H_A H_B} - \ket{V_A V_B}) \ket{R_A R_B} \big]. \ee
By adjusting the concave mirror, $\phi = 0$ can be obtained. Then
the state is equivalent to  the cluster state
$\ket{\psi}=(\ket{0000}+\ket{0011}+\ket{1100} - \ket{1111}),$ if the
replacements $\ket{H/V}\mapsto \ket{0/1}$ and $\ket{L/R}\mapsto
\ket{0/1}$ are done. Since only two photons are required, the count
rate in the present experiment is around $1.2 \cdot 10^4$ per
second, which is about four orders of magnitude higher than in
experiments that produced the same cluster state using four photons
\cite{PhysRevLett.95.210502,walther-2005-434}.

In order to measure the state, the elements as in
Fig.~\ref{hypersetup}(b) have been used. A measurement without
any beamsplitter as in (i) first reads out whether the photon
is in $L/R$ and also the polarization of the photon. Hence
it can be used to measure $Z_{\rm spat}X_{\rm pol}$ at one
photon. Using a beamsplitter as in (ii) leads to a Hadamard
transformation on the spatial modes, such that
$X_{\rm spat} Z_{\rm pol}$ can be measured.
The problem is that such a setup is sensitive to small path length
changes, hence it requires a significant amount of adjustment and is
only stable for relatively short times.\footnote{Only recently,
using a Sagnac-like interferometer, stable interferometers for this
problem have been introduced \cite{gao-2008}.} Therefore, despite of
the high counting rate, an analysis of the state should require only
few measurements.

One can use the witness
\be
\mathcal{W}= 2\eins - \frac{1}{2}
(
X_1 X_2 \eins_3 Z_4+
X_1 X_1 Z_3 \eins_4 +
\eins_1 \eins_2 Z_3 Z_4+
\eins_1 Z_2 X_3 X_4 +
Z_1 \eins_2 X_3 X_4 +
Z_1 Z_2 \eins_3 \eins_4
)
\ee
which is just the witness from Eq.~(\ref{CN}) adapted to the form of the
cluster state used here. This witness can be measured with the settings
$X_1 X_2 Z_3 Z_4$ and $Z_1 Z_2 X_3 X_4$.
This corresponds to the setups in  Fig.~\ref{hypersetup}b (i)
and (ii). The experimental data give
\be
\mean{W} = -0.766\pm 0.004 \mbox{ and } F \geq  \frac{1-\mean{W}}{2} =0.883 \pm 0.002,
\ee
clearly confirming the four-qubit entanglement. Furthermore, this 
cluster state has then been used to demonstrate two-qubit gates.

The nature of multipartite entanglement in experiments with 
hyper-entangled states deserves some more comments. First, 
it should be stressed that due to the hyper-entanglement 
not all qubits are spatially separated, hence the entanglement 
refers to different degrees of freedom, and not different 
particles. Second, in multiphoton experiments as in 
Ref.~\cite{gao-2008} the process of postselection has a 
different effect now. Ideally, in any experiment, one aims to 
generate some entangled state $\vr$ which can then be distributed 
and used it as a resource for some task to be determined later. 
In usual postselection experiments, a state $\vr_{\rm ini}$ is prepared, 
which contains different numbers of photons. Postselection results in a 
``state'' $\vr_{\rm ps},$ which, however, is destroyed and cannot be 
stored. However, as usual postselection is a local projection, entanglement
of $\vr_{\rm ps}$ implies entanglement of $\vr_{\rm ini},$ which may 
therefore be considered as the (in principle) distributable and 
storable resource \cite{vanenk-2007-75}. 
When hyper-entanglement is used, the postselection acts 
collectively on two qubits at the same time, and there is no direct 
connection between the multi-qubit entanglement of $\vr_{\rm ps}$
and the one of $\vr_{\rm ini}.$ For many applications, however, 
this does not matter.

Finally, it is interesting that if one wants to detect entanglement 
only between the different degrees of freedom  of hyper-entangled 
particles, stabilizer witnesses can again be used \cite{vallone-2008}.


\subsection{Eight-qubit W states in ion traps}
\label{sectionionexp}

In this Section we explain an experiment in which eight qubits have
been entangled using trapped ions \cite{haeffner-2005-438}. Apart
from being the most advanced experiment to generate multi-particle
entanglement in ion traps, it has also some interesting  aspects
from a theoretical point of view. First, a witness different from
the projector-based one was used, second, the usage of local filters
enabled the detection of entanglement even in a very noisy
environment.

\subsubsection{Description of the experiment}

Let us first shortly describe the setup. Detailed overviews on
quantum information processing with ion traps and their physical
principles can be found in
Refs.~\cite{RevModPhys.75.281,blattnature,haeffner-2008,
eschnervarenna} and for a more detailed description of the present
experiment see Refs.~\cite{haeffner-2008,korbicz:052319}.

\begin{figure}[t]
\begin{minipage}{0.48\textwidth}
\centerline{\psfig{figure=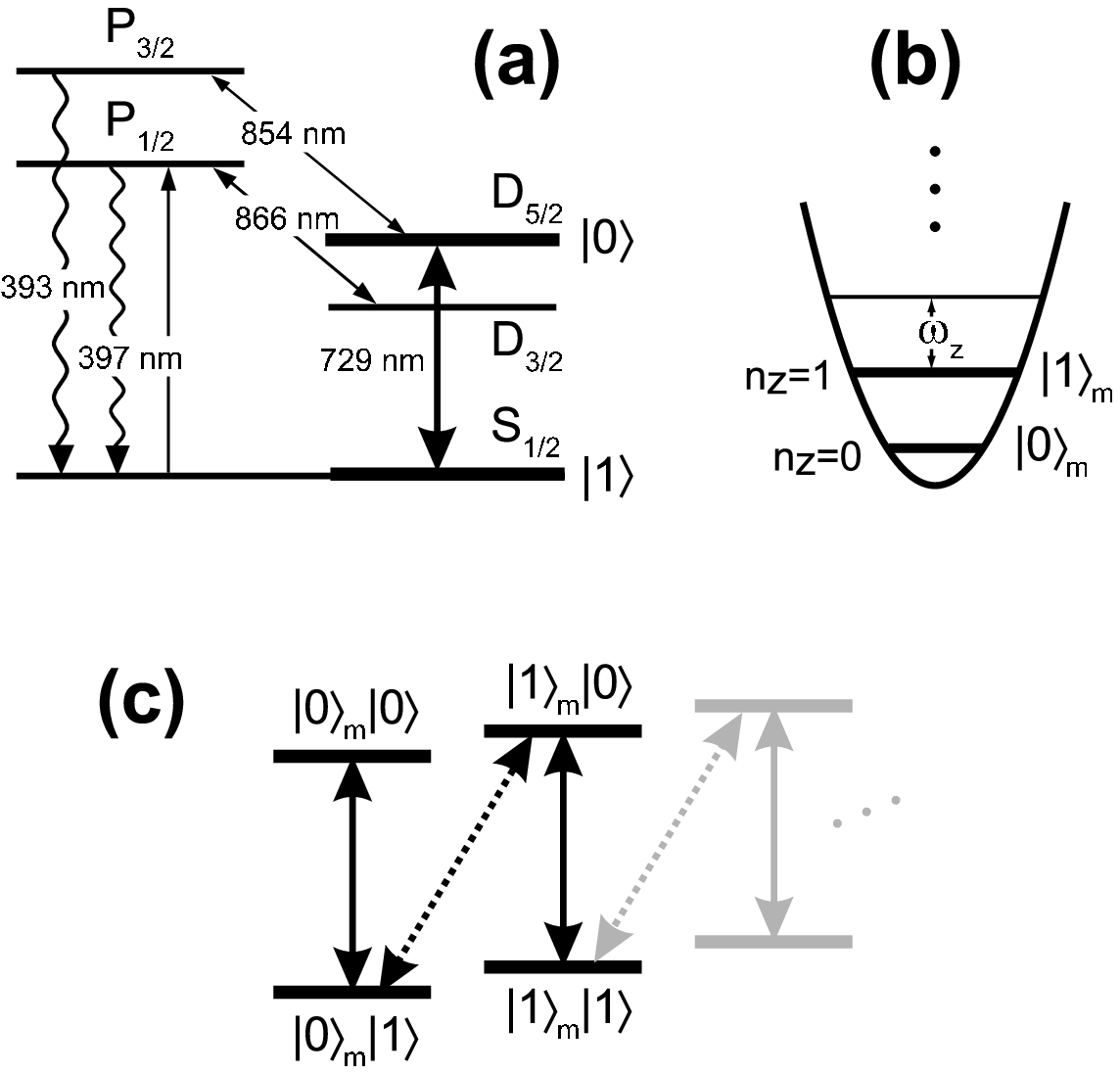,width=0.8\columnwidth}}
\caption{ \label{levelscheme} (a) Level scheme of the $^{40}$Ca$^+$
ion. The D$_{5/2}$ and the S$_{1/2}$ levels are used as the basis
states of the qubit. (b) Schematic view of the motional levels
describing the bus mode. (c) Combined level scheme of the motional
qubit and the internal qubit. The figure is taken from
Ref.~\cite{korbicz:052319}. }
\end{minipage}
\begin{minipage}{0.04\textwidth}
\mbox{ }
\end{minipage}
\begin{minipage}{0.48\textwidth}
\centerline{\psfig{figure=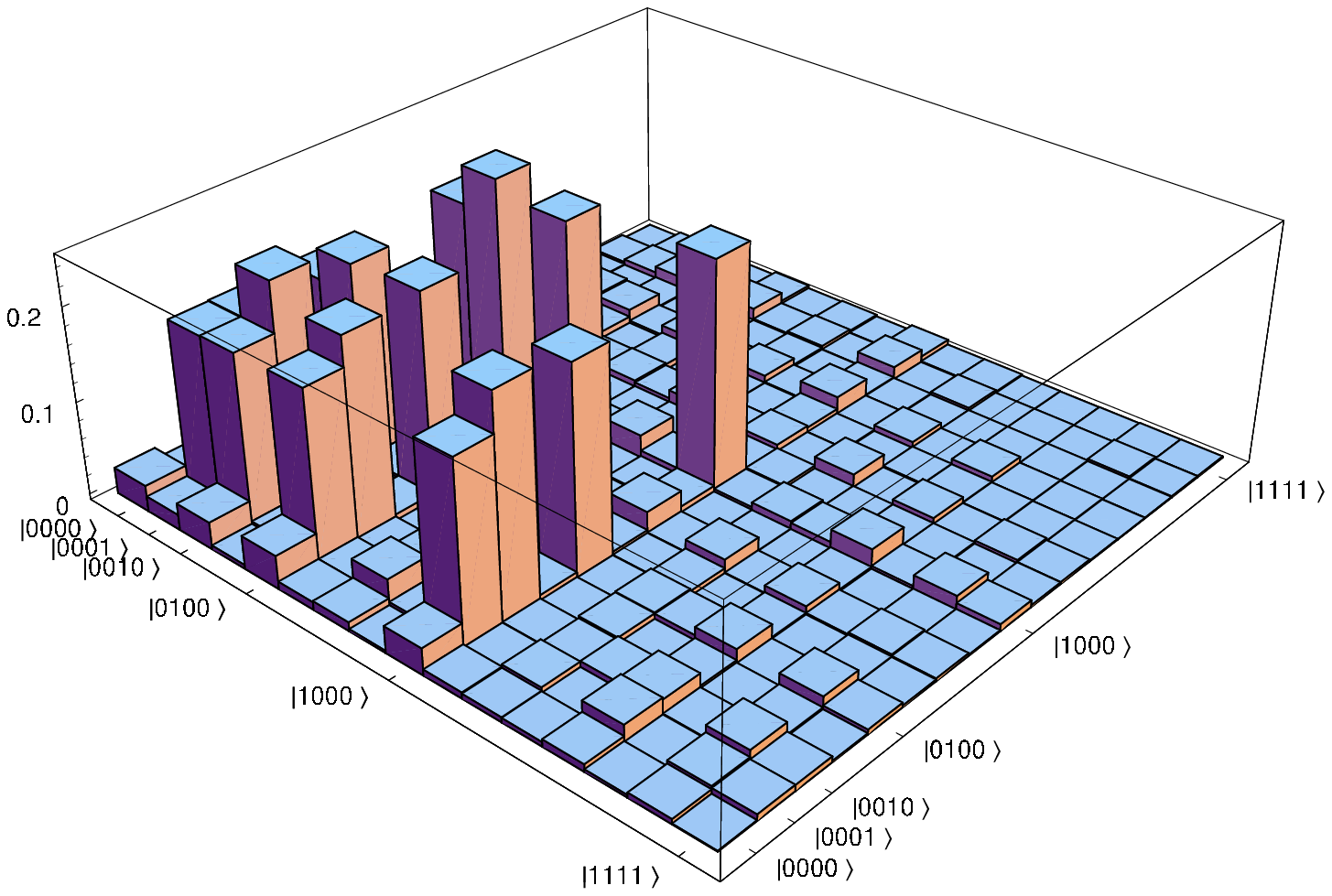,width=0.98\columnwidth}}
\caption{Absolute values of the reconstructed density matrix
for the four-qubit W state.
\label{tomobild1}
}
\end{minipage}

\end{figure}

In the experiment strings of up to eight $^{40}$Ca$^+$ ions
were trapped in a linear Paul trap. The D$_{5/2}$ ($=\ket{0}$)
and the S$_{1/2}$ ($= \ket{1}$) levels  of the ion are used as
the basis states of the internal qubit [see Fig.~\ref{levelscheme}(a)].
The ions can be individually addressed via laser pulses on the
$\ket{0}  \leftrightarrow \ket{1}$ transition.

For the interaction between the qubits, the vibrational excitations
of the ion string are used. These excitations arise from a harmonic
oscillator potential [see Fig.~\ref{levelscheme} (b)], and the two
lowest levels $\ket{0}_m$ and $\ket{1}_m$ are referred to as the
motional qubit. Via driving the $\ket{0}_m\ket{1} \leftrightarrow
\ket{1}_m \ket{0}$ transition (see Fig.~\ref{levelscheme} (c)), each
ion can individually interact with the motional qubit, allowing
finally two-qubit gates \cite{PhysRevLett.74.4091,
sorensen-1999-82,molmer-1999-82,
ciraczollernature404,mintert-2001-87}.

The idea to prepare the $N$-qubit W states
\be
\ket{W_N} = \frac{1}{\sqrt{N}} (\ket{00...01} + \ket{00...10} +
... + \ket{10... 00})
\ee
can be sketched as follows. First, the state $\ket{0}_m \ket{10...0}$
is created, where all but the first internal qubit are in the $\ket{0}$
state. Then, the motional qubit is partly excited and distributed over
all the ions. That is, the state is changed according to the sequence
\begin{align}
\ket{0}_m \ket{10...0} &\mapsto
\frac{1}{\sqrt{N}}\ket{0}_m \ket{100...0} + \frac{N-1}{\sqrt{N}}\ket{1}_m \ket{000...0}
\nonumber
\\
& \mapsto \frac{1}{\sqrt{N}}\ket{0}_m \ket{100...0} +
\frac{1}{\sqrt{N}}\ket{0}_m \ket{010...0}
+ \frac{N-2}{\sqrt{N}}\ket{1}_m \ket{000...0}
\nonumber
\\
&\mapsto ...
\nonumber
\\
& \mapsto \frac{1}{\sqrt{N}}\ket{0}_m \ket{100...0}
+  \frac{1}{\sqrt{N}}\ket{0}_m \ket{010...0} + ... +
\frac{1}{\sqrt{N}}\ket{0}_m \ket{000...1}.
\end{align}
Details on the required pulse sequences are given in
Refs.~\cite{haeffner-2005-438,korbicz:052319}. This scheme allows to
generate W states in a time of 500 - 1000 $\mu$s, for comparison,
the life time of the D$_{5/2}$ level is $\tau=1.16$ s. It should be
noted that W states in ion traps can also be prepared in different
ways \cite{retzker-2007-75}.

In order to characterize the experimentally generated state
$\vr_{\rm exp},$ state tomography has been performed. For that,
all possible correlations of the type
$\sigma_i \otimes \sigma_j \otimes ... \otimes \sigma_k$
have been measured. Repeating each correlation measurement
at least a hundred times requires $100 \cdot 3^N$ repetitions
of the state generation, for $N=8$ qubits this amounts to a total
measurement time of 10 hours.

Due to the statistical nature of the measurement, the raw data from
the correlation measurements do not result directly in a density
matrix that is positive semidefinite. Therefore, using the results
of Ref.~\cite{hradil-iteration} a maximum-likelihood approximation
to the  data has been performed. In Ref.~\cite{hradil-iteration}, it
was shown that the maximum-likelihood estimation is a fixed point of
an iterative map, where the iteration depends on the measurement
data. This fixed point was numerically determined by iterating the map, starting
from a maximally mixed state until convergence was reached. Due to
the exponential growing size of the density matrix this iteration
requires  a considerable effort, for the eight qubit case it
requires more than one day. As a result, valid density matrices have
been obtained, for the case $N=4$ the absolute values of it are
plotted in Fig. \ref{tomobild1}.\footnote{The numerical values of
the density matrices are  given in the online material of
Ref.~\cite{haeffner-2005-438}.} From these matrices, the fidelity of
the W states can be calculated, see Table \ref{fidtable}. In
addition, given the estimated density matrix one can generate with a
Monte Carlo simulation further data sets, which were used to
calculate the error bars.

\begin{table}
\caption{Fidelities and mean values for the witnesses of
the $N$-qubit W states for different values of $N$.
\label{fidtable}}
\centering{
\begin{tabular}{|c||c|c|c|c|c|}
\hline
$\;\;N\;\;$ & 4 &5&6&7&8
\\
\hline
$F$ & $0.846 \pm 0.011$ & $0.759 \pm 0.007$ & $0.788 \pm 0.005$ &
$0.763 \pm 0.003$ & $0.722 \pm 0.001$
\\
\hline
$\mean{\WW}$ & $-0.460 \pm 0.031$ & $-0.202 \pm 0.027$ & $-0.271 \pm 0.031$ &
$-0.071 \pm 0.032$ & $-0.029 \pm 0.008$
\\
\hline
\end{tabular}
}
\end{table}

\subsubsection{Analysis of the states with witnesses and local filters}
\label{sectionionexpwit}

A first tool for the analysis of the states is the fidelity-based
witness  $\WW = \tfrac{N-1}{N} \eins - \ketbra{W_N}.$ However, as
can be seen from the fidelities in Table \ref{fidtable}, this
witness does not detect any entanglement for $N \geq 5.$ To improve
this witness, one starts with the following observation: The
projector onto the W state is a matrix that is acting on the
subspace with one excitation ($=\ket{1}$) only. Therefore, if one
calculates the overlap with a biseparable state
$\ket{\phi}=\ket{a}\ket{b}$ it suffices to  assume that $\ket{a}$
and $\ket{b}$ have maximally one excitation, as any contribution
with two ore more $\ket{1}$ in $\ket{a}$ or $\ket{b}$ has a
vanishing overlap with the state $\ket{W_N}.$ From this
consideration, one can directly see that \be \WW = \frac{N-1}{N}
\PP_{\leq 2} - \ketbra{W_N}, \label{firstimprovement} \ee where
$\PP_{\leq 2}$ is the projector onto the space with maximally two
excitations, is a valid witness.\footnote{For example, for $N=3$ we
have $\PP_{\leq 2}= \eins - \ketbra{111}$.}

For larger $N,$ the witness in Eq.~(\ref{firstimprovement}) is
already a significant improvement, as the difference between
$\PP_{\leq 2}$ and $\eins$ increases exponentially. However, one can
make the more general ansatz \be \WW = \alpha \PP_{0}+ \beta
\PP_{1}+ \gamma \PP_{2} - \ketbra{W_N}, \label{secondimprovement}
\ee where the $\PP_{i}$ are now projectors onto the subspaces with
exactly $i$ excitations. It remains to determine the allowed values
for $\alpha, \beta$ and $\gamma$ in order to guarantee that $\WW$ is
indeed non-negative on all biseparable states
$\ket{\phi}=\ket{a}\ket{b}.$ From the symmetry of the W state, one
can conclude that it suffices to prove non-negativity for $\ket{a}$
of the form $\ket{a} = a_0 \ket{00...00} + a_1 (\ket{00...01}+ ...
+ \ket{10...00})$ and similarly for $\ket{b}$ \cite{haeffner-2005-438}. 
Therefore, the
resulting conditions  on $\alpha, \beta$ and $\gamma$ are given by
simple optimization problems of a polynomial in four variables with
two normalization constraints, which can numerically simply be
solved for any $N.$

For given experimental data, one can choose then the tuples
$(\alpha,\beta,\gamma)$ that yield the most negative value. In
practice, the further condition $\alpha=\gamma$ was
fixed.\footnote{In Ref.~\cite{haeffner-2005-438} the witness was
found in a different way, leading to a different notation. However,
it is equivalent to the construction in
Eq.~(\ref{secondimprovement}) with $\alpha=\gamma$. For the
experimental data of Ref.~\cite{haeffner-2005-438}, a possible
additional variation over $\alpha,\gamma$ would not yield a big
improvement.} In order to make the values comparable, all witnesses
have been normalized such that $Tr[\WW (\openone/2^N)] = 1,$ i.e.,
the mean value for the maximally mixed state equals one. The witness
values in Table \ref{fidtable} for $4 \leq N \leq 7$ have been
obtained in this way.

For the case $N=8,$ however, this construction still does not yield
a witness with a negative expectation value. In order to prove still
the presence of entanglement, one can use the help of {local filtering
operations} \cite{gisin-1996,verstraete-2003-68,leinaas:012313}. Such a
local filter is a map 
\be
\vr \mapsto \tilde{\vr} = \NN \cdot
(\mathcal{F}_1 \otimes \mathcal{F}_2 \otimes ... \otimes \mathcal{F}_8)
\vr
(\mathcal{F}_1^\dagger \otimes \mathcal{F}_2^\dagger \otimes ... \otimes \mathcal{F}_8^\dagger)
\ee
where the $\mathcal{F}_i$ are arbitrary invertible $2 \times 2$ matrices
and $\NN$ denotes the normalization (see also Section \ref{jamiolsection}). 
As the $\mathcal{F}_i$ are invertible,
the state $\vr$ is entangled (or biseparable) if and only if $\tilde{\vr}$
is entangled (or biseparable) and it suffices to detect the entanglement in
the filtered state. Equivalently, one may also consider the class of witnesses
\be
\tilde{\WW} = \tilde{\NN}
(\mathcal{F}_1^\dagger \otimes \mathcal{F}_2^\dagger \otimes ... \otimes \mathcal{F}_8^\dagger)
\WW
(\mathcal{F}_1 \otimes \mathcal{F}_2 \otimes ... \otimes \mathcal{F}_8)
\ee
and try to find filters such that the mean value $\mean{\tilde{\WW}}$ gets negative.
Here, one also has to renormalize the filtered witness with a factor $\tilde{\NN}$
[e.g., by fixing $Tr(\tilde{\WW})=Tr({\WW})$] in order to make the witnesses' values
comparable.

The optimization over all filters $\mathcal{F}_i$ can be performed
by first optimizing $\mathcal{F}_1,$ then $\mathcal{F}_2,$ etc.,
until a negative mean value can be found. With this technique, the
mean value for the eight-qubit witness could be decreased
significantly, and finally a negative value has been found.

It should be noted that the filter technique is not restricted to
the case where complete information about the state via state
tomography is available. As shown in Refs.~\cite{ToolBox,gao-2008}
some witnesses that require only few measurements can be improved by
special filters that leave the required measurement settings
invariant. Finally, it should be added that Ref.~\cite{nha-2008}
gave an interesting scheme to implement witnesses as in
Eqs.~(\ref{WN}, \ref{firstimprovement}) in linear optics
experiments.

\subsection{Estimation of entanglement measures}
\label{estimationofentanglementmeasures}

In the previous Sections we introduced methods for the
detection of entanglement in experiments. More generally,
one can ask how entanglement in experiments can be
{\it quantified.} This leads to the problem, how to 
quantify entanglement, if only incomplete information 
about the state is given \cite{PhysRevA.59.1799}.

In this Section, we will explain methods to quantify 
entanglement using entanglement witnesses.
It should be noted that there are other proposals for
the determination or estimation of entanglement measures
in experiments, they are described in Sections \ref{walbornexp}
and \ref{sectionmaps}.
We will first describe a general method to derive a lower
bound on a generic entanglement measure from the mean value
of a witness \cite{guhne-2006-1,eisert-2006-1,guhne-2008-77}.
Then we will discuss results for specific
witnesses and entanglement measures.

\subsubsection{A general method using the Legendre transform}
\label{ageneralmethodusingthelegendretransform}

Let us consider the following situation: In an experiment, an
entanglement witness $\WW$ has been measured, and the mean value
$\mean{\WW}=Tr(\vr \WW)=w$ has been found. The task is to derive
from this single value a quantitative statement about the
entanglement present in the quantum state. That is, one aims to
derive statements like \be Tr(\vr \WW)=w \;\; \Rightarrow \;\;
E(\vr) \geq f(w), \label{task} \ee where $E(\vr)$ denotes some
convex entanglement measure or another convex figure of merit, 
which one is interested in (e.g. the fidelity). Clearly, one aims 
to derive an optimal bound $f(w)$ and an estimate is optimal, 
if there is a state $\vr_0$
with $Tr(\vr_0 \WW)=w$ and $E(\vr_0) = f(w).$

In order to derive such lower bounds, one can use the
Legendre transform \cite{rockafellar,wernerskript} of $E$ for the witness
$\WW$, defined via the maximization
\be
\hat{E}(\WW) = \sup_\vr \{Tr(\WW \vr) - E(\vr)\}.
\label{legtrafo}
\ee
As this is defined as the maximum over all $\vr,$ we have
for any fixed $\vr$ that
$\hat{E}(\WW) \geq  Tr(\WW \vr) - E(\vr),$ hence
$
E(\vr) \geq Tr(\WW \vr) - \hat{E}(\WW).
\label{firstbound}
$
The point of this rewriting is that the first term on the right hand
side is the given measurement data, while the second term can be
computed. Therefore, a measurable bound on $E(\vr)$ has been
obtained.

\begin{figure}[t]
\centerline{\includegraphics[width=0.4\columnwidth]{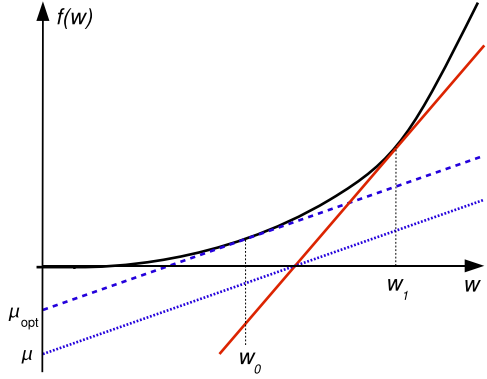}}
\caption{Geometrical interpretation of the Legendre transform.
The dotted line is a general affine lower bound, and the dashed
line is an optimized bound with the minimal $\mu.$ By varying
the slope, one finds an affine lower bound, which is tight for
a given $w_1$ (solid line). This figure is taken from
Ref.~\cite{guhne-2008-77}.
\label{legendrebild}
}
\end{figure}

More generally, one can use the fact that $Tr(\vr \WW)=w$ is
equivalent to knowing $Tr(\vr \lambda \WW)= \lambda w$ for
any $\lambda.$ Therefore, optimizing over all $\lambda$ yields
\be
E(\vr) \geq
\sup_\lambda \{ \lambda Tr(\WW \vr) - \hat{E}(\lambda\WW)\}.
\label{secondbound}
\ee
This is already the optimal bound in Eq.~(\ref{task}). This follows
from the geometrical interpretation, of the Legendre transform
(see Fig.~\ref{legendrebild}): For a given convex function $f(w)$
an affine lower bound $g(w)= \lambda w - \mu$ with a given fixed
slope $\lambda$ is optimal, if $\mu_{\rm opt}=\sup_w [\lambda w - f(w)],$
which is exactly the Legendre transform. Such a lower bound is optimal
for one $w_0.$ To obtain the optimal bound for any given
$w_1$ one optimizes over the slope $\lambda,$  corresponding to
Eq.~(\ref{secondbound}). Since $f$ is convex, we obtain for
each $w$ the tight linear bound.

Note that it is not needed that $\WW$ is an entanglement
witness, we may consider an arbitrary observable instead. Also
one may consider a set of observables $\vec{\WW} = \{\WW_1, ..., \WW_n\}$
at the same time, by introducing a vector $\vec{\lambda} =\{\lambda_1,...,\lambda_n\}$
and replacing  $\lambda \WW$ by $\sum_k \lambda_k \WW_k.$

In any case, the main problem in this scheme lies in the calculation of
the Legendre transform in Eq.~(\ref{legtrafo}). The difficulty of
this task clearly depends on the witness $\WW$ and on the measure
$E(\vr)$ chosen. For this problem, the following results have been obtained:

\begin{enumerate}

\item For entanglement measures defined via the convex roof extension
[see Eq.~(\ref{convexroofdefinition})] the optimization can be simplified to
\be
\hat{E}(\WW) = \sup_{\ket{\psi}}
\{
\bra{\psi}\WW\ket{\psi} - E(\ket{\psi})
\},
\ee
i.e., it suffices to optimize over pure states only \cite{guhne-2006-1}.

\item For the entanglement of formation [see Eq.~(\ref{eofdefinition})] one can design
a simple iterative algorithm that can perform the calculation
numerically \cite{guhne-2006-1}. For special witnesses of the type
$\WW = \ketbra{\psi}^{T_B}$ the Legendre transform can be determined
analytically \cite{eisert-2006-1}. Concerning the concurrence, one
can find a similar algorithm as for the entanglement of formation
\cite{guhne-2008-77}.

\item For the geometric measure of entanglement, one can also construct an
iterative algorithm for general witnesses. Further one can compute
the Legendre transform for a witness like
$\WW = \alpha \eins - \ketbra{\psi}$ analytically
\cite{guhne-2006-1}. From this one can show that if an experimentally generated
state has a fidelity $F$ with the target state $\ket{\psi},$ and the geometric
measure of $\ket{\psi}$ is $E,$ then for $F \geq 1-E$ the geometric measure is
bounded by\footnote{Clearly, for
$F\leq 1-E$ the fidelity is compatible with a separable state, hence the only
lower bound is the trivial bound $E \geq 0$}
\be
E_G \geq \max_{k=0,1}\Big\{
1- F - E  + 2 E F +
\frac{1}{2} \sqrt{\frac{E(E-1)}{F(F-1)}}
\big[(-1)^k (2F-1)^2 -1 \big]
\Big\}.
\ee
For some other witnesses one can analytically estimate the
Legendre transform \cite{eisert-2006-1,guhne-2008-77}.

\end{enumerate}

\subsubsection{Other relationships between witnesses and entanglement
measures}
\label{furtherestimationmethods}

In order to investigate  the relationship between witnesses and
entanglement measures from a different viewpoint, one can ask
whether the negative expectation value can directly serve for
an entanglement quantification. Therefore, one considers the
quantity \cite{brandao-2005-72}
\be
E(\vr) = \max\{0, \max_{\WW \in \mathcal{C}} [-Tr(\vr \WW)]\}
\ee
where the optimizations is restricted over $\mathcal{C},$ which
is a certain compact subset of all witnesses. This $E(\vr)$ is by
definition already convex.

Then, by choosing $\mathcal{C}$ appropriately, one can obtain
different $E(\vr)$ that can be used for the entanglement
quantification: If  $\mathcal{C}$ is the set of witnesses with $\WW
\leq \eins$ then $E(\vr)$ is the generalized robustness (see Section
\ref{sectionrobustness}) and if $\mathcal{C}$ is the set of
witnesses with $Tr(\WW) = d$, then $E(\vr)$ is the random robustness
\cite{brandao-2006-4,brandao-2005-72}. Also the distance to the set
of separable states in Hilbert-Schmidt norm, the negativity and the
concurrence can be expressed in a similar way
\cite{bertlmann-2002-66,brandao-2005-72}. All these connections
allow for a simple estimation of entanglement quantifiers in
experiments, as any appropriately normalized witness gives a lower
bound \cite{cavalcanti-2006-89}. Experimentally, this has been used
in Ref.~\cite{cavalcanti-2008-78} to probe the boundary of the set
of separable states.

Let us add that witnesses as in Eqs.~(\ref{pptwitness}, \ref{ccnrwitness})
can also be used to estimate the violation of the PPT or CCNR criterion,
which can further be used to estimate the concurrence or the entanglement
of formation as in Eq.~(\ref{cafformel}) \cite{chen-2005-95a,chen-2005-95b}.
Also other special witnesses can be used: One can relate the violation of a
CHSH inequality for two qubits to the entanglement of formation 
\cite{verstraetewolf}, use witnesses for the reduction criterion 
to estimate the concurrence \cite{mintert:052302} (see Section 
\ref{walbornexp}) and can also use the extended reduction map to 
estimate the concurrence \cite{breuer-2006-39}. For special cases 
one can also estimate the negativity from incomplete
tomographical information \cite{audenaert-2006-8,wunderlichplenio-2009}. 
A comparison of some of these methods can be found in 
Ref.~\cite{guhne-2008-77}.

Finally, a completely different approach tries to estimate 
entanglement from incomplete information via Bayesian updating. 
If one assumes some {\it a priori} probability distribution $p(\vr)$ 
for the quantum states and measures an observable with the result 
$d$ [with a priori probability $p(d)=\int_\vr p(\vr) p(d|\vr)$], one 
can update the probability distribution via the Bayesian rule, 
\be
p(\vr|d)  = \frac{p(d|\vr) p(\vr)}{p(d)}.
\label{bayes}
\ee
 This can then be iterated. Consequently, one can obtain for
the negativity a probability distribution $p[N(\vr)]$ from
incomplete data \cite{lougovski-2008}. Similar methods can 
be used for state tomography \cite{blumekohout-2006}.

\subsection{Entanglement witnesses in quantum key distribution}
\label{sectionqkd}
Let us finally discuss another application of entanglement witnesses,
namely the security analysis of quantum key distribution (QKD)
\cite{RevModPhys.74.145,dusek-2006-49,woottersbook, scarani-2008-review}.
There are many different protocols for quantum key distribution.
They can be divided into two classes: 
First, in {\it entanglement based} (EB) schemes an untrusted source
distributes an entangled state $\vr$, and Alice and Bob make
measurements on it. From these measurement data they obtain the
secret key, using classical post-processing of the data (e.g., key
distillation or advantage distillation).  An example is the Ekert
protocol, where the source ideally distributes two-qubit singlet
states $\ket{\psi} = (\ket{01} - \ket{10})/\sqrt{2}$
\cite{PhysRevLett.67.661}.

In {\it prepare \& measure} (P\&M) schemes like the BB84 protocol
\cite{bennett84a, bennett92b, bruss98a}, Alice prepares randomly
states from a certain set $\{\ket{\psi_i}\}$ and sends them to Bob,
who makes measurements. Then, using classical communication they
obtain the secret key. At first sight, these two types of protocols
do not seem to be closely related, but, as shown in
Ref.~\cite{PhysRevLett.68.557} one can always view a P\&M scheme as
an EB scheme: The state preparation on Alice's side can be interpreted
as if a bipartite entangled state 
\be \ket{\Psi} = \sum_{i} \ket{i}_A \ket{\psi_i}_B 
\ee 
is prepared in Alice's lab, then one
part is sent to Bob, and Alice measures in the computational basis
$\ket{i}_A.$ This is practically equivalent to an EB scheme,
however, the possibilities of an eavesdropper are slightly more
limited: In the usual EB scheme, Eve may have access to the source
and can hence affect the complete $\vr,$ while in the P\&M scheme
Eve can only access the part sent to Bob, i.e., the reduced state
$\vr_A$ of Alice is fixed.

What is the role of entanglement in these protocols? In the EB protocols,
it is natural to ask whether the presence of entanglement is necessary
for the possibility to create a secret key from the measurement data.
Indeed, Ref.~\cite{curty:217903} showed the following result:

{\bf Theorem.} If the measurement data obtained in an EB scheme are
compatible with a separable state, then no secret key can be extracted
from that data and the QKD scheme is not secure. The same holds for
P\&M schemes, if one considers the corresponding EB scheme.

Concerning the contrary implication, it has been shown in
Ref.~\cite{acin-2005-94-qkd} that a state is entangled
iff there exist measurements which result in probabilities
which contain secret correlations. Hence, if Alice and Bob
can verify entanglement, their correlations contain secrecy,
however, as there might be bound information, they might not
be able to distill a secret key.

Note that the theorem requires that the entanglement can be proven
with the {\it measured} data. Since many protocols Alice and Bob do not
perform state tomography of the distributed state, this leads to the
question, how one can prove the entanglement of a state, if only
the expectation values of some observables are given.

It is not difficult to see that if mean values $\mean{A_1}, ... ,
\mean{A_k}$ of some observables are given, then entanglement of the
underlying state can be proven, if and only if  there is a witness
that is a linear combination of the $A_i$ that detects it.  This can
be seen from the fact that in the space of the mean values of the
$A_i$ the separable states form again a convex set and this set is
characterized by linear functions of the $\mean{A_1}, ... ,
\mean{A_k}.$ Therefore, the task remains to characterize the optimal
witnesses for a given set of observables.

As an example let us consider the Ekert protocol. There, Alice and Bob
measure randomly the polarization in the x- and z-direction. As they can
communicate via classical communication, one can assume  that the mean values of
\begin{align}
A_1 &= \sigma_x \otimes \eins; \;\;\;
A_2 = \eins \otimes \sigma_x ; \;\;\;
A_3 = \sigma_x \otimes \sigma_x;\;\;\;
A_4 = \sigma_x \otimes \sigma_z;
\nonumber
\\
A_5 &=\sigma_z \otimes \eins; \;\;\;
A_6 = \eins \otimes \sigma_z; \;\;\;
A_7 = \sigma_z \otimes \sigma_z;\;\;\;
A_8 = \sigma_z \otimes \sigma_x;
\end{align}
are known, while other correlations like $\sigma_y \otimes \sigma_x$
are unknown.\footnote{For the P\&M protocol, the reduced density
matrix of Alice is fixed, as Eve cannot access it. Therefore, one
can assume that also $\mean{\sigma_y \otimes \eins}$ is known. This
does, however, not affect the optimal witnesses obtained later. See
Ref.~\cite{curty04suba} for a discussion.} Using the decomposition
of a witness into Pauli matrices, one can first prove that all
entanglement witnesses $\WW$ that can be measured with these
observables fulfill $\WW = \WW^T = \WW^{T_A} = \WW^{T_B}.$

Then, it has been shown in Refs.~\cite{curty:217903,curty04suba}
that the optimal witnesses for entanglement detection in this class
are given by \be \WW = \ketbra{\psi} + \ketbra{\psi}^{T_B}, \ee
where $\ket{\psi}$ is an entangled state with real coefficients.
Therefore, it suffices to check these witnesses, and if they do not
detect any entanglement, the QKD implementation is provably not
secure.

Similar ideas have then be considered for other QKD protocols, often using the EVM matrix
[see Eq.~(\ref{evmdef})]. This includes
the B92 protocol \cite{curty04suba}, QKD with continuous variables
\cite{rigas-2006-73,lorenz:042326} and entanglement detection using
Stokes parameters \cite{haseler-2008-77}.

\section{Further methods of entanglement detection}

In this Section, we explain other methods for entanglement detection
and quantification beside witnesses and Bell inequalities. We first
discuss methods that have already been implemented in experiments,
then we consider theoretical proposals that are still waiting for an
implementation.

\subsection{Entanglement criteria using entropies}
\label{bovinoexp}

Here, we describe an entanglement detection method that requires the
measurement of some nonlinear properties of a quantum state. To
start, let us recall the majorization criterion for separability
[see Eq.~(\ref{majorizationequation})]. It states that for a
separable bipartite state the global state is more disordered than
the local (reduced) states. If we measure the disorder by the linear
(or Tsallis) entropy $S(\vr) = 1 - Tr(\vr^2),$ this implies that for
separable states 
\be Tr(\vr_A^2) \geq Tr(\vr^2) \mbox{ and }
Tr(\vr_B^2) \geq Tr(\vr^2) \label{bovino1} 
\ee has to hold
\cite{PhysRevA.54.1838,vollbrecht-2002,horodecki-alpha}.\footnote{For 
generalizations of such inequalities see 
Refs.~\cite{augusiak-2008-77A,augusiak-2008-77B}.} For two
qubits, it can be shown that this inequalities detect strictly more
states than the CHSH inequality \cite{horodecki-alpha}, however,
there are some entangled states that are not detected by it.

From an experimental point of view, it is not obvious how
Eq.~(\ref{bovino1}) can be tested, as it is a nonlinear function of
the quantum state. However, in Ref.~\cite{bovino:240407} an elegant
measurement scheme has been implemented that we will describe
now.\footnote{For a general scheme to measure nonlinear properties
without state tomography see Ref. \cite{ekert-2002-88,
PhysRevA.68.032306}.}

Let us first discuss the determination of $Tr(\vr^2).$
This can be written on two copies of the state as
$Tr(\vr^2) = Tr(\vr \otimes \vr V),$ where
$V=\sum_{ij} \ket{ij}\bra{ji}$ is the swap operator
between the two copies. Using the fact that $V=P_S - P_A = \eins - 2 P_A$,
where $P_S$ ($P_A$) is the projector onto the symmetric
(antisymmetric) subspace of the two-copy system, we have
$Tr(\vr^2) = Tr[\vr \otimes \vr (P_S-P_A)].$ Decomposing
$P_S$ and $P_A$ into local projectors onto the symmetric (or
antisymmetric) subspaces of Alice's or Bob's local
two-copy system, one can write it further as
\be
Tr(\vr^2) = Tr([\vr \otimes \vr] X) \mbox{ with }
X = P_S^a \tilde{\otimes} P_S^b
-P_A^a \tilde{\otimes} P_S^b
-P_S^a \tilde{\otimes} P_A^b
+P_A^a \tilde{\otimes} P_A^b,
\ee
where $P_S^a$ denotes the projector onto the symmetric space of
Alice's two copies, etc., $\otimes$ denotes the tensor product
between the two copies, and $\tilde{\otimes}$ denotes the tensor
product between Alice and Bob. Clearly,
$Tr(\vr_{A}^2) =
 Tr([\vr \otimes \vr] [(P_S^a - P_A^a) \tilde{\otimes} \eins])$
can be written in a similar fashion as an expectation value on a
two-copy state.

\begin{figure}[t]
\setlength{\unitlength}{0.04\columnwidth}
\begin{center}
\begin{picture}(20,7)
\thinlines
\put(3.8,0.5){\includegraphics[width=0.48\columnwidth]{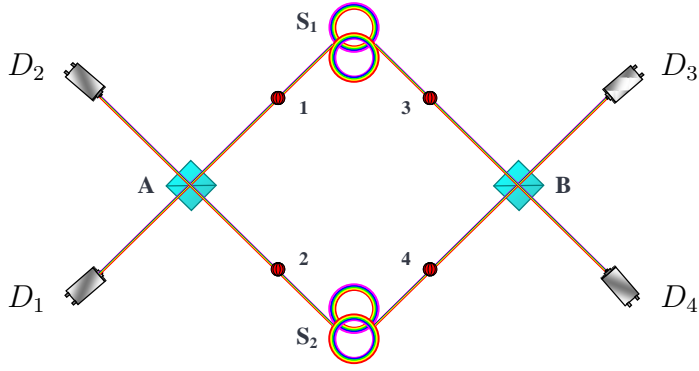}}
\put(2.8,1.8){\mbox{\large $D_1$}}
\put(2.8,6.5){\mbox{\large $D_2$}}
\put(16,1.8){\mbox{\large $D_4$}}
\put(16,6.5){\mbox{\large $D_3$}}
\end{picture}
\end{center}
\caption{Schematic view of the setup of the experiment in
Ref.~\cite{bovino:240407}. $S_1$ and $S_2$ denote the sources
of the entangled photons, $A$ and $B$ two beamsplitters and
$D_1, ..., D_4$ the four output ports. The figure is taken
from Ref.~\cite{bovino:240407}.
\label{bovinobild1}
}
\end{figure}

For the experimental measurement of these quantities, two
beamsplitters are required (see Fig. \ref{bovinobild1}). Consider
first the left beamsplitter (BS) A. If two identical photons 1 and 2
arrive at the BS the photon statistics of the BS will result in a
coalescence of the photons, i.e., they will either be both in the
output port $D_1$ or $D_2$ \cite{pan-2008}. More precisely, if the
two photons are in a symmetric polarization state, they will
coalesce, while if they are in an antisymmetric polarization state,
they will anti-coalesce and  there will be one photon in each of the
output ports. Therefore, the probability of coalescence equals the
mean value of $P_S^a,$ and the probability of anti-coalescence
corresponds to $P_A^a.$

In the experiment, the coincidence probabilities of coalescence
and anti-coalescence for Alice and Bob, $p_{cc},p_{ac}, p_{ca}$
and $p_{aa}$ have been measured. In this language, the inequality
$Tr(\vr_A^2) \geq Tr(\vr^2)$ is equivalent to $p_{ca} \geq p_{aa}.$
Experimentally, $p_{ac} = 0.0255 \pm 0.008$ and $p_{aa} = 0.2585 \pm 0.008$
have been found, clearly violating the inequalities for separable
states.

There are, however, some caveats with the interpretation of
such data. First, in the generation of entangled photon pairs
in the setup of Ref.~\cite{bovino:240407} it happens that the source
$S_1$ emits four photons, while $S_2$ emits none. This  is not a
rare event (it has the same probability as the case that each source
emits two photons) and may alter the probabilities  $p_{cc},p_{ac}, p_{ca}$
and $p_{aa}.$ However, these events could be
taken into account by varying the relative phase in the pump beams of
the sources $S_1$ and $S_2.$

A more fundamental problem lies in the usage of two copies of the
state $\vr$ \cite{vanenk-2009neu}. From a theoretical point there 
is no problem to write the separability conditions as inequalities 
for two copies. Experimentally, however, it is not clear that the 
two instances of the state really are the same and uncorrelated. 
This can not be verified without state tomography \cite{vanenk-2007-75}. 
Especially, classical correlations between the two systems may lead to a
systematic error in some entanglement detection schemes, see
Refs.~\cite{vanenk-2006,vanenk-2009neu} for a discussion.

A scheme similar to the one described can also be used for the
investigation of entanglement in optical lattices
\cite{alves:110501, palmer:042335}. Namely, any fully separable
state fulfills inequalities similar to Eq.~(\ref{bovino1}), and also
a transformation acting as a beamsplitter can, in principle be
implemented. It should be noted, however, that the resulting scheme
is unable to detect genuine multipartite entanglement, only full
separability can be excluded.

\subsection{Estimating the concurrence by using several copies}
\label{walbornexp}

Similar ideas as in the previous subsection can
be used to investigate the entanglement in  experiments quantitatively,
by giving lower bounds on the concurrence.
To start, recall that the squared concurrence of a pure state is given by
\begin{equation}
C^2(\ket{\psi}) = 2 [1-Tr(\vr_A^2)].
\label{CCC}
\end{equation}
In a similar way as in the previous section one can write
therefore the concurrence as
\begin{equation}
C^2(\ket{\psi})
= 4
\bra{\psi}\otimes\bra{\psi}[P_A^a \tilde \otimes \eins] \ket{\psi}\otimes\ket{\psi}
= 4
\bra{\psi}\otimes\bra{\psi} [P_A^a \tilde \otimes P_A^b] \ket{\psi}\otimes\ket{\psi},
\label{CCC2}
\end{equation}
where the second equality comes from the fact that $P_A^a \tilde{\otimes}P_S^b$
acts on the global antisymmetric subspace only, and its expectation value should
therefore vanish, if the two copies of $\ket{\psi}$ are indeed identical
\cite{walborn:032338}.

Therefore, by measuring the (local) projector onto the antisymmetric subspace
on Alice's part only, one can get a quantitative statement about the entanglement
contained in the state. This has been implemented using hyper-entangled photons
\cite{mintert-nature, walborn:032338} and a proposal for the implementation
using ion traps or atoms in cavities exists \cite{romero-2007-75}. It should be 
noted, however, that this approach requires the assumption that the generated state 
is pure, which is difficult
to guarantee in any experiment without doing state tomography
\cite{vanenk-2006}.\footnote{Moreover, since one explicitly assumes that 
the generated state is pure, one could also just determine $\vr_A$ by state 
tomography on Alice's side, and then  compute the concurrence or the entanglement 
of formation.}

Interestingly, however, expressions similar to Eq.~(\ref{CCC2})
could be used to obtain lower bounds on the concurrence for mixed
states. As shown in Ref.~\cite{mintert-2006-}, a lower bound on the
concurrence can be obtained by
\begin{equation}
C(\rho)^2\ge {\rm Tr}(\rho \otimes\rho V) \mbox{ with }
V = 4  (P_A^a \tilde \otimes P_A^b  - P_A^a \tilde{\otimes}P_S^b).
\label{CCC3}
\end{equation}
The difference to the previous expression in Eq.~(\ref{CCC2}) is the
sign of the term $P_A^a \tilde{\otimes}P_S^b$, which therefore,
could be interpreted as taking account of the mixedness of the
states. For an experimental implementation, still assumptions about
the measurements on two copies have to be discussed
\cite{vanenk-2009neu,vanenk-2006}. In theory, however, Eq.~(\ref{CCC3})  already
allows to estimate the convex roof of the concurrence, and the lower
bounds are reasonable for states that are close to pure states, for 
investigations of the tightness see Refs.~\cite{mintert-2006-, walborn:032338, 
guhne-2008-77, borras-mbbound}. Also upper bounds
on the concurrence can be achieved from measurements on two copies 
\cite{zhang-2008-78-neu}.

Moreover, Eq.~(\ref{CCC3}) can be generalized to
\begin{equation}
C(\rho) C(\sigma) \ge {\rm Tr}(\rho \otimes \sigma V),
\label{CCC3neu}
\end{equation}
where $\vr$ and $\sigma$ are different states \cite{mintert:052302}.
This inequality has been experimentally tested using photons 
\cite{schmid:260505}. In this experiment, two two-photon states
$\vr$ and $\sigma$
have been generated via down-conversion, then the nonlocal 
observable $V$ has been measured using a controlled-phase gate 
between two of the four photons. Using $C(\vr)\leq 1$ one obtains 
bounds on $C(\sigma)$  and vice versa, however, assumptions about 
the fidelity of the phase gate have to be discussed.
Furthermore, Eq.~(\ref{CCC3}) shows that 
\be 
\WW = - \frac{1}{C(\sigma)} Tr_2(\eins \otimes\sigma V) 
= - \frac{2}{C(\sigma)} (\sigma - \eins \otimes \sigma_B)
\ee 
is a quantitative entanglement witness that can be used for the
estimation of the concurrence. This witness corresponds to the
reduction criterion [see Eq.~(\ref{reductionequation}) and the
discussion after Eq.~(\ref{pptwitness})], hence a lower bound on the
concurrence can be obtained, iff the state violates this criterion.

Finally, note that one may define a multipartite generalization of
the concurrence $C_m$ as the sum of the bipartite concurrence over
all bipartitions, 
\be C^2_m(\ket{\psi}) = \frac{1}{2^{N-1}}
\sum_{bp} C^2_{bp}(\ket{\psi}), 
\ee 
where the sum goes over all
$2^N-2$ bipartitions \cite{mintert-2005-95}. This quantity does not 
measure genuine multipartite correlations, but again expressions 
similar to Eqs.~(\ref{CCC2}, \ref{CCC3}) can be derived, since bounds 
for each $C^2_{bp}(\ket{\psi})$ are available from Eq.~(\ref{CCC3})
\cite{aolita-2006-97,aolita-2008-78}.

\subsection{Variance based criteria and nonlinear witnesses}
\label{sectionvariances}

Entanglement witnesses and Bell inequalities can be viewed as
inequalities for mean values of observables, where a violation
indicates entanglement. Therefore, it is natural to ask whether it
is possible to derive similar conditions using the variances of
observables. These conditions would then also be nonlinear in the
state $\vr,$ hence one may consider them as nonlinear witnesses. One
the one hand, it might be interesting to use nonlinear witnesses as
an improvement of a given linear witness. On the other hand,
nonlinear witnesses are also interesting from a geometrical point of
view, as they approximate the convex set of separable states better.

In the literature, entanglement criteria using variances were first
known for the case of continuous variables and infinite dimensional
systems (\cite{duan-2000-84, PhysRevLett.84.2726,
PhysRevA.67.052315, vogelprl,mancini-2002-88,PhysRevA.68.062310}, 
for reviews see Refs.~
\cite{braunstein:513,eisert-2003-1,adesso-2007-40,wang-2007-448}).
For the case of discrete systems, the first variance based criteria
have been derived in
Refs.~\cite{PhysRevLett.88.230406, giovannetti-2003-67}. 
A very interesting approach
was then proposed by Hofmann and Takeuchi \cite{hofmann-2003-68}.
These, so-called {\it local uncertainty relations} (LURs) turned
later out to be a very interesting tool to investigate the
separability problem, so we will explain them now.

\subsubsection{Local uncertainty relations}
\label{sectionlurs}

First, recall that the variance of an observable $M$ in the 
state $\vr$
is given by
$
\va{M}_{\varrho} :=
\mean{(M-\mean{M}_{\varrho})^2}_{\varrho} =
\mean{M^2}_{\varrho}-\mean{M}^2_{\varrho}
$
and is concave in the state \cite{hofmann-2003-68},
i.e., if  $\varrho=p \varrho_1 + (1-p)\vr_2$
is a convex combination of two states, then
$ \va{M}_{\varrho}
\geq
 p \va{M}_{\varrho_1} + (1-p )\va{M}_{\varrho_2}.$
If $\vr=\ketbra{\psi}$ describes a pure state, the variance of $M$ is
zero iff $\ket{\psi}$ is an eigenstate of $M.$

Let us assume that we have observables $A_i$ for Alice and $B_i$ for
Bob, so that the  observables in each of these two sets do not share
a common eigenstate. Then, there must be positive numbers $U_A$ and
$U_B$ such that 
\be \sum_{i=1}^n \va{A_i}\geq U_A \;\;\;\mbox{ and }
\;\;\; \sum_{i=1}^n \va{B_i}\geq U_B \label{lur1} 
\ee 
holds, for all
states on Alice's (resp. Bob's) system. Under this condition, it was
shown in Ref.~\cite{hofmann-2003-68} that for separable states and
for the observables $M_i := A_i \otimes \Eins +\Eins \otimes B_i$
the inequality
\begin{equation}
\sum_{i=1}^n\va{M_i}_{\varrho} \geq U_A + U_B
\label{lur2}
\end{equation}
holds and violation of it implies entanglement. Physically, this may be
interpreted as if the separable states inherit the uncertainty relations
from the reduced states.

{From} an experimental point of view it is interesting, that the LURs
can be viewed as nonlinear witnesses improving an optimal linear witness.
To see this, let us consider an example \cite{guehne-2004-734}.
For a single-qubit system
is is easy to see that for the Pauli matrices the uncertainty relation
$\sum_{i=x,y,z} \va{\sigma_i} \geq 2$ holds. Defining
$M_i=\sigma_i \otimes \eins + \eins\otimes \sigma_i$ this yields the
LUR $\sum_{i=x,y,z} \va{M_i} \geq 4.$
Writing down this explicitly, one obtains the condition
\begin{equation}
\mean{
\eins \otimes \eins +
\sigma_x \otimes \sigma_x+
\sigma_y \otimes \sigma_y+
\sigma_z \otimes \sigma_z
}
-\frac{1}{2}
\sum_{i=x,y,z} \mean{\sigma_i \otimes \eins + \eins\otimes \sigma_i}^2
\geq 0,
\label{lur5}
\end{equation}
which has to hold for all separable states. This is a quite
remarkable inequality for the following reason: The first part, which
is linear in the expectation values is known to be an {\it optimal}
entanglement witness [see Eq.~(\ref{witdec})]. {From} this witness
some quadratic terms are subtracted. Note that an experimental
measurement of these terms would require just the same measurements
as the measurements that are required anyway for the measurement of
the witness in Eq.~(\ref{witdec}). Thus, in this case, the LUR can
be viewed as a nonlinear witness that improves an optimal linear
witness.\footnote{It is further interesting that for any pure product
state $\ket{\psi}= \ket{a}\ket{b}$ equality holds in
Eq.~(\ref{lur5}), which shows that the nonlinear witness is curved
around the whole set of separable states.}

In general, the entanglement criteria resulting from the LURs
are very strong: they can detect bound entanglement
\cite{hofmann-2003-68bound},
indeed, it can be shown that they are stronger than the CCNR criterion
\cite{guhne-2006-74}.\footnote{This can be easily seen from the fact
that LURs always allow to improve the witnesses for the CCNR criterion,
see Eq.~(\ref{ccnrwitness}).} Finally, it  turned out, that they are
equivalent to the covariance matrix criterion described above
\cite{guhne-2007-99,gittsovich-2008} (see Eq.~\ref{cmccriterion}).
Also, statements similar to the LURs can be extended to nonlocal
observables \cite{guehne-2004-92}, or other formulations of the
uncertainty principle, like entropic uncertainty relations
\cite{giovannetti-2004-70,guehne-2004-70} or the
Landau-Pollak uncertainty relations \cite{devicente-2005-71}. 
The local uncertainty relations and the variance based criteria
can also be used to give lower bounds on entanglement measures, 
similarly to Section \ref{estimationofentanglementmeasures} 
\cite{devicente-2007-75,devicente-2007-75-erratum,devicente-2008-41,zhang-2007-76}.
Concerning an experimental implementation, some of the variance
based criteria have been implemented in Ref.~\cite{wang-2007-78}.

\subsubsection{Nonlinear entanglement witnesses}
\label{sectionnonlinearwitnesses}

The LURs show that in some cases it is possible to construct
nonlinear witnesses. However, it is easy to see that the recipe of
the LURs can not be applied to arbitrary witnesses. Therefore,
the question arises whether it is possible to find a general recipe
to improve a witness by some nonlinear terms.

Indeed, as shown in Ref.~\cite{guhne-2006-96}, for bipartite witnesses
this is the case.
The recipe is the following: Let us consider the standard witness from
the PPT criterion
\be
\WW = \ketbra{\psi}^{T_B}
\ee
Let us define $X = \ket{\psi}\bra{\phi}$ with an arbitrary $\ket{\phi}$
and consider the functional
\be
\mathcal{F}(\vr) = \mean{\WW} -
\frac{1}{s(\phi)} \mean{X^{T_B}}\mean{(X^{T_B})^\dagger},
\ee
where $s(\phi)$ denoted the maximal squared Schmidt coefficient of
$\ket{\phi},$ which is a bound on the maximal overlap between
$\ket{\phi}$ and product states. Then we have for a pure product state
$\ket{\eta}= \ket{a}\ket{b}$ that
$\mathcal{F}(\ketbra{\eta}) =
|\braket{\tilde\eta}{\psi}|^2[1-|\braket{\tilde\eta}{\phi}|^2/s(\phi)] \geq 0,$
where $\ket{\tilde{\eta}}= \ket{a}\ket{b^*}$ is the vector corresponding to
the projector $\ketbra{\eta}^{T_B} = \ketbra{\tilde{\eta}} = \ketbra{a b^*}.$
By convexity, this positivity holds also for all separable states, proving that
$\mathcal{F}(\vr) $ is indeed a nonlinear entanglement witness, which
improves the witness $\WW.$

There are also some other constructions of nonlinear witnesses bases
on the same idea \cite{guhne-2006-96,guhne-2007-67} and due to the
Choi-Jamio{\l}kowski isomorphism [see Section \ref{jamiolsection}
and Eq.~(\ref{nl15})] all these methods can be extended to {\it
arbitrary} bipartite witnesses. This has also some consequences for
the geometry of separable states, as it proves that the set of
separable states has no facets \cite{guhne-2007-67}.\footnote{A face
of a convex set is a hyperplane at the boundary, and a facet is a
face of maximal dimension. For example, a three-dimensional cube has
its six surfaces as facets, while the edges are faces, but not
facets \cite{ioannou-2006-73}. Therefore, the set of separable
states is not as simple as in Fig.~\ref{eierbild1}(b).} Also, one
can define sequences of nonlinear witnesses by an iteration
procedure, such that the nonlinear witnesses become stronger and
stronger in each step and finally detect everything that is detected
by the corresponding positive, but not completely positive map
\cite{moroder-2008}.

At the moment, there are not much results known concerning nonlinear
witnesses for multipartite entanglement. The method from above can
not be extended to witnesses for genuine multipartite entanglement.
Some nonlinear witnesses for multipartite entanglement have already
been proposed in 2002 by Uffink 
\cite{PhysRevLett.88.230406} and recently some criteria have been
developed, which can also be viewed as some nonlinear conditions on
the state (see Ref.~\cite{seevinck-2007} and Section
\ref{section:multipartitesepcrit}).

\subsection{Estimation of positive maps}
\label{sectionmaps}

Given the importance of positive maps for the theoretical
characterization of entanglement, one may wonder whether
it is possible to use them directly for entanglement detection.
Indeed, several proposals for this task exist. Although
their implementation is beyond the current experimental
capabilities, they offer interesting insights in measuring
certain nonlinear properties of the density matrix without doing
state tomography.

As the transposition is not a completely positive map, it is not
possible to perform it directly in a lab. However, as shown in
Ref.~\cite{horodecki:127902} one can circumvent this problem by a
so-called structural physical approximation (SPA) of the
transposition: if a sufficient amount of white noise is added, the
total map becomes completely positive. Especially, it was shown in
Ref.~\cite{horodecki:127902} that in a $d\times d$ system the map
\be \Lambda(\vr) = \frac{d}{d^3+1} \eins_A \otimes \eins_B +
\frac{1}{d^3+1}\vr^{T_B} \ee is completely positive and can be
implemented. SPAs of other 
positive maps have been discussed in Ref.~\cite{korbicz:062105}.
The noise term  ${d}/({d^3+1}) \cdot \eins_A \otimes
\eins_B$ only shifts the spectrum of the partial transposition
$\vr^{T_B}$, hence it remains to measure the eigenvalues of
$\sigma=\Lambda(\vr)$ to evaluate the PPT criterion.

To do so, it suffices to measure $Tr(\sigma), Tr(\sigma^2), ..., Tr
(\sigma^{d^2}),$ as from these $d^2$ quantities all $d^2$
eigenvalues can be determined \cite{ekert-2002-88,
PhysRevA.68.032306, PhysRevA.68.052101}. The  quantity
$Tr(\sigma^k)$ can be determined similarly as in
Section~\ref{bovinoexp} by measuring a multiple of the shift
operator on a $k$-fold copy. Alternatively, one may use the spectrum
estimation technique from \cite{PhysRevA.64.052311}. Note that
knowing the spectrum of $\sigma,$ one can then determine the
negativity as an entanglement measure.

As shown in Ref.~\cite{PhysRevLett.90.167901} one can determine the 
concurrence in a similar way:
As explained in Section \ref{sectionconcurrence} the concurrence for two 
qubits can be computed from the
eigenvalues of $\vr \tilde{\vr}$ with
\be
\tilde{\vr} = (\sigma_y \otimes \sigma_y) \vr^T (\sigma_y \otimes \sigma_y).
\ee
To determine $Tr[(\vr\tilde{\vr})^k]$ requires $2k$ copies, where on half of
them first the transposition is implemented via the SPA, and then the unitary
transformation $(\sigma_y \otimes \sigma_y)$ is performed. Then, the eigenvalues
of $\vr \tilde{\vr}$ can be determined similar as before.

The previous proposals try to characterize entanglement by
implementing an interesting positive map {\it physically.} However,
as noted in Ref.~\cite{carteret:040502} this is not mandatory. The
$k$-th moments $Tr[(\vr^{T_B})^k]$ are invariants under local
unitary transformations, which can be measured using quantum
networks \cite{leifer-2004-69}. They can also be measured directly
by making measurements on $k$ copies of the state
\cite{grassl-1998-58,cai-2008}. Similar ideas can be used to determine 
the concurrence or the 3-tangle of mixed states \cite{carteret-2003}
or to evaluate the separability criteria given by arbitrary positive, 
but not completely positive maps \cite{horodecki-2006-74-maps}.

Finally, for the case of two qubits, similar methods can be used to
decide separability using a single, nonlinear entanglement witness.
As shown in Ref.~\cite{augusiak-2008-77}, for two qubits one has
that 
\be 
\vr \mbox{ is entangled } \Leftrightarrow det({\vr^{T_B}})
< 0. 
\label{tollegleichung}
\ee 
This equivalence reflects the fact that for two-qubits the
partially transposed state can have maximally one negative
eigenvalue \cite{sanpera-1998-58}. The quantity $det({\vr^{T_B}})$
is a fourth order polynomial of the matrix elements of $\vr$ and can
hence be measured by a single witness on a fourfold copy of $\vr$
\cite{augusiak-2008-77}. Similarly, for rotationally invariant states
on a $2 \times d$-system, the PPT criterion is necessary and sufficient
for separability \cite{schliemann-2003-68}, and if $d$
is even, Eq.~(\ref{tollegleichung}) can again be used for entanglement detection 
\cite{augusiak-2007-363}.

\section{Entanglement detection with collective measurements}
\label{Sec_collmeas}

In the following Section we discuss how to detect entanglement in a
many-qubit physical system if the qubits cannot be individually
accessed. For such systems entanglement criteria are needed that are
based on the measurement of collective observables.\footnote{Note
that in the literature, the notion of a ``collective measurement''
is sometimes differently used to describe measurements on several
copies (as in Section \ref{bovinoexp}).} This includes spin
squeezing inequalities as well as measurements of the magnetic
susceptibility or the energy in spin models. We also discuss
entanglement detection in optical lattices of cold atoms. In future,
when larger and larger coherent quantum systems will be prepared,
such criteria will surely gain large importance. It should be noted,
however, that several criteria in this Section do not detect
genuine multipartite entanglement, they only rule out full
separability.

\subsection{Spin squeezing inequalities}
\label{Sec_collspinent}

In this Section we describe entanglement detection with
collective spin-$\tfrac{1}{2}$ observables, which is relevant
to multi-qubit systems in which the qubits cannot be accessed
individually. Even for systems in which the qubits can be accessed
individually, such criteria can be advantageous since they need few
measurements.

In a multi-qubit system, the quantities that can be measured
collectively are the components of the collective angular momentum
\begin{equation}
J_{l}:=\frac{1}{2}\sum_{k=1}^N \sigma_{l}^{(k)},
\end{equation}
where $l=x,y,z$ and $\sigma_{l}^{(k)}$ are the Pauli spin matrices,
and their moments $\exs{J_{l}^m}$ where $m=2,3,4,...$ This includes
also the variances $\va{J_k}=\ex{J_k^2}-\ex{J_k}^2.$
Clearly, measuring only the expectation values of collective
observables is not
enough for entanglement detection: Such measurements do not give
information on entanglement between the spins. The minimum
requirement for detecting entanglement is measuring the first and
second order moments of collective angular momenta
\cite{PhysRevA.69.052327}.

\subsubsection{Spin squeezing and entanglement.}

The first criterion based on such ideas was the spin squeezing criterion.
On the one hand, Ref.~\cite{KitagawaSpinSqueezing1993} defined spin squeezing
in analogy with squeezing in quantum optics. Spin squeezing, 
according to the definition of Ref.~\cite{KitagawaSpinSqueezing1993}, 
means the following: In general, the variances of the angular momentum 
components
are bounded by the uncertainty relation
\begin{equation}
\va{J_y}\va{J_z}\ge \frac{1}{4}\vert\exs{J_x}\vert^2.\label{eqKU}
\end{equation}
If $\va{J_z}$ is smaller than the standard quantum limit
$\vert\exs{J_x}\vert/2$ then the state is called spin 
squeezed.\footnote{Note that for Eq.~(\ref{eqKU}) we assume 
that the mean spin points to the $x$ direction. Without this requirement, 
even a pure state in which all spins point into the same direction could 
violate Eq.~(\ref{eqKU}) with an appropriate
choice of the $x$, $y$, and $z$ coordinate axes \cite{KitagawaSpinSqueezing1993}.} 
In practice this means that a spin squeezed state has a large mean spin and
a small variance in a direction orthogonal to the mean spin.
On the other hand, in Refs.~\cite{PhysRevA.46.R6797,PhysRevA.50.67}, 
spin squeezing was defined from the point of view of spectroscopy.
States fulfilling this definition are useful for reducing
spectroscopic noise or increasing the accuracy of atomic clocks.\footnote{The definitions of spin squeezing in Ref.~\cite{KitagawaSpinSqueezing1993} and in Refs.~\cite{PhysRevA.46.R6797,PhysRevA.50.67} are slightly different.
For a discussion about that, see Sec. IV.A in Ref.~\cite{hammerer-2008}. For ananalysis concerning the two-qubit case, see Ref.~\cite{PhysRevA.68.064301}.}

It has already been pointed out in Ref.~\cite{KitagawaSpinSqueezing1993}
that the collective spin can be squeezed due to quantum correlations
between the particles. Indeed, spin squeezing can be shown to be
connected to entanglement: If an $N$-qubit state violates the
inequality \cite{sorensen-nature}
\begin{equation}
\frac{\va{J_z}}{\exs{J_x}^2+\exs{J_y}^2}\ge \frac{1}{N},
\label{motherofallspinsqueezinginequalities}
\end{equation}
then the state is entangled, i.e., not fully separable. The degree
of violation can also be used to characterize how strong the
entanglement is \cite{PhysRevLett.86.4431}. It was found in
Ref.~\cite{PhysRevA.68.012101} that if a symmetric state violates
\EQ{motherofallspinsqueezinginequalities} then its two-qubit reduced
density matrix is entangled, that is, the state is two-qubit
entangled. However, in general, the violation of
\EQ{motherofallspinsqueezinginequalities} does not indicate
two-qubit entanglement, as one can find counterexamples
\cite{2008arXiv0806.1048T}.

A typical spin squeezing experiment starts with a fully polarized
sample $\ket{\Psi}_{\rm init}:=\ket{\tfrac{1}{2}}_x^{\otimes N}.$
Such a state saturates \EQ{motherofallspinsqueezinginequalities}.
During the entangling dynamics the large spin decreases only slightly,
while the uncertainty in an orthogonal direction decreases 
considerably.\footnote{For an exhaustive survey of the topic and a list
of references, see the excellent review in
Ref.~\cite{hammerer-2008}.} Thus, the resulting state violates 
\EQ{motherofallspinsqueezinginequalities}.
Entanglement can be generated by direct
interaction between the particles, as in the proposal of
Ref.~\cite{sorensen-nature}. In another scheme, the atomic ensemble
interacts with light \cite{kuzmich1998aqn}. The light is then
measured, projecting the atoms into an entangled state. For an
ensemble strongly polarized into the $x$ direction one has
$\exs{J_x}\approx \tfrac{N}{2}$ and $\exs{J_y}\approx0.$ Thus, the
entanglement condition \EQ{motherofallspinsqueezinginequalities}
can be reformulated as $ \va{J_z} \ge \tfrac{N}{4}.$

\subsubsection{Singlet criterion}
\label{sec_singletcrit} After the discovery of the criterion
Eq.~(\ref{motherofallspinsqueezinginequalities}), it was realized
that other generalized spin squeezing criteria can also be
constructed that detect various types of entangled states and still
need only collective measurements. Such an entanglement condition
was presented in Ref.~\cite{PhysRevA.69.052327}. It contains the
variances of all the three angular momentum components: For
separable states
\begin{equation}
(\Delta J_x)^2+(\Delta J_y)^2+(\Delta J_z)^2 \ge \frac{N}{2}
\label{singletcrit}
\end{equation}
holds. States violating this condition have small angular momentum:
\EQ{singletcrit} detects entanglement in the vicinity of many-qubit
singlet states, which were considered in Sec.~\ref{sectionsinglets}.\footnote{
Note that the criterion \EQ{singletcrit} can straightforwardly generalized for $N$ spin-$j$ 
particles. In this case on the right hand side there must be $Nj$ \cite{PhysRevA.69.052327}.} 
Such states give zero for the left hand side of \EQ{singletcrit}. One
of such states is the chain of two-qubit singlet states. Another example
is the ground state of the isotropic anti-ferromagnetic Heisenberg chain.
Interestingly, all pure product states saturate the inequality.
\EQ{singletcrit} has a good noise tolerance, it detects a singlet state
mixed with white noise as entangled if
$
p_{\rm noise}<\tfrac{2}{3}.
$
Again,  violation of \EQ{singletcrit} does not imply two-qubit
entanglement. Moreover, it can detect PPT entangled states (bound
entangled states) in the thermal states of several spin models
\cite{optimalspsq,2008arXiv0806.1048T}.

In addition, the violation of the criterion gives information about
the number of spins that are unentangled with the rest
\cite{PhysRevA.71.010301,2008arXiv0806.1048T}: Let us consider a
pure state for which the first $M$ qubits are not entangled with
other qubits while the rest of the qubits are entangled with each
other $ \ket{\Psi}=(\otimes_{k=1}^{M} \ket{\psi_k}) \otimes
\ket{\psi}_{M+1,...,N}. $ For such a state, based on the theory of
entanglement detection with uncertainties, one has
\cite{guehne-2004-92}
\begin{equation}
\va{J_x}+\va{J_y}+\va{J_z} \ge \frac{M}{2}. \label{JxyzM}
\end{equation}
If a mixed state $\vr:=\sum_k p_k \ketbra{\Psi_k}$ violates
Eq.~(\ref{JxyzM}) then at least one of the components
$\ketbra{\Psi_k}$ must have $M$ or more spins that are entangled
with other spins. If the left-hand side of Eq.~(\ref{JxyzM}) is
smaller than $\tfrac{1}{2}$ then the state cannot be created by
mixing states that have one or more unentangled spins.

\subsubsection{Inequalities detecting entanglement close to Dicke states}
A third type of inequality for entanglement detection appeared in
Ref.~\cite{DickeEntanglementJOSAB2007}: For separable states
\begin{equation}
\exs{J_x^2}+\exs{J_y^2} \le
\frac{N}{2}\left(\frac{N}{2}+\frac{1}{2}\right)
\label{CritSymDickeState}
\end{equation}
holds. This inequality detects entanglement close to the $N$-qubit
Dicke state with two excitations. It detects such a state as
entangled if for the added white noise $ p_{\rm noise}<\tfrac{1}{N}.
$ Thus, the robustness to noise is decreasing with $N.$ Moreover,
the violation of this inequality implies the presence of two-qubit
entanglement. This can be seen as follows: The average two-qubit
density matrix is
\begin{equation}
\vr_{\rm av2}:=\frac{1}{N(N-1)}\sum_{k\ne l}\vr_{kl}.
\label{rhoav2}
\end{equation}
Here $\vr_{kl}$ represents the two-qubit reduced density matrix
obtained from the original one after tracing out all qubits but
qubits $k$ and $l.$ Eq.~(\ref{CritSymDickeState}) can be re-expressed
with $\vr_{\rm av2}$ as
\begin{equation}
\exs{\sigma_k^{(1)} \sigma_k^{(2)}+\sigma_l^{(1)}
\sigma_l^{(2)}}_{\vr_{\rm av2}}\le 1. \label{CritSymDickeState2}
\end{equation}
Eq.~(\ref{CritSymDickeState2}) can be violated only if $\vr_{\rm
av2}$ is entangled [see Eq.~(\ref{cschwarz})].

Criteria, similar to Eq.~(\ref{CritSymDickeState}) can also be used
to detect genuine multi-qubit entanglement. Such criteria for $N=3$
and $4$ qubits, respectively, are
\begin{eqnarray}
\exs{J_x^2}+\exs{J_y^2} &\le & 2 + \sqrt{5}/2\approx 3.12
\mbox{ for $N=3$},
\label{crit3q}
\\
\ex{J_x^2}+\ex{J_y^2} &\le & \frac{7}{2}+\sqrt{3}\approx 5.23
\mbox{ for $N=4$}.
\label{4qubit}
\end{eqnarray}
Entanglement was detected based on Eq.~(\ref{CritSymDickeState})
around a four-qubit Dicke state with two excitations in a four-qubit
photonic system \cite{DickeExperimentPRL2007}.
Although in such a system the qubits are individually accessible,
using such a criterion was still advantageous since only two
measurement settings (i.e., the
$\{\sigma_x,\sigma_x,\sigma_x,...,\sigma_x\}$ and
$\{\sigma_y,\sigma_y,\sigma_y,...,\sigma_y\}$ settings ) were needed
for the detection. The experiment will be explained
in Section~\ref{sectiondickeexp}.

\subsubsection{Detecting two- and three-qubit entanglement}
Finally, in
Refs.~\cite{PhysRevLett.95.120502,PhysRevLett.95.259901,korbicz:052319}
spin squeezing criteria were constructed detecting two-qubit entanglement
of the  reduced two-qubit state. It was proven that if
\begin{equation}
\left[
\ex{J_x^2}+\frac{N(N-2)}{4}\right]^2
\ge\left[
\ex{J_y^2}+\ex{J_z^2}-\frac{N}{2}\right]^2+(N-1)^2\ex{J_x}^2
 \label{Korbicz}
\end{equation}
is violated then the state is two-qubit entangled. This criterion
can be simplified to
\begin{equation}
1-\frac{4\exs{J_z}^2}{N^2} \ge \frac{4 \va{J_z}}{N}
\label{Korbicz_sym}
\end{equation}
for symmetric states. A symmetric state is two-qubit entangled 
iff it violates Eq.~(\ref{Korbicz_sym}) for some choice of the 
$z$-axis.\footnote{Ref.~\cite{vidal-2006-73} conjectures a 
formula with collective observables for the two-qubit concurrence 
of symmetric states.}
Moreover, criteria detecting genuine
three-qubit entanglement of the W and the GHZ classes, respectively,
were also presented that need the measurement of third order moments
of the collective angular momenta \cite{korbicz:052319}. The
previous ideas have already been applied to investigate the data of
a ion trap experiment (see Section \ref{sectionionexp})
\cite{korbicz:052319}. Such criteria are especially applicable to W
states, as W states maximize two-qubit concurrence among symmetric
states \cite{PhysRevA.62.062314,PhysRevA.62.050302}.

\subsubsection{Optimal spin squeezing inequalities}
After presenting several generalized spin squeezing inequalities the
questions arises: can one find a unifying framework for all these
inequalities? In Ref.~\cite{optimalspsq} this has been done. It
turned out, that the first and second order moments of collective
angular momenta for fully separable states fulfill simple
constraints, namely
\begin{subequations}
\begin{eqnarray}
\exs{J_x^2}+\exs{J_y^2}+\exs{J_z^2} &\le& \frac{N(N+2)}{4},
\label{theorem1a}
\\
\va{J_x}+\va{J_y}+\va{J_z} &\ge& \frac{N}{2}, \label{Jxyzineq_singlet}
\\
\exs{J_k^2}+\exs{J_l^2}-\frac{N}{2} &\le& (N-1)\va{J_m},
\label{Jxyzineq_spsq2}
\\
(N-1)\left[\va{J_k}+\va{J_l}\right] &\ge& \exs{J_m^2}+\frac{N(N-2)}{4}.
\label{Jxyzineq_spsq3} \;\;\;\;\;\;
\end{eqnarray}
\label{Jxyzineq}
\end{subequations}
While Eq.~(\ref{theorem1a}) is
valid for all quantum states, Eqs.~(\ref{Jxyzineq}.b-d) can be
violated. Any state violating them is entangled. It was also proven,
that this set of generalized spin squeezing inequalities is
complete: For certain cases (e.g., $\exs{J_x}=\exs{J_y}=\exs{J_z}=0$
and $N$ is even) or for large $N,$ a state that does not violate
Eqs.~(\ref{Jxyzineq}b-d) cannot be detected by any other
entanglement condition based only on the expectation values of the
first and second moments of $J_k.$

\begin{figure}
\begin{minipage}{0.45\textwidth}
\centerline{ \epsfxsize3.3in \epsffile{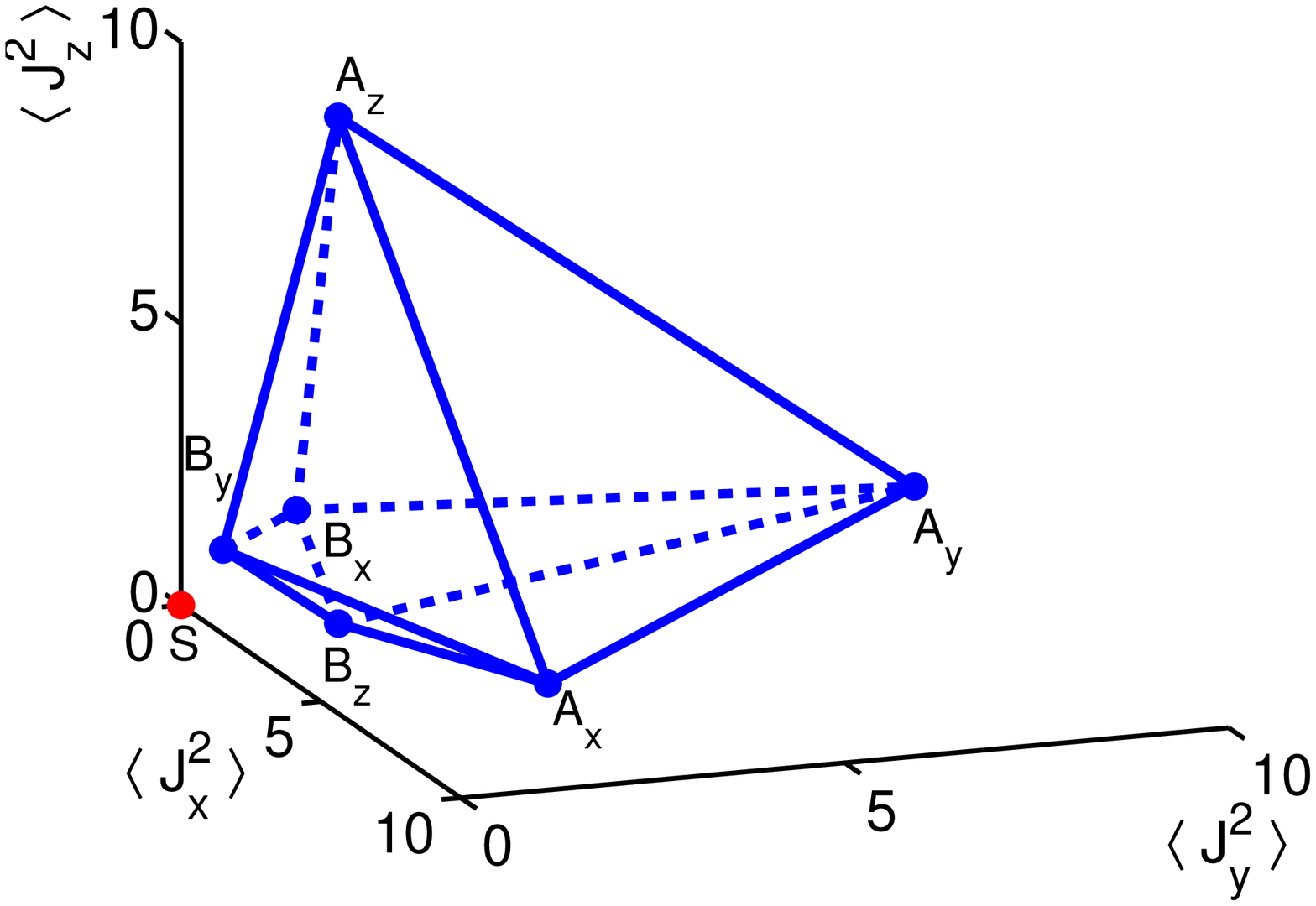}}
\caption{The polytope of separable states corresponding to
Eq.~(\ref{Jxyzineq}) for $N=6$ and for
$\vec{J}=0.$  The point $S$ corresponds to a many body singlet state.
The Figure is taken from Ref.~\cite{optimalspsq}\label{J2xyz}.}
\end{minipage}
\begin{minipage}{0.08\textwidth}
\mbox{ }
\end{minipage}
\begin{minipage}{0.45\textwidth}
\includegraphics[width=8.0cm,clip]{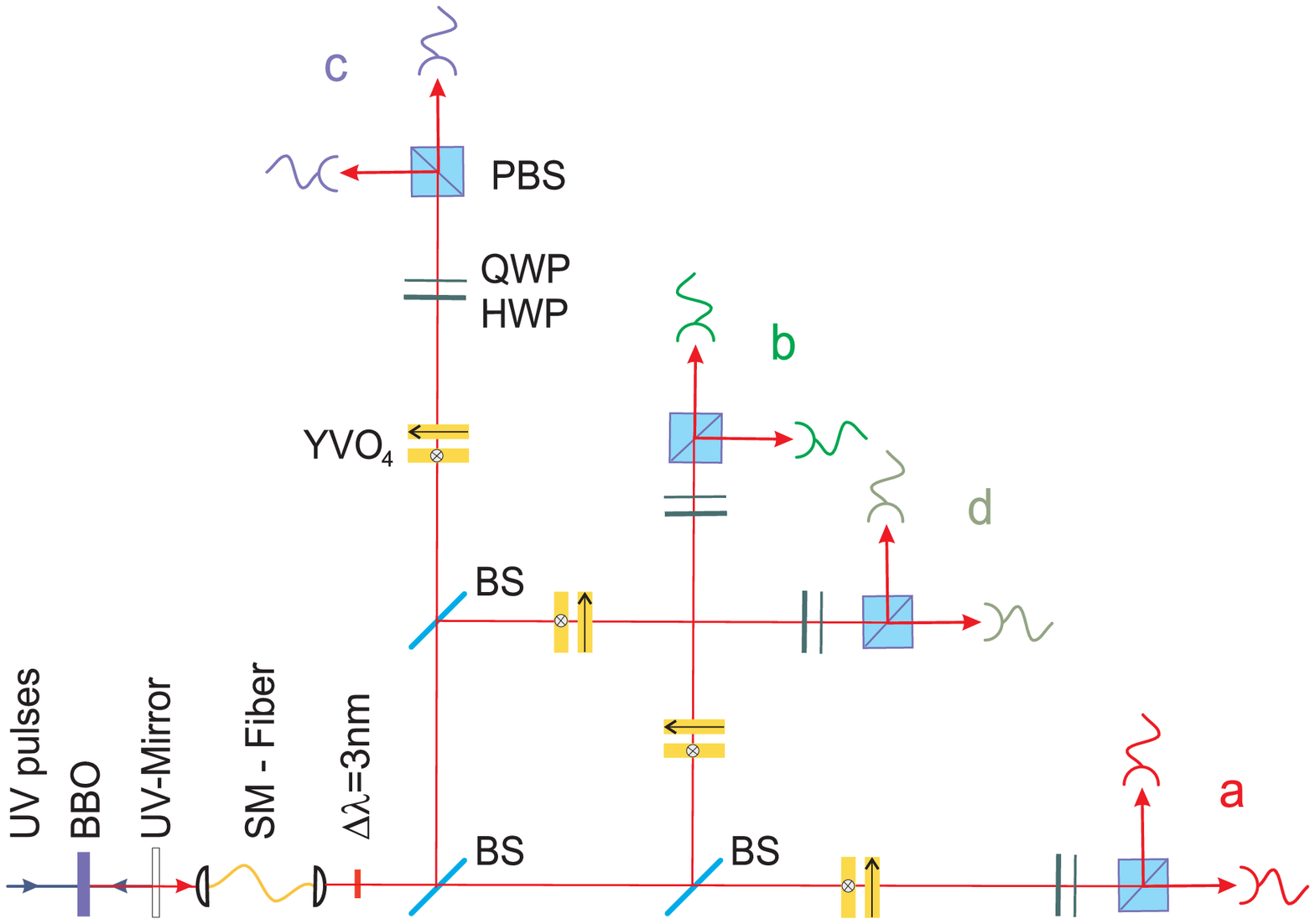}
\caption{Experimental setup for the analysis of the four-photon
polarization-entangled state $\ket{D_{2,4}}.$ The state is
detected after the symmetric distribution of four photons onto the
spatial modes $a$, $b$, $c$ and $d$ via non-polarizing beam
splitters (BS). The Figure is taken from
Ref.~\cite{DickeExperimentPRL2007}.\label{dicke_setup}}
\end{minipage}
\end{figure}

For any value of $\vec{J}:=(\exs{J_x},\exs{J_y},\exs{J_z})$ these
eight inequalities define a polytope in the three-dimensional
$(\exs{J_{x}^2},\exs{J_{y}^2},\exs{J_{z}^2})$-space. Separable
states lie inside this polytope. If one of the inequalities is
violated, then the state is on the outside and is hence entangled.
For the case $\vec{J}=0$ and $N=6$ the polytope is depicted in
Fig.~\ref{J2xyz}. Such a polytope is completely characterized by its
extremal points. They are given by
\begin{align}
A_x &:=\left[ \frac{N^2}{4}-\kappa(\exs{J_y}^2+\exs{J_z}^2),
\frac{N}{4}+\kappa\exs{J_y}^2, \frac{N}{4}+\kappa\exs{J_z}^2
\right], \nonumber
\\
B_x&:= \left[ \exs{J_x}^2+\frac{\exs{J_y}^2+\exs{J_z}^2}{N},
\frac{N}{4}+\kappa \exs{J_y}^2, \frac{N}{4}+\kappa\exs{J_z}^2
\right],
\end{align}
where $\kappa:=(N-1)/N.$ The points $A_{y/z}$ and $B_{y/z}$ can be
obtained from these by permuting the coordinates.

Interestingly, violation of these spin squeezing inequalities does
not necessarily imply entanglement of the reduced two-qubit states.
As all the expectation values of first and second moments of $J_k$
can be evaluated knowing the average two-qubit reduced density
matrix $\vr_{\rm av2}$ defined in Eq.~(\ref{rhoav2}), one could
naively expect that these criteria detect the entanglement of the
reduced two-qubit density matrix. However, somehow surprisingly
these criteria are also able to detect entangled states for which
the reduced states are not entangled.

This phenomenon is related to the representability problem
\cite{RevModPhys.35.668}, i.e., to the problem of finding
multipartite quantum states that have a given set of states as
reduced states. For a given $\vr_{\rm
av2},$ one has to answer the question whether there is a
separable $N-$qubit state that
has $\vr_{\rm av2}$ as the average two-qubit reduced state? If the
answer is no then the system is entangled. Interestingly, it is also
possible that $\vr_{\rm av2}$ is separable, however, there is not an
$N$-qubit separable quantum state that has it as a reduced state. In
this case we can conclude that the system is an entangled state even
if $\vr_{\rm av2}$ is separable. If $\vr_{\rm av2}$ is separable and
of the form
\begin{equation}
\vr_{\rm av2}=\sum_k p_k \vr_k \otimes \vr_k \label{av2}
\end{equation}
then there always can be found an $N$-qubit separable state that has
this state as its reduced state $ \vr_{N}=\sum_k p_k \vr_l^{\otimes
N}. $ Note that separable states in the symmetric subspace can
always be written in the form of Eq.~(\ref{av2}). On the other hand,
if $\vr_{\rm av2}$ is not symmetric then it is possible that it is
separable, however, there is not an $N-$qubit symmetric separable
state that has it as its reduced two-qubit state. One can write down
the condition Eq.~(\ref{Jxyzineq_singlet}) with the operator
expectation values of $\vr_{\rm av2}$ as
\begin{equation}
\exs{\sigma_x \otimes \sigma_x + \sigma_y \otimes \sigma_y +
\sigma_z \otimes \sigma_z}_{\vr_{\rm av2}}\ge \frac{N}{N-1}
\sum_{k=x,y,z}\exs{\sigma_k\otimes\mathbbm{1}}_{\vr_{\rm
av2}}^2-\frac{1}{N-1}. \label{xxyyzz}
\end{equation}
If this condition is violated then Eq.~(\ref{Jxyzineq_singlet})
detects the state as entangled. However, Eq.~(\ref{xxyyzz}) is not a
condition for detecting the entanglement of $\vr_{\rm av2}.$ That
is, there are separable $\vr_{\rm av2}$ that violate
it.\footnote{For example, for the $T=0$ ground state of the
Hamiltonian $H=J_x^2+J_y^2+J_z^2$ a simple calculation shows that
$\exs{\sigma_x \otimes \sigma_x + \sigma_y \otimes \sigma_y +
\sigma_z \otimes \sigma_z}_{\vr_{\rm av2}}=-\tfrac{3}{4(N+1)}$ and
$\exs{\sigma_k\otimes\mathbbm{1}}_{\vr_{\rm av2}}=0$, but $\vr_{\rm
av2}$  can be shown to be unentangled for $N\ge 4$ for even $N$
\cite{PhysRevA.71.010301}.}

\subsection{Experimental observation of a four-qubit symmetric Dicke state}
\label{sectiondickeexp}

In this Section we explain an experiment aiming to create a
four-photon symmetric Dicke state with two excitations,
$\ket{D_{2,4}}$ in a photonic system \cite{DickeExperimentPRL2007}.
On the one hand, the experiment is interesting since the state $\ket{D_{2,4}}$ is 
highly entangled while the experimental setup is quite simple. On the other
hand, the generalized spin squeezing criteria can be used for the
detection of entanglement with few measurement settings.

The Dicke state $\ket{D_{2,4}}$ has the form
\begin{eqnarray}
\notag
\ket{D_{2,4}}=
\frac{1}{\sqrt{6}}(\ket{HHVV}+\ket{HVHV}+\ket{VHHV}+
\ket{HVVH}+\ket{VHVH}+\ket{VVHH}), \label{eq_W2}
\end{eqnarray}
where $\ket{H}$ and $\ket{V}$ denote linear horizontal ($H$) and
vertical ($V$) polarization of a photon in the four spatial modes,
denoted by $a$, $b$, $c$, and $d.$ This is the symmetrization of a
product state with two $H$ and two $V$ photons.

The setup can be seen in Fig.~\ref{dicke_setup}. The four
photons are created by the second order emission of collinear
type-II parametric down-conversion. The BBO crystal is used for the
down-conversion, while the UV-mirrors separate the down-converted
photons from the pump beam. As a result, in the single-mode fiber
four photons appear, two in state $\ket{V}$, the other two in state
$\ket{H}.$ Now, the four photons must be distributed into four
spatial modes. This is done by beam splitters. Birefringent effects
of these non-polarizing beam splitters is compensated by pairs of
perpendicularly oriented YVO$_4$ crystals in all the four modes.
Note that the Dicke state can also be prepared with a different setup,
which flexibly allows to create an entire family of multi-photon states
\cite{wieczorek:010503}.

In the experiment, first the density matrix was determined by
measuring all correlations of the type
$\exs{\sigma_k\otimes\sigma_l\otimes\sigma_m\otimes\sigma_n}$ where
$k,l,m,n\in\{0,x,y,z\}.$ For that $3^4=81$ measurement settings are
needed. This makes it possible to compute the fidelity of the
prepared state: $F_{\rm exp} = 0.844 \pm 0.008.$ Note that the
fidelity is high since the generation of the state was quite
straightforward with parametric down-conversion and post-selection
and no additional gates were needed. In comparison, for creating
a cluster state, it is necessary to realize a phase gate. The
larger complexity of the quantum circuit results in a
decrease of the fidelity of the prepared state
\cite{PhysRevLett.95.210502}.

The fidelity was higher than the bound for biseparable states, $0.6$
[see the witness Eq.~(\ref{W_Dicke})]. Thus, the fidelity computed
from the density matrix already signals the presence of genuine
multipartite entanglement. Instead of complete tomography, $21$
settings would also be enough to get the fidelity and detect genuine
multipartite entanglement \cite{DickeExperimentPRL2007}. However, in
the experiment a more efficient criterion, namely Eq.~(\ref{4qubit})
was also tested, which needed only two settings. For the left hand
side of Eq.~(\ref{4qubit}) a value of $5.58 \pm 0.02$ was obtained
that clearly demonstrated genuine multipartite entanglement.

In the second part of the experiment, one of the photons was
measured. This leads to various symmetric states, in
particular states of the type
\begin{equation}
\ket{\psi}=\alpha \ket{W_3}+\beta \ket{\overline{W}_3},
\end{equation}
where $\ket{W_3}$ is the three-qubit W state and
$\ket{\overline{W}_3}$ is a state obtained from $\ket{W_3}$ by
inverting all the qubits. In the experiment, the criterion
Eq.~(\ref{crit3q}) was used to detect the genuine multipartite
entanglement of these states. After the loss of one of the photons,
from the $\ket{D_{2,4}}$ state we get a mixture of $\ket{W_3}$ and
$\ket{\overline{W}_3}$ states. This state is also genuine
multipartite entangled, that indicates the persistency of the
entanglement of the $\ket{D_{2,4}}$ state. Criterion
Eq.~(\ref{crit3q}) was then used to detect entanglement of this
state.

\subsection{The Hamiltonian as a witness}
\label{sectionenergywit}

In this and the following Section we discuss how entanglement can be
detected by measuring fundamental quantities of a physical system,
e.g., the energy of a spin chain \cite{PhysRevA.71.010301,
brukner-2004-, PhysRevA.70.062113,wu2005eoa}. This allows to connect
thermodynamical quantities of condensed  matter systems to their
entanglement properties \cite{vedralnature,amico:517}. We will also
discuss how these ideas can be extended to detect multipartite
entanglement and how other fundamental observables beside energy can
also be used. It should be noted, however, that in these approaches
the conclusion that a given state was entangled very often depends
on the assumption that the considered spin model indeed describes
the physical system.

There are two basic approaches in entanglement detection: (i) One
would like to detect entanglement in the vicinity of a given quantum
state. Then some entanglement condition, for instance an entanglement 
witness, is designed for this aim. (ii) It is also possible to proceed 
in the
opposite way and look for a fundamental observable of a physical
system that is easy to measure. Then one can examine what quantum
states can easily be detected by such an observable. When following
the second path, one constructs witness operators of the form \be
\WW_O:=O-\inf_{\Psi \in S} \big[ \exs{\Psi|O|\Psi}\big], \label{WO}
\ee where $S$ is the set of separable states, "$\inf$" denotes
infimum, and $O$ is a fundamental quantum operator of a spin system
that is easy to measure. Note that in the general case $\inf_{\Psi
\in S} \exs{\Psi|O|\Psi}$ is difficult, if not impossible, to
compute.

A natural observable for the second approach is the Hamiltonian of
the system. First of all, it is typically easy to measure. In some
systems it can be measured directly. In other systems the
expectation value of the Hamiltonian can be measured indirectly by
two-body correlation measurements since the Hamiltonian is typically
the sum of few two-body correlation terms. Second, it turns out that
it is simple to find the minimum for separable states for many spin
model Hamiltonians. This makes it possible to investigate
entanglement in systems in thermal equilibrium, which appear many
areas of physics. In thermal equilibrium the state of the system is
given by \be \vr_T\propto\exp(-H/k_BT),\label{rhoT} \ee where $T$ is
the temperature and $k_B$ is the Boltzmann constant. For simplicity
we will set $k_B=1$. Using Eq.~(\ref{WO}) a temperature bound,
$T_E$, can be found such that when $T<T_E$ then the system is shown
to be entangled.

\begin{figure}
\centerline{\epsfxsize=1.7in\epsffile{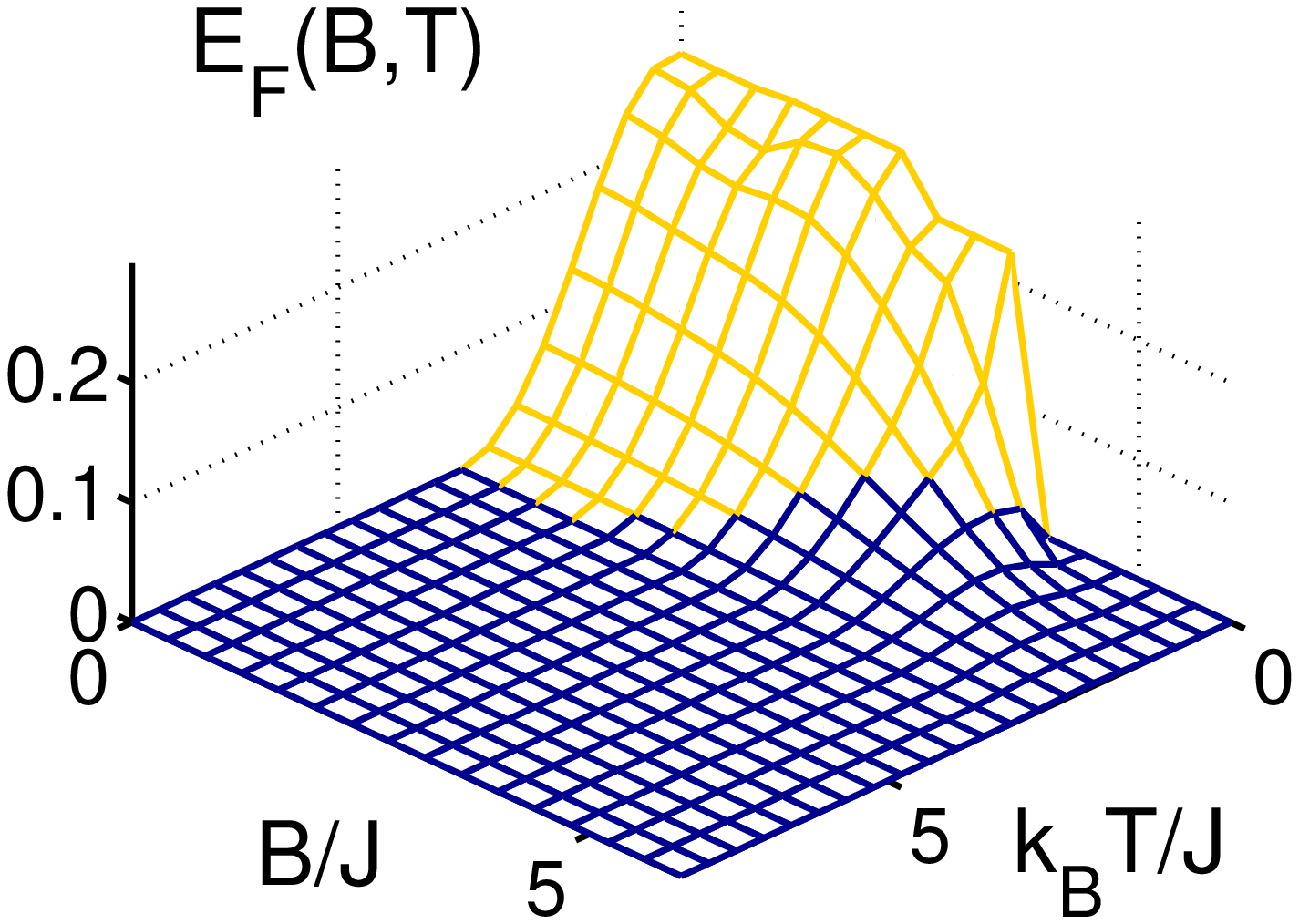}
\epsfxsize=1.8in\epsffile{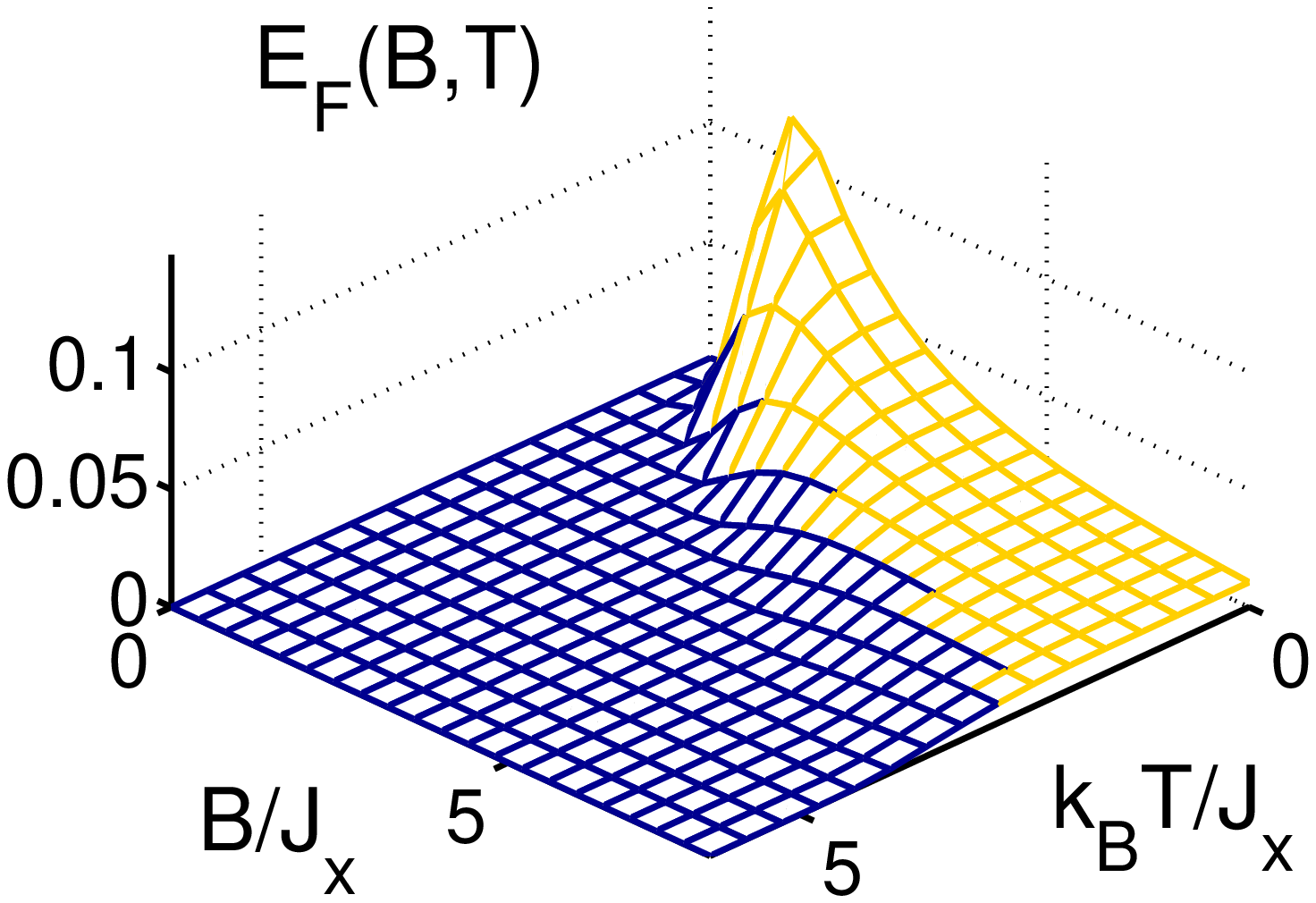}}
\hskip4.0cm(a)\hskip4.0cm (b) \caption{(a) Heisenberg chain of $8$
spins. Nearest-neighbor entanglement (quantified by the entanglement
of formation) as a function of magnetic field $B$ and temperature
$T$. (b) The same for an Ising spin chain. Here $k_B$ is the
Boltzmann constant, $J$ and $J_x$ are coupling constants. Light
color indicates the region where entanglement is detected by
Hamiltonian as a witness. The Figure is taken from
Ref.~\cite{PhysRevA.70.062113}.} \label{fig_HeisenbergIsing}
\end{figure}

To give a simple example, consider the one-dimensional Heisenberg
Hamiltonian on $N$ spins with periodic boundary conditions,
\be
H_H = \sum_k
\sigma_x^{(k)} \sigma_x^{(k+1)} +
\sigma_y^{(k)} \sigma_y^{(k+1)} +
\sigma_z^{(k)} \sigma_z^{(k+1)}.
\ee
With an argument as in  Eq.~(\ref{cschwarz}) one can directly see that
for a fully separable state $\mean{H_H} \geq -N$ has to hold. As the
ground state energy for large $N$ for the Heisenberg model equals
$E_0 = -[4\ln(2) -1]N \approx -1.773 N$ the thermal state at low
temperatures (namely $k_B T \leq 3.18$) has to be entangled 
\cite{PhysRevA.73.052319}.

Similar  ideas have been applied for the  detection of entanglement
in various spin models such as the XY-, Heisenberg- and Ising spin
lattices with an external field in different various dimensions
\cite{PhysRevA.71.010301,brukner-2004-,
PhysRevA.70.062113,1367-2630-8-8-140,
hide-2008,hide-2007-76,1367-2630-7-1-229,PhysRevA.73.052319,nakata-2008,
markham-2008-81}. An example is given in
Fig.~\ref{fig_HeisenbergIsing}. The figures show the entanglement of
formation for the reduced two-qubit state as a function of the
external field and the temperature. Light color indicates that the
energy of the system is smaller than the minimum for separable
states thus the Hamiltonian as a witness detects it as entangled.
Note that if the entanglement is not very small, the Hamiltonian 
as a witness detects the thermal state as entangled. Thus, in such 
systems measurement of a single observable, i.e., the energy, seems 
to be sufficient for entanglement detection. Moreover, for systems 
in thermal equilibrium it is possible to find witnesses that are 
different from the Hamiltonian and detect all entangled thermal states
\cite{wu2005eoa}.

Beside entanglement in general, one can also use similar ideas for
detecting various forms of multipartite entanglement. As explained
in Section \ref{sectiongeneralentanglementclasses} 
[Eq.~(\ref{producibilitydefinition})] one can ask whether
multipartite entanglement is necessary to form a given state
\cite{seevinck-2007,1367-2630-7-1-229,PhysRevA.73.052319}. 
Then, a state $\ket{\psi}$ contains only $k$-party
entanglement (or is $k$-producible) if one can write the
state $\ket{\psi}$ as a tensor product
$
\ket{\psi}=\ket{\phi_1}\otimes\ket{\phi_2}\otimes ...\otimes
\ket{\phi_m},
$
where the states $\ket{\phi_i}$ are states on
maximally $k$-qubits. In this definition, a 
two-producible state does not contain any 
multipartite entanglement, since it suffices to
generate the two-qubit states $\ket{\phi_i}$ to arrive at the state
$\ket{\psi}.$ In addition, one says that a state {contains
genuine $k$-party entanglement} if it is not producible by
$(k-1)$-party entanglement. This definition can be extended 
to mixed states via convex combinations.

For spin chains of {\it macroscopic} size, it is in general very
difficult to prove that the total state is genuine $N$-partite
entangled via energy measurements. This is due to the fact that the
notion of genuine $N$-partite entanglement is extremely sensitive to
the properties of a single qubit. Indeed, in order to prove genuine
multipartite entanglement, one has to exclude the possibility, that
one single qubit can be separated from the remaining $N-1$ qubits.
However, multipartite entanglement in the {\it reduced} states of
small numbers of qubits can easily be detected. Moreover, if the
reduced state is multipartite entangled then the state is not
two-producible. Based on these ideas, in
Refs.~\cite{1367-2630-7-1-229,PhysRevA.73.052319,APB06,richertguehne} 
energy bounds for two-, three- and four-producibility are presented for
various spin models such as the Heisenberg- and the XY model, and
also for two-dimensional models with frustration.

\subsection{Entanglement detection with susceptibility
measurements}

\label{Sec_suscept}

Finally, let us consider the magnetic susceptibility as another
thermodynamic quantity. In particular, the generalized spin
squeezing inequality \EQ{singletcrit} can be reformulated using
susceptibility measurements \cite{SusceptWiesnak2005}. Let us
consider a system with a Hamiltonian $H$ and add a magnetic
interaction $H_{I} := \sum_{l=x,y,z} B_l J_l$ to it, where $\vec{B}$
is the strength of the magnetic field. It has been shown in
Ref.~\cite{SusceptWiesnak2005} that if we define the magnetic
susceptibility along the directions $l=x,y,z$ as \be\chi_l :=
\left(\frac{\partial \exs{J_l}} {\partial
B_l}\right)\bigg\vert_{\vec{B}=0}, \ee and if $H$ commutes with
$J_k$ for $k=x,y,z$ then for separable states
\begin{equation}
\chi_x + \chi_y + \chi_z \geq \frac{N}{2kT}. \label{susc_crit}
\end{equation}
Here it is assumed that the system is in thermal equilibrium. As one
can directly check [using $\chi_l=\tfrac{1}{kT}\va{J_l}$]
\EQ{susc_crit} is equivalent to \EQ{singletcrit}, giving the spin
squeezing inequality a new physical interpretation.

There have been experiments for detecting entanglement with
susceptibility measurements in solid state systems. In
Ref.~\cite{brukner:012110}, it is explained that the magnetic
susceptibility of copper nitrate measured in 1963 shows the presence
of entanglement. Recently, experiments in low-dimensional spin
systems have been carried out with pyroborate MgMnB$_2$O$_5$ and the
warwickite MgTiOBO$_3$, systems with spin 5/2 and 1/2, respectively
\cite{rappoport:054422}.

Eq.~(\ref{susc_crit}) has also been used for looking for
entanglement in small molecular spin-clusters theoretically.
Ref.~\cite{bose:022314} determined the bound temperature below which
the system is entangled. The four spins interact via
nearest-neighbor, diagonal, and four-spin Heisenberg interactions.
Ref.~\cite{vertesi:134404} considered
thermal entanglement in the nanotubular system Na$_2$V$_3$O$_7$ and
determined also the bound temperature for entanglement.

\subsection{Entanglement detection in optical lattices}
\label{optlatt}

Optical lattices of cold atoms are important examples for
large-scale quantum information processing.\footnote{For a review on
quantum information processing in optical lattices of cold atoms see
Refs.~\cite{treutlein2006qip,zoller-2004}.} In particular, in a
three-dimensional lattice of $10^5$ two-state bosonic atoms an array
of one-dimensional cluster states has been generated
\cite{mandel2003ccm}. The entangling-disentangling dynamics, which
created the cluster state and then restored the initial product
state, was observed through the appearance and disappearance of the
interference fringes. In this experiment the lattice sites could not
be addressed individually, thus entanglement detection in such
systems is a relevant example of entanglement detection with
collective measurements.

The simplest model of such a one-dimensional array of trapped
two-state bosonic atoms is the following. Let $a_k$ and $b_k$ be the
destruction operators corresponding to the two species at site $k.$
(Here, species "a" and "b" mean atoms in states $\ket{a}$ and  $\ket{b},$ respectively.) 
A typical Hamiltonian for this system is \cite{1367-2630-5-1-376}
\bea H_{\rm lattice} &:=&
 -J_a \sum_k \big(a_ka_{k+1}^\dagger+a_k^\dagger a_{k+1}\big)
 -J_b \sum_k \big(b_kb_{k+1}^\dagger+b_k^\dagger b_{k+1}\big)
 -\Omega \sum_k \big(a_kb_k^\dagger + a_k ^\dagger b_k\big)\nonumber\\&+&
 \tfrac{1}{2}U_{aa} a_k^\dagger a_k^\dagger a_k a_k
 +\tfrac{1}{2}U_{bb} b_k^\dagger b_k^\dagger b_k b_k
 +U_{ab}a_k a_k^\dagger b_k b_k^\dagger,
 \label{Hlatt}
\eea where $J_k$ are the tunnel couplings between the sites,
$\Omega$ is the Rabi frequency for the transition
$\ket{a}\rightarrow\ket{b},$ and $U_k$ are the on-site interactions
between atoms. The term $a_k a_{k+1}^\dagger+a_k^\dagger a_{k+1}$
denotes the tunneling of an atom in state $\ket{a}$ between sites
$k$ and $k+1.$ Moreover, $a_kb_k^\dagger+a_k ^\dagger b_k$ denotes a
process in which an atom at site $k$ is turning from state $\ket{a}$
to state $\ket{b}$ and vice versa. This can be the result of
manipulating atoms with a laser pulse. The sign of the on-site
interactions $U_a$ and $U_b$ between atoms of the same specie
control whether atoms favor to be at the same site or they tend to
be distributed equally over the lattice. The sign of $U_{ab}$
determines whether atoms with different internal states favor to be
at the same site.

In the realization of Ref.~\cite{mandel2003ccm}, the atoms cannot be
individually addressed. In the experiment, after producing the
desired state, one lets the cloud expand to many times of its
original size. Atoms with a large momentum can get further from
their original position than atoms with a small momentum. Thus, from
the distribution of the particles after the expansion, one can
obtain information on the distribution in momentum space before the
expansion. Knowing the distribution in the momentum space, one can
obtain the distribution in real space by a Fourier transform. This
way, in principle, one can obtain for the prepared state such
quantities as \be \exs{Q_x^{a}}:=\sum_m \exs{a_m^\dagger a_{m+x}}
\mbox{  and } \exs{Q_x^{ab}}:=\sum_m \exs{a_m^\dagger a_{m+x}
b_{m+x}^\dagger b_m}. \ee

In one of the realizations of quantum information processing, most
of the lattice sites have a single atom. The qubit is encoded in the
internal states of the atoms. Schemes can aim to detect specific
states in the experiment:

\begin{enumerate}

\item {\it Cluster states.} In a lattice of such two-state atoms, typically the atoms can
interact with the nearest neighbors in a one-dimensional chain.
While in some systems such an interaction is part of the Hamiltonian
\cite{brennen1999qlg}, in the experiment of
Ref.~\cite{mandel2003ccm} it was efficiently engineered by applying
two different optical lattices for trapping the two species
\cite{jaksch1999eav}. Thus, cluster states can be realized, since
they could be created from product states with a nearest-neighbor
Ising dynamics. However, first and second moments of the collective
observables $J_k$ are not sufficient for detecting the entanglement
of cluster states \cite{PhysRevA.69.052327}. One proposal in this
case  is to use dynamics before the detection of collective observables
\cite{PhysRevA.69.052327}. In this way, a simple scheme can be
constructed that can be used to study the decohering cluster state
and find a good lower bound for its entanglement lifetime. However,
since now the dynamics is part of the entanglement detection, its
accuracy must be included (or assumed) in the calculations.

\item {\it Singlet states.} Instead of cluster states, two-particle singlets can be
generated in a lattice of double-wells \cite{rey2007pad}. Such
states can be readily detected by the criterion
Eq.~(\ref{Jxyzineq_singlet}), based on second order moments, and its
generalization for multi-qudit systems \cite{PhysRevA.69.052327}.
This is also true for various proposals for realizing multi-particle
singlet states as ground states of Heisenberg spin lattices with
spins $\tfrac{1}{2}$ or higher \cite{eckert2007qnd}.

\end{enumerate}
Entanglement detections schemes not designed for particular states
are also available. Such proposals are as follows:

\begin{enumerate}

\item In Ref.~\cite{brennen-2003-3}, an approach is described using two copies of a multi-qubit pure
state to compute a multi-particle entanglement measure based on the
single-qubit purities. The two copies can be in two one-dimensional
optical lattices that interact with each other during the
entanglement detection.

\item In Ref.~\cite{alves:110501}, a method is suggested that works if
one possesses two copies of the mixed multi-qubit state of the
lattice (see also Section \ref{bovinoexp}). It is based on the fact that for
separable states the purity of reduced states is larger than the
purity of the state. If for a quantum state this is not the case,
then it is entangled.

\item In Ref.~\cite{vollbrecht-2007-98}, it is shown how to obtain a
bound for the fidelity of the two-qubit states with respect to some
maximally entangled state, measuring correlations between the
momentum distributions, that is, measuring $\exs{Q_x^{kl}}$ for
$k,l=a,b.$ Knowing the fidelity with respect to a maximally
entangled state allows to obtain a bound on the two-qubit
entanglement of formation.

\end{enumerate}

In the other realization of quantum information processing one does
not employ the internal states for storing information but the
presence or absence of atoms at the lattice sites encodes the single
qubits \cite{parades2004tgg}. That is $\ket{0}$ corresponds to an
empty site, while $\ket{1}$ describes a site with a single atom. The
system can be described by the Bose-Hubbard model, that is,
Eq.~(\ref{Hlatt}) with only specie "a". In the limit of very strong
repulsive interaction between the particles, there is at most a
single particle per site. Entanglement conditions for such systems
are the following:

\begin{enumerate}

\item In Ref.~\cite{PhysRevA.71.010301}, a simple entanglement
condition is described that needs the measurement of two-site
correlations of the type $\exs{a_ka_{k+1}^\dagger+a_k^\dagger
a_{k+1}}$, where $a_k$ is the destruction operator corresponding to
the site $k.$ This criterion detects entanglement close to the
ground state of the system.

\item In Ref.~\cite{vollbrecht-2007-98}, an approach is presented
that exploits particle number superselection rules. These prohibit
to have single lattice site states that are the superposition of
states with different particle numbers. Under such conditions, all
states that can be prepared locally are diagonal in the basis of
$\ket{0}$ and $\ket{1}.$ Thus, if a state is nondiagonal, then it
implies the presence of a resource, which is, however, less powerful
than entanglement defined independently from superselection rules.
Nevertheless, the state detected cannot be prepared locally with
atoms, thus if the state is nondiagonal, the presence of quantum
correlations are demonstrated. Any non-flat momentum distribution
indicates entanglement in this sense, which has been observed in
several experiments \cite{vollbrecht-2007-98}. Moreover, knowing
$\exs{Q_x}$ makes it possible to obtain a lower bound on the
entanglement of formation of the two-qubit reduced state.

\end{enumerate}

\section{Conclusion}

As we have seen, many methods for the detection of
entanglement have been proposed, ranging from Bell
inequalities to spin squeezing inequalities. Each of
them has its advantages and disadvantages and the
question which of them is preferable depends on the
given experimental implementation. But one has
always to be careful that the used entanglement 
detection scheme really detects the desired type 
of entanglement. Furthermore, possible hidden assumptions 
in an entanglement verification procedure have to be taken into account.

For the future, it can be foreseen that due to the progress in the
experimental techniques more and more qubits can be entangled.
Especially in ion traps and in photonic experiments using
hyper-entanglement significant progress can be expected. In these
experiments the creation of genuine multipartite entanglement will
probably remain the main goal, as this ensures that the experiment
presents something qualitatively new compared to previous ones.
However, it will become more and more important to design
entanglement verification schemes that require less experimental
effort, as measurements become more demanding.

On the other hand, as such systems become larger, it can be expected
that in many of these experiments the qubits will not be
individually accessible anymore. Thus, entanglement detection
schemes based only on collective measurements will be needed. In
such experiments, the detection of genuine multipartite entanglement
of the total state is not in all cases realistic. However, as we
have seen, one can aim to investigate how many qubits are entangled,
or to quantify the entanglement in the state.

In this way, quantum information science and in particular
entanglement theory can play a crucial role in the technological
development of quantum control and quantum engineering. Finally,
quantum control over larger and larger systems might also help
to answer fundamental questions concerning quantum theory, such
as the appearance of a classical macroscopic world based on a
quantum microworld.

\vspace{0.3cm}
\noindent
{\bf Acknowledgements}
\vspace{0.3cm}

This work has benefitted a lot from discussions and collaborations
with many different people in the past. Therefore, we would like
to thank
A.~Ac\'{\i}n,
R.~Blatt,
H.J.~Briegel,
D.~Bru{\ss},
A.~Cabello,
J.I.~Cirac,
M.~Curty,
W.~D\"ur,
J.~Eisert,
S.~van~Enk,
W.-B.~Gao,
G.~Giedke,
J.J.~Garc\'{\i}a-Ripoll,
O.~Gittsovich,
H.~H{\"a}ffner,
K.~Hammerer,
P.~Hyllus,
N.~Kiesel,
M.~Kleinmann,
C.~Knapp,
T.~Konrad,
B.~Kraus,
S.~Iblisdir,
B.~Jungnitsch,
M.~Lewenstein,
C.-Y.~Lu,
N.~L\"utkenhaus,
M.W.~Mitchell,
T.~Moroder,
M.~Murao,
S.~Niekamp,
J.-W.~Pan,
M.~Piani,
D.~Porras,
M.~Reimpell,
J.~Richert, 
C.~Roos,
A.~Sanpera,
C.~Schmidt,
M.~Seevinck,
C.~Simon,
E.~Solano,
H.~Weinfurter,
R.~Werner,
W.~Wieczorek, 
and 
M.M.~Wolf
for such discussions. 
We further thank
R.~Augusiak,
N.~Brunner, 
P.~Hyllus, 
M.~Seevinck,
W.~Wieczorek
and many other colleagues
for comments on the manuscript.
Especially, we
would like to express our thanks to Mohamed Bourennane, not only
for discussions about the subject, but also for motivating us to
write this review.

This work has been supported by the FWF (START Prize), the
EU (SCALA, OLAQUI, QICS), the National Research Fund of Hungary
OTKA (Contract No. T049234), the Hungarian Academy of Sciences
(J\'anos Bolyai Programme) and  the Spanish MEC (Ramon y Cajal Programme,
Consolider-Ingenio 2010 project ''QOIT'').


\bibliographystyle{apsrev}

\bibliography{reviewlib}

\end{document}